\documentclass[aps,superscriptaddress,nofootinbib,floatfix,notitlepage]{revtex4-1}
\usepackage[utf8x]{inputenc}
\pdfoutput=1
\usepackage{graphicx}
\usepackage{hyperref}
\usepackage{gensymb}
\usepackage{bm,physics}
\usepackage{comment}
\usepackage{wasysym}
\usepackage{float}
\usepackage{multirow}
\usepackage{tablefootnote}

\newcommand{\MEANMOD}[1]{\left \langle #1 \right \rangle}

\newcommand{\be}{\begin{equation}}
\newcommand{\ee}{\end{equation}}

\begin{document}

\title{Global analysis of charm mixing parameters and determination of the CKM angle \texorpdfstring{$\mathbf{\gamma}$}{gamma}}

\author{F. Betti}
\affiliation{School of Physics and Astronomy, University of Edinburgh, \\ 
James Clerk Maxwell Building, Peter Guthrie Tait Road, \\ 
Edinburgh, EH9 3FD, United Kingdom}
\author{M.~Bona}
\thanks{Member of the UTfit Collaboration}
\affiliation{Department of Physics and Astronomy, Queen Mary University of London, \\  London, United Kingdom}
\author{M.~Ciuchini}
\thanks{Member of the UTfit Collaboration}
\affiliation{INFN,  Sezione di Roma Tre, Via della Vasca Navale 84, I-00146 Roma, Italy}
\author{D.~Derkach}
\thanks{Member of the UTfit Collaboration}
\affiliation{HSE University, Myasnitskaya ulitsa, 20, 101000 Moscow, Russia}
\author{R.~Di Palma}
\affiliation{INFN,  Sezione di Roma Tre, Via della Vasca Navale 84, I-00146 Roma, Italy}
 \affiliation{Dipartimento di Matematica e Fisica, Universit{\`a} di Roma Tre, Via della Vasca Navale 84, 
   I-00146 Roma, Italy} 
\author{A.L.~Kagan}
\affiliation{Department of Physics, University of Cincinnati, Cincinnati, Ohio 45221,USA}
\author{V.~Lubicz}
\thanks{Member of the UTfit Collaboration}
\affiliation{INFN,  Sezione di Roma Tre, Via della Vasca Navale 84, I-00146 Roma, Italy}
 \affiliation{Dipartimento di Matematica e Fisica, Universit{\`a} di Roma Tre, Via della Vasca Navale 84, 
   I-00146 Roma, Italy} 
\author{G.~Martinelli}
\thanks{Member of the UTfit Collaboration}
\affiliation{INFN, Sezione di Roma, Piazzale A. Moro 2, I-00185 Roma, Italy}
\author{M.~Pierini}
\thanks{Member of the UTfit Collaboration}
\affiliation{CERN, CH-1211 Geneva 23, Switzerland}
\author{L.~Silvestrini}
\thanks{Member of the UTfit Collaboration}
\affiliation{INFN, Sezione di Roma, Piazzale A. Moro 2, I-00185 Roma, Italy}
\author{C.~Tarantino}
\thanks{Member of the UTfit Collaboration}
\affiliation{INFN,  Sezione di Roma Tre, Via della Vasca Navale 84, I-00146 Roma, Italy}
\affiliation{Dipartimento di Matematica e Fisica, Universit{\`a} di Roma Tre, Via della Vasca Navale 84,
  I-00146 Roma, Italy} 
\author{V.~Vagnoni} 
\thanks{Member of the UTfit Collaboration}
\affiliation{INFN, Sezione di Bologna,  Via Irnerio 46, I-40126 Bologna, Italy}

\begin{abstract}
  We present an updated global analysis of beauty decays sensitive to the angle $\gamma$ of the Cabibbo-Kobayashi-Maskawa matrix and of $D$-meson mixing data in the framework of approximate
  universality, in which CP violation in $D-\overline{D}$ mixing is described in terms of two universal weak phases corresponding to dispersive and absorptive contributions. We extract  the fundamental theoretical parameters
  determining absorptive and dispersive contributions to $D$ meson mixing and CP
  violation, together with the angle $\gamma$. The results for the charm mixing parameters are $x_{12} \simeq x = (0.401 \pm 0.043)\%$ and $y_{12} \simeq y = (0.610 \pm 0.017)\%$, while the two CP-violating phases are given by $\phi_2^M = (0.13 \pm 0.70)\degree$ and $\phi_2^{\Gamma} = (2.1 \pm 1.6)\degree$. The angle $\gamma$ is found to be $\gamma = (65.7 \pm 2.5)\degree$, in excellent agreement with the indirect determination from the Unitarity Triangle analysis.
\end{abstract}
 
\maketitle

\section{Introduction}
Charm mixing occurs through Flavour Changing Neutral Currents (FCNC) that are absent at the tree-level in the Standard Model (SM) and suppressed by the Glashow-Iliopoulous-Maiani (GIM) mechanism \cite{Glashow:1970gm} and by the hierarchical structure of the Cabibbo-Kobayashi-Maskawa (CKM) matrix \cite{Cabibbo:1963yz,Kobayashi:1973fv}. Thus, new physics (NP) effects  due to heavy particles may appear in the neutral $D$ system \cite{Grossman:2006jg, Ciuchini:2007cw, Golowich:2007ka, Fajfer:2007dy, UTfit:2007eik, Artuso:2008vf, Brod:2011re, Pirtskhalava:2011va, Bhattacharya:2012ah, Cheng:2012wr, Franco:2012ck, Brod:2012ud, Giudice:2012qq, Feldmann:2012js, Keren-Zur:2012buf, Grossman:2019xcj, Dery:2019ysp, Buras:2021rdg, Schacht:2021jaz, Schacht:2022kuj} through observables such as the amount of CP violation.
The latter could be used as an interesting benchmark of the SM since it is suppressed by the fourth power of the sine of the Cabibbo angle $\theta_C$. \\
The experimental study of neutral $D$ mixing has been
continuously progressing in the past two decades \cite{E791:1996klq, CLEO:2005oam, BaBar:2004grg, BaBar:2007fup, Belle:2008qhk, Belle:2008qhk,  BaBar:2008xkf, BaBar:2010nhz, LHCb:2015lgi, BaBar:2016kvp, CLEO:2012fel, CLEO:2012obf,  CDF:2013gvz,  Belle:2014ydf, LHCb:2019mxy, LHCb:2022cak, Malde:2015mha, BESIII:2024nnf, BESIII:2024zco, LHCb:2015lnk, BESIII:2025vbt, BESIII:2025ypr,  LHCb:2016zmn, Betti:2021mpf, Libby:2014rea, Evans:2016tlp, BESIII:2021eud, BESIII:2022qkh, Pajero:2022vev, Belle:2006ipk, BaBar:2007kib, LHCb:2025kch, LHCb:2024hyb,  E791:1999bzz, FOCUS:2000kxx, CLEO:2001lgl,Belle:2009xzl,BaBar:2012bho, BESIII:2015ado,Belle:2015etc,LHCb:2018zpj,Belle:2019xha, LHCb:2022gnc, CDF:2014wyb, LHCb:2015xyd, LHCb:2017ejh, LHCb:2019dom, LHCb:2021vmn,   LHCb:2016nxk, LHCb:2022lry, LHCb:2014kcb, LHCb:2016csn, LHCb:2019hro, BaBar:2007tfw, Belle:2008ddg, DiCanto:2012ufu, CDF:2011ejf,  LHCb:2024hyb}. The most precise estimates of the charm parameters are obtained from the rates of Cabibbo Favoured (CF) (\emph{i.e.} proportional to $V_{cs}V_{ud}^*$) or Doubly Cabibbo Suppressed (DCS) (\emph{i.e.} proportional to $V_{cd}V_{us}^*$) $D$ decays \cite{Belle:2006ipk, BaBar:2007kib, LHCb:2025kch, LHCb:2024hyb}, from charm decays to CP eigenstates \cite{E791:1999bzz, FOCUS:2000kxx, CLEO:2001lgl,Belle:2009xzl,BaBar:2012bho, BESIII:2015ado,Belle:2015etc,LHCb:2018zpj,Belle:2019xha, LHCb:2022gnc, CDF:2014wyb, LHCb:2015xyd, LHCb:2017ejh, LHCb:2019dom, LHCb:2021vmn,   LHCb:2016nxk, LHCb:2022lry, LHCb:2014kcb, LHCb:2016csn, LHCb:2019hro, BaBar:2007tfw, Belle:2008ddg, DiCanto:2012ufu, CDF:2011ejf} and from  Dalitz plot analyses of the three-body modes $K^0_S \pi^+\pi^-$, $K^0_S K^+K^-$ or $\pi^+\pi^-\pi^0$ \cite{ BaBar:2010nhz, LHCb:2015lgi, Belle:2014ydf, BaBar:2016kvp, LHCb:2019mxy, LHCb:2022cak}. In
particular, LHCb has achieved impressive results during the LHC Run II, including the observation of direct CP asymmetries in the Singly Cabibbo Suppressed (SCS) decays of $D$ mesons to $K^+K^-$ and $\pi^+\pi^-$ \cite{LHCb:2019hro, LHCb:2022lry} and the very recent measurements of CP-violating observables in $D^0 \to K^{\pm} \pi^{\mp}$ \cite{LHCb:2024hyb}. \\
This remarkable increase in the experimental accuracy
calls for an update of our previous works on $D$ meson mixing \cite{UTfit:2012ich, UTfit:2014hez}, generalizing the so-called ``superweak assumption'', in which the weak phase describing CP violation in the interference between $D$ meson decays with and without absorptive mixing is set to zero, to the more suitable ``approximate universality'' scenario introduced in Ref.~\cite{Kagan:2020vri} to be detailed below. \\
Charm mixing also plays an important role in the extraction of the CKM angle $\gamma = \arg (- V_{ud} V_{ub}^*  V_{cb} V_{cd}^*)$ from tree-level $B$ decays \cite{Belle:2006cuz, Belle:2011ac, Belle:2013dtr,  LHCb:2015dlc, LHCb:2017egy, LHCb:2016bsl, LHCb:2020hdx, LHCb:2020vut, LHCb:2021mmv, Belle:2023lha, Belle:2023yoe, LHCb:2024oco, LHCb:2022nng, LHCb:2024ett,  Belle:2010xyn, Belle:2019uav, LHCb:2020yot, Belle:2021efh, LHCb:2023yjo, LHCb:2023lib, LHCb:2023kpr, LHCb:2023ayf, LHCb:2018zap, LHCb:2020qag, LHCB-PAPER-2024-020, BaBar:2005jis, BaBar:2006slj, Belle:2006lts, Belle:2011dhx, BaBar:2007dro, BaBar:2008qcq, BaBar:2009dzx, BaBar:2010hvw, BaBar:2010otv, BaBar:2011rud, CDF:2009wnr, CDF:2011xrp, Belle:2006lys, BaBar:2010uep, Belle:2015roy}, calling for a simultaneous combination of charm and beauty measurements, as pioneered by the LHCb collaboration~\cite{LHCb:2021dcr}. Our results can be compared with the latest LHCb \cite{LHCb-CONF-2024-004} estimates, the  Belle/Belle II analysis~\cite{Belle-II:2024eob} for $\gamma$ and HFLAV~\cite{HFLAV:2022pwe} for the charm parameters. 
The consistency of the fit is checked by comparing  $\gamma$ estimates obtained by combining measurements involving different types of  $B$ mesons separately: charged $B^{\pm}$ or neutral $B^0$ and $B^0_s$. The impact of the global fit on the charm parameters is assessed by comparing it with the results of a fit to $D$ observables only.\\ 
The paper is organized as follows: the dispersive-absorptive formalism for $D$ meson mixing in the framework of approximate universality is briefly summarized in Sec.~\ref{Sec:Formalism}. The expressions of the charm observables are described in Sec.~\ref{Sec:CHARM_OBS}, while the beauty observables are introduced in Sec.~\ref{Sec:Beauty_Obs} and their dependence on the parameters of interest of this study is reported in Apps.~\ref{Sec:GLW} and~\ref{Sec:ADS}. 
We present and discuss the most relevant results in Sec.~\ref{Sec:Results}. The posterior probability intervals of the fit parameters are given in App.~\ref{Sec:all_results}. A summary can be found in Sec.~\ref{Sec:Conclusions}.

\section{Charm Mixing Formalism}
\label{Sec:Formalism}
In this section, we present a parametrization of the quantities needed to describe charm  decays and $D$ mixing underlying all the observables considered in the combination. \\ We indicate with $M$ ($\overline{M}$) a generic $B$ ($\overline{B}$) or $D$ ($\overline{D}$) meson, while its decay amplitudes to CP conjugate final states $f$ and $\overline{f}$ are denoted as 
\be
\begin{aligned}
    \mathcal{A}_M^f = \bra{f} \mathcal{H} \ket{M}, \quad \quad & \quad \quad \mathcal{A}^{\overline{f}}_M = \bra{\overline{f}} \mathcal{H} \ket{M}, \\ 
    \mathcal{A}^f_{\overline{M}} = \bra{f} \mathcal{H} \ket{\overline{M}},  \quad \quad & \quad \quad \mathcal{A}^{\overline{f}}_{\overline{M}} = \bra{\overline{f}} \mathcal{H} \ket{\overline{M}},
\end{aligned}
\label{Eq:decay_amplitudes}
\ee
where $\mathcal{H}$ is the relevant effective Hamiltonian involved in the process.  \\
The time evolution of a linear combination of $D^0$ and $\overline{D^0}$ mesons follows the Schrödinger equation, with a $2 \times 2$ non-Hermitian Hamiltonian $\mathbf{H}$ (see \emph{e.g.}~\cite{Nir:1992uv,Branco:1999fs, Grossman:2009mn,Kagan:2009gb}) that in terms of its dispersive $\mathbf{M}$ and absorptive $\mathbf{\Gamma}$ components is given by
\be
\mathbf{H} = \mathbf{M} - i/2 \mathbf{\Gamma},
\label{Eq:hamiltonian}
\ee
so that its matrix element between charm states can be written as 
\be
\langle D^0 \vert \mathbf{H}  \vert \overline{D^0}  \rangle = H_{12} = M_{12} - i/2 \Gamma_{12}.
\label{Eq:H_matrix_element}
\ee
Adopting the long-lived (L) and short-lived (S) meson notation, the Hamiltonian eigenstates read 
\be
\ket{D_{\text{L,S}}} = p \vert D^0 \rangle \pm q \vert \overline{D^0} \rangle,
\label{Eq:ham_eigenstates}
\ee
with $p$ and $q$ complex coefficients satisfying $\left \vert p \right \vert^2 + \left \vert q \right \vert^2 = 1$. Then, the differences between masses and decay widths of the Hamiltonian eigenstates are parametrized through the usual quantities
\be
x = \frac{m_{\text{S}} - m_{\text{L}}}{\Gamma}, \quad \quad \quad \quad \quad y = \frac{\Gamma_{\text{S}} - \Gamma_{\text{L}}}{2 \Gamma},
\label{Eq:x,y}
\ee
with $\Gamma$ being the averaged $D^0$ lifetime. \\ 
$D$ meson mixing is described in terms of the matrix elements of $\mathbf{H}$ by the parameters
\be
\phi_{12} = \arg \bigg( \frac{M_{12}}{\Gamma_{12}} \bigg), \qquad  x_{12} = 2\frac{|M_{12}|}{\Gamma} \qquad \mathrm{and }\qquad  y_{12} = \frac{|\Gamma_{12}|}{\Gamma}.
\label{Eq:mixing_parameters}
\ee
The phase $\phi_{12}$  governs CP violation in pure mixing, while $x_{12}$ and $y_{12}$ are CP-conserving.  \\ 
The CP-violating quantity $|q/p|-1$ and the observables in Eq.~(\ref{Eq:x,y}) are related to the mixing parameters in Eq.~(\ref{Eq:mixing_parameters}) as 
\be
|x| = x_{12},  \qquad \qquad    \left \vert \frac{q}{p} \right \vert -1 =  \frac{x_{12} y_{12}}{x_{12}^2 + y_{12}^2} \sin \phi_{12},   \qquad \qquad |y| = y_{12},
\label{Eq:connection_qop,x,y}
\ee
up to negligible corrections quadratic in $\sin \phi_{12}$. \\ 
In the SM, it is possible to decompose the dispersive and absorptive parts of $H_{12}$ in terms of U-spin amplitudes as follows \cite{Kagan:2020vri}:
\be
\xi_{12}^{\text{SM}} = \frac{(\lambda^s_{uc} - \lambda^d_{uc})^2}{4} \xi_2 + \frac{(\lambda^s_{uc} - \lambda^d_{uc})\lambda^b_{uc}}{2} \xi_1 + \frac{(\lambda^{b}_{uc})^2}{4} \xi_0, \qquad  \xi = M, \Gamma,
\label{Eq:gamma_decomposition}
\ee
where $\lambda^{j}_{kl} = V_{kj}^* V_{lj}$, while $M_{0,1,2}$ and $\Gamma_{0,1,2}$ are the $\Delta U_3 = 0$ components of the $\Delta U = 0$,~$1$,~$2$ multiplets, respectively. Thus, the order of magnitude of $\xi_{0,1,2}$ is related to the U-spin breaking parameter $\varepsilon$, such that $ \xi_n \approx \mathcal{O}(\varepsilon^n)$. \\ In the approximate universality scenario, CP violation in the interference between decays with and without dispersive and absorptive mixing can be described through two final state-independent weak phases $\phi_{2}^{M, \Gamma}$, defined as
\be
\phi_2^{\xi} = \arg \frac{\xi_{12}}{\frac{1}{4}(\lambda^s_{uc} - \lambda^d_{uc})^2 \xi_2}, \qquad  \xi = M, \Gamma.
\label{Eq:phi2_M,Gamma}
\ee
Furthermore, the difference between $\phi_2^{M}$ and $\phi_{2}^{\Gamma}$ is exactly $\phi_{12}$, meaning that all types of indirect CP violation concerning $D$ meson mixing, namely CPV in mixing and in the interference between mixing and decay, are entirely determined by knowledge of these two universal weak phases. 
\\ Employing in the definition of Eq.~(\ref{Eq:phi2_M,Gamma}) the expressions of $M_{12}^{\text{SM}}$ and $\Gamma_{12}^{\text{SM}}$ given in  Eq.~(\ref{Eq:gamma_decomposition}) and neglecting the third (smallest) contribution, the orders of magnitude of the two CP-violating phases in the SM can be roughly estimated to be \cite{Kagan:2020vri}
\be
 (\phi_2^{M}) \sim (\phi_{2}^{\Gamma})  \simeq \left \vert \frac{\lambda^b_{uc}}{\theta_C}  \right \vert \sin \gamma \times \frac{\Gamma_1}{\Gamma_2}  \sim \left \vert \frac{\lambda^b_{uc}}{\theta_C}  \right \vert \sin \gamma \times \frac{1}{\varepsilon} \sim (2.2 \times 10^{-3}) \times \bigg(\frac{0.3}{\varepsilon}\bigg).
\label{Eq:SM_estimate_phi2MG}
\ee
From Eqs.~(\ref{Eq:mixing_parameters}) and~(\ref{Eq:gamma_decomposition}), we get the relation $\left \vert \Gamma_2 \right \vert \simeq y_{12}  \ \Gamma / \left \vert \lambda^s_{uc} \right \vert^2$, which can be employed in Eq.~(\ref{Eq:SM_estimate_phi2MG}) to provide an upper bound on $\left \vert \phi_2^\Gamma \right \vert$ as \cite{Kagan:2020vri}
\be
\left \vert \phi_2^\Gamma \right \vert \simeq \frac{\left \vert \lambda^b_{uc} \lambda^s_{uc}  \sin \gamma \right \vert}{y_{12}}  \times \frac{ \left \vert \Gamma_1 \right \vert}{\Gamma} < (5 \times 10^{-3}) \times \bigg( \frac{0.61\%}{y_{12}} \bigg),
\label{Eq:alter_phi2G}
\ee
where we used the conservative assumption $\left \vert 4\Gamma_1/(\Gamma_0 - \Gamma_2) \right \vert < 1$, a ratio which is nominally of $\mathcal{O}(\varepsilon)$. \\ 
For a given final state $f$, CP violation in the interference between decays with and without mixing  can be equivalently described by the weak phase of the observable 
\be
\lambda^{f}_D \equiv \frac{q}{p} \frac{\mathcal{A}_{\overline{D}}^f}{\mathcal{A}_D^f},
\label{Eq:lambda_f}
\ee 
which becomes universal in the framework of approximate universality and we denote by $\phi_2$. The posterior of  $\phi_2$ is determined from the charm parameters as 
\be
\tan 2 \phi_2 = - \bigg( \frac{x_{12}^2 \sin 2 \phi_2^M + y_{12}^2 \sin 2 \phi_2^{\Gamma}}{x_{12}^2 \cos 2 \phi_2^M + y_{12}^2 \cos 2 \phi_2^{\Gamma}}  \bigg). 
\label{Eq:phi2}
\ee
Notice that $\phi_2$ corresponds to the phase $\phi$ extracted in global analyses under the assumption of final state independence (see \emph{e.g.} refs. \cite{LHCb:2021dcr,LHCb-CONF-2024-004}). \\ 
In the superweak assumption, dispersive mixing is supposed to be the dominant source of CPV in the charm system, corresponding to the approximation where 
$\phi_2^\Gamma = 0$ and $\phi_{12} = \phi_2^M$ or, equivalently, $\phi = \phi_2 =  \arctan \frac{1-|q/p|^2}{1+|q/p|^2}\frac{x}{y}$. Then, going from the superweak assumption, adopted in our previous works~\cite{UTfit:2012ich, UTfit:2014hez}, to approximate universality, used in this work, involves moving from a description in terms of a single independent CPV parameter ($\phi_{12}$ or $\phi_2$) to a scenario with two free CPV parameters ($\phi_2^{M}$ and $\phi_2^{\Gamma}$ or $\phi_2$ and $|q/p|-1$).

\section{Charm observables}
\label{Sec:CHARM_OBS}
The charm observables used in our combination rely on time-integrated and time-dependent studies of $D$ meson decays, reconstructed from two- and multi-body final states. The final states can be either CP eigenstates or non CP eigenstates originated from CF or DCS decays. We parametrize the decay amplitudes through the ratios of their magnitudes and  strong phases and provide expressions for the observables in Tab.~\ref{Tab:D_obs} in terms of charm decay, mixing and CP-violating parameters in the framework of approximate universality for the neutral $D$ system, following Ref.~\cite{Kagan:2020vri}.

\subsection{CF/DCS decays to \texorpdfstring{$\mathbf{K^{\pm} X}$}{Kpm}}
\label{Sec:2bodyCF/DCS}
\begin{table}[]
        \centering
        \begin{tabular}{c|c|c|c|c|c}
        \hline \hline & & & & & \\ [-2.5 ex]
        \textbf{Observables} & $\mathbf{D^0}$ \textbf{decays} & \textbf{Ref.} & \textbf{Observables} & $\mathbf{D^0}$ \textbf{decays} & \textbf{Ref.} \\ [0.75 ex] \hline \hline  & & & & &  \\ [-2.5 ex]
        $(x^2 +y^2)/2$ & $D \to K l \overline{\nu}_l$ & \cite{E791:1996klq, CLEO:2005oam, BaBar:2004grg, BaBar:2007fup, Belle:2008qhk} & 
        $x'^{\pm}_{K\pi\pi^0}$,  $y'^{\pm}_{K\pi\pi^0}$ & $D \to K \pi \pi^0$ & \cite{BaBar:2008xkf}  \\ [0.75 ex] \hline & & & & &  \\ [-2.5 ex]
         $x$, $y$ & $D \to K^0_S XX$ & \cite{BaBar:2010nhz} &  $x$, $y$ & $D \to K^0_S \pi\pi$ & \cite{LHCb:2015lgi} \\ [0.75 ex] \hline & & & & &  \\ [-2.5 ex] 
        $x$, $y$ & $D \to \pi^+\pi^-\pi^0$ & \cite{BaBar:2016kvp} & $x$, $y$, $|q/p|$, $\phi_2$ & $D \to K^0_S \pi \pi$ & \cite{Belle:2014ydf}   \\ [0.75 ex] \hline & & & & &  \\ [-2.5 ex] 
       $(r_D^{K\pi})^2$, $x^2$, $y$,   &  &    &   $\frac{\mathcal{B}(D^0 \to K^0_s K^+ \pi^-)}{\mathcal{B}(D^0 \to K^0_s K^- \pi^+)}$, & &  \\ [0.75 ex]  
         $\cos \delta_{D}^{K\pi}$, & $D \to K \pi$ & \cite{CLEO:2012fel}  &   $\delta_{D}^{K^0_SK\pi}$,  & $D \to K^0_S K \pi$  &  \cite{CLEO:2012obf}  \\ [0.75 ex] 
            $\sin \delta_{D}^{K\pi}$ & & &   $\kappa_{D}^{K^0_SK\pi}$  &   &  \\ [0.75 ex] \hline & & & & &  \\ [-2.5 ex]
             $F_+^{4\pi}$ & $D \to 4\pi$ & \cite{Malde:2015mha, BESIII:2024zco}  & $F_+^{X X \pi^0}$ & $D \to  XX  \pi^0$  & \cite{Malde:2015mha, BESIII:2024nnf}   \\ [0.75 ex] \hline & & & & &  \\ [-2.5 ex] 
              $\frac{\mathcal{B}(D \to K^0_s K^+ \pi^-)}{\mathcal{B}(D^0 \to K^0_s K^- \pi^+)}$ & $D \to K^0_S K \pi$ & \cite{LHCb:2015lnk} &  $F_+^{KK\pi\pi}$ & $D \to K^+K^-\pi^+\pi^-$ & \cite{BESIII:2025ypr} \\ [0.75 ex] \hline & & & & &  \\ [-2.5 ex] 
                $x_{\text{CP}}$, $y_{\text{CP}}$, $\Delta x$, $\Delta y$ & $D \to K^0_S \pi \pi$ & \cite{LHCb:2019mxy, LHCb:2022cak} & $ (x^2 + y^2)/4$ & $D \to K3\pi$ & \cite{LHCb:2016zmn}  \\ [0.75 ex]  \hline & & & & & \\ [-2. ex] 
                 $r_D^{K3\pi}$, $\delta_D^{K3\pi}$, $\kappa_D^{K3\pi}$,  & $D \to K3\pi$ &  \cite{Libby:2014rea, Evans:2016tlp, LHCb:2016zmn, BESIII:2021eud} & $A_D(f^+)$, $A_D(\pi \pi \pi^0)$, & $D \to f^+$, $D \to K \pi$ & \cite{BESIII:2022qkh}  \\ [0.75 ex] 
             $r_D^{K\pi \pi^0}$, $\delta_D^{K\pi\pi^0}$, $\kappa_D^{K\pi\pi^0}$ & $D \to K \pi \pi^0$ & &  $r_D^{K\pi} \cos \delta_D^{K\pi}$, $r_D^{K\pi} \sin \delta_D^{K\pi}$ & $D \to XX \pi^0$ &    \\ [0.75 ex]  \hline & & & & & \\ [-2. ex]
            $\tilde{y}_{\text{CP}}$ & $D \to XX$ & \cite{E791:1999bzz, FOCUS:2000kxx,CLEO:2001lgl,Belle:2009xzl,BaBar:2012bho, BESIII:2015ado,Belle:2015etc,LHCb:2018zpj,Belle:2019xha,  LHCb:2022gnc} & $\Delta Y^{KK} - \Delta Y^{\pi\pi}$, $\Delta Y$ & $D \to X X$ & \cite{LHCb:2015xyd, LHCb:2017ejh, LHCb:2019dom, LHCb:2021vmn}  \\ [0.75 ex] \hline & & & & &  \\ [-2.5 ex] 
             $(r_D^{K\pi})^2$, $c^{(')}_{K\pi}$,  $\Delta c^{(')}_{K\pi}$, $\Delta \Tilde{c}^{(')}_{K\pi}$ & $D \to K \pi$ & \cite{LHCb:2024hyb} & $(r_D^{K\pi})^2$, $(x'^{\pm}_{K\pi})^2$,  $y'^{\pm}_{K\pi}$ & $D \to K \pi$ & \cite{Belle:2006ipk,BaBar:2007kib, LHCb:2025kch} \\ [0.75 ex] \hline & & & & & \\ [-2. ex]
           $A_D^{\text{CP}}(KK)$, $\MEANMOD{\tau}_E^{KK}$ & $D \to K^+K^-$ & \cite{LHCb:2014kcb, LHCb:2016nxk, LHCb:2022lry} &  $\Delta A_D^{\text{CP}}$, $\MEANMOD{\tau}^{KK}_{E}$, $\MEANMOD{\tau}^{\pi\pi}_{E}$     & $D \to X X$  & \cite{LHCb:2014kcb, LHCb:2016csn, LHCb:2019hro} \\ [0.75 ex]  \hline & & & & & \\ [-2. ex]
             $\Delta Y^{KK}$, $\Delta Y^{\pi\pi}$ & $D\to XX$ & \cite{CDF:2014wyb}& $A_D^{\text{CP}}(XX)$, $\MEANMOD{\tau}_E^{KK}$, $\MEANMOD{\tau}_E^{\pi\pi}$ & $D\to XX$ & \cite{BaBar:2007tfw, Belle:2008ddg, DiCanto:2012ufu, CDF:2011ejf} 
             \\ [0.75 ex] \hline & & & & & \\ [-2. ex]
           $\frac{\mathcal{B}(D^0 \to K^+\pi^-)}{\mathcal{B}(D^0 \to K^-\pi^+)}$, $\frac{\mathcal{B}(D^0 \to K^+\pi^-\pi^0)}{\mathcal{B}(D^0 \to K^-\pi^+\pi^0)}$ & $D \to K\pi$, $D\to K\pi\pi^0$ & \cite{BESIII:2025vbt} &       &   & 
            \\ [0.75 ex]  & & & & & \\ [-2. ex]
             $\frac{\mathcal{B}(D^0 \to K^+\pi^-\pi^+\pi^-)}{\mathcal{B}(D^0 \to K^-\pi^+\pi^+\pi^-)}$ & $D\to K3\pi$ &       &   & 
             \\ [0.75 ex]   
                \hline \hline
        \end{tabular} 
                \caption{Charm observables used in the combination. We use $XX$ to indicate both $\pi^+\pi^-$ and $K^+K^-$ states, while $K\pi$ stands for $K^{\mp} \pi^{\pm}$. We refer to the multi-body final states $K^{\mp} \pi^{\pm} \pi^+ \pi^-$ and $\pi^+\pi^-\pi^+\pi^-$ as $K3\pi$ and $4\pi$, respectively. The state $f^+$ stands for a  CP-even final state. A description of the observables can be found in   Secs.~\ref{Sec:2bodyCF/DCS}, \ref{Sec:CP_eigenstates} and \ref{Sec:Multibody}.}
        \label{Tab:D_obs}
        \end{table}
In this section, we consider CF/DCS decays to $K^{\pm} X$, such as $D^0 \to K^{\pm} \pi^{\mp}$. We denote with $f$ the final state of the CF decay of the $D^0$ meson (\emph{e.g.} $K^- \pi^+$) and by ``wrong-sign" (WS) and ``right-sign" (RS) the time-dependent DCS (\emph{i.e.} $D^0 \to \overline{f}$ and $\overline{D^0} \to f$) and CF (\emph{i.e.} $D^0 \to f$ and $\overline{D^0} \to \overline{f}$) decays respectively. We parametrize the ratios of decay amplitudes in Eq.~(\ref{Eq:decay_amplitudes}) as
\be
r_{D}^f  e^{-i \delta_{D}^f} = \frac{\mathcal{A}^{\overline{f}}_D}{\mathcal{A}^f_D} = \frac{\mathcal{A}^{f}_{\overline{D}}}{\mathcal{A}^{\overline{f}}_{\overline{D}}},
\label{Eq:rf}
\ee
where we have neglected possible new weak phases in CF/DCS decays (weak phases are fully negligible in the SM). \\ 
The observables used in the fit for this type of final states are extracted from the ratios of the WS and RS decay rates~\cite{ LHCb:2025kch, BaBar:2007kib, Belle:2006ipk}, given by 
\be
\begin{aligned}
\frac{\Gamma(D^0 \to \overline{f}, t)}{\Gamma(D^0 \to f, t)} = & \  (r_D^f)^2 + r_D^f y_f'^+ \tau + \frac{(x_f'^+)^2 + (y_f'^+)^2}{4} \tau^2, \\ 
\frac{\Gamma(\overline{D^0} \to f, t)}{\Gamma(\overline{D^0} \to \overline{f}, t)} = & \  (r_D^f)^2 + r_D^f y_f'^- \tau + \frac{(x_f'^-)^2 + (y_f'^-)^2}{4} \tau^2,
\end{aligned}
\label{Eq:WS/RS}
\ee
up to second order in the charm mixing and CP-violating parameters and with $\tau = \Gamma t$. The coefficients $x_f'^{\pm}$ and $y_f'^{\pm}$ can be written as a rotation of an angle $\phi_2$ of their CP-conserving limits, which we denote with $x'_f$ and $y'_f$, respectively, as 
\be
\begin{pmatrix}
    x_f'^{\pm} \\ 
    y_f'^{\pm} 
\end{pmatrix} = - \left \vert \frac{q}{p} \right \vert^{\pm1} \begin{pmatrix}
    \cos \phi_2 & \pm \sin \phi_2 \\ 
    \mp \sin \phi_2 & \cos \phi_2 
\end{pmatrix} \begin{pmatrix}
    x'_f \\ 
    y'_f
    \end{pmatrix},
\ee
with 
\be
\begin{aligned}
x_f' = & \       x \cos \delta_D^f + y \sin \delta_D^f, \\ 
y_f' =& \  y \cos \delta_D^f - x \sin \delta_D^f.
\end{aligned}
\label{Eq:ypm_xpm}
\ee
The linear and quadratic terms in Eq.~(\ref{Eq:WS/RS}) can be decomposed into 
their CP-conserving ($c_f$, $c'_f$) and CP-violating parts ($\Delta c_f$, $\Delta c'_f$), as follows: 
\be
\begin{aligned}
y_f'^{\pm} =   c_f \pm   & \  \Delta c_f, \\ 
\frac{(x_f'^\pm)^2 + (y_f'^\pm)^2}{4} = & \  c_f' \pm \Delta c_f',
\end{aligned}
\label{Eq:cf,dcf_formal}
\ee
where the coefficients on the right-hand side of Eq.~(\ref{Eq:cf,dcf_formal}) are given in terms of the charm mixing parameters as
\be
\begin{aligned}
    c_f = x_{12} \cos \phi_2^M \sin \delta_D^f - y_{12} \cos \phi_2^{\Gamma} \cos \delta_D^f, \qquad & \qquad \Delta c_f = - x_{12} \sin \phi_2^M \cos \delta_D^f - y_{12} \sin \phi_2^\Gamma \sin \delta_D^f, \\ 
    c'_f = \frac{x_{12}^2 + y_{12}^2}{4} + \frac{(r_D^f)^2}{4} (y_{12}^2 - x_{12}^2), \qquad & \qquad \Delta c'_f = \frac{1}{2} x_{12} y_{12} \sin \phi_{12}. 
\end{aligned}
\label{Eq:cf,dcf}
\ee
The separation of the CP-violating from the CP-conserving contributions entering the  WS/RS ratios of $D^0 \to K^{\pm} \pi^{\mp}$ decays has been performed for the first time this year  by LHCb~\cite{LHCb:2024hyb}, combining both the Run 1 and Run 2 results. However, while the Run 1 measurements can be used directly to extract the coefficients in Eq.~(\ref{Eq:cf,dcf}), the Run 2 data use experimental information from $D^0 \to K^+K^-$ decays to correct a detection asymmetry. This operation leaves invariant $c_f$ and $c'_f$, while it changes the expressions of the CPV observables as 
\be
\begin{aligned}
    \Delta \Tilde{c}_f = & \  \Delta c_f - c_f a_D^{KK} - 2 r_D^f \Delta Y^{KK}, \\ 
    \Delta \Tilde{c}'_f = & \  \Delta c'_f - 2 c'_f a_D^{KK} - 2 r_D^f c_f \Delta Y^{KK},
\end{aligned}
\label{Eq:Dctilde}
\ee
and the parameters $r_D^f$ are replaced by $r_D^f (1\mp a_D^{KK})$ in Eq.~(\ref{Eq:WS/RS}). 
Here, $a_D^{KK}$ is the direct CP asymmetry in $D^0 \to K^+K^-$ decays and we introduced the parameter $\Delta Y^{KK}$. They were measured in refs.~\cite{LHCb:2014kcb, LHCb:2016nxk, LHCb:2022lry, BaBar:2007tfw, Belle:2008ddg, DiCanto:2012ufu, CDF:2011ejf} and~\cite{CDF:2014wyb, LHCb:2015xyd, LHCb:2017ejh, LHCb:2019dom, LHCb:2021vmn},  and are defined in terms of decay amplitudes and decay rates in Eqs.~(\ref{Eq:rf_CP}) and~(\ref{Eq:time_dependent_cp_asymmetry}), respectively, when describing the observables involving SCS decays to CP eigenstates.  
\\ We consider in the combination also observables coming from the analysis of quantum correlated  $D^0- \overline{D^0}$  pairs (see \emph{e.g.}~\cite{BESIII:2022qkh}). Here, one of the $D$ mesons decays to the so-called tagging mode, while the other decays to the signal mode, typically $K^-\pi^+$. Then, the following asymmetries are measured: 
\be
A_{D}(f^+) = \frac{\mathcal{B}(D_{f^+} \to K^{-} \pi^+ ) - \mathcal{B}(D_{f^-} \to K^-\pi^+)}{\mathcal{B}(D_{f^+} \to K^-\pi^+ ) + \mathcal{B}(D_{f^-} \to K^-\pi^+)}, 
\label{Eq:asymmetry_quantum_correlated}
\ee
where $\mathcal{B}(D_{f^{\pm}} \to K^-\pi^+ )$  are the branching fractions of  neutral charm states decays to $K^-\pi^+$ when tagged by CP-even or multi-body modes, also known as quasi-CP eigenstates, $D_{f^+}$ or CP-odd modes $D_{f^-}$. For example, we could have $f^+ = \pi^+\pi^- (\pi^0)$ and $f^- = K^0_S \pi^0$. In the CP-conserving limit, the asymmetry in Eq.~(\ref{Eq:asymmetry_quantum_correlated}) depends on the charm parameters as 
\be
A_{D}(f^+) = \frac{F_{+}^{f^+} (y - 2 r_D^{K\pi} \cos \delta_D^{K\pi})}{1 + (r_D^{K\pi})^2 + (1 - F_{+}^{f^+})(y + 2 r_D^{K\pi} \cos \delta_D^{K\pi} )}, 
\label{Eq:asymmetry_quantum_correlated_charm}
\ee
up to $\mathcal{O}(x, y, (r_D^{K\pi})^2)$ corrections. Here, we have introduced  the so-called CP-even fraction  $F_{+}^{f^+}$ of the mode $f^+$, which is a non-negative real number needed for quasi-CP eigenstates that reaches its maximum of one for CP-even eigenstates. A definition of $F_{+}^{f^+}$  in terms of decay amplitudes is given in Sec.~\ref{Sec:Multibody}, in the context of multi-body final states.

\subsection{SCS decays to CP eigenstates}
\label{Sec:CP_eigenstates}
SCS decays to CP eigenstates $f$, such as $D^0 \to \pi^+ \pi^-$ or $D^0 \to K^+ K^-$, are described in terms of just two of the four amplitudes in Eq.~(\ref{Eq:decay_amplitudes}) and their magnitudes can be parametrized as 
\be
 1 - a_D^f = \left \vert \frac{\mathcal{A}_{\overline{D}}^f}{\mathcal{A}_D^f} \right \vert,
\label{Eq:rf_CP}
\ee
where $a_D^f$ is the direct CP asymmetry, already introduced in Eq.~(\ref{Eq:Dctilde}). CP-conserving and CP-violating  ratios of time-dependent decay rates are measured. The first type is obtained by calibrating the signal through  the CF mode $K^{\mp} \pi^{\pm}$ as 
\be
\frac{\Gamma(D^0 \to f, t) + \Gamma(\overline{D^0} \to f, t)}{\Gamma(D^0 \to K^-\pi^+, t) + \Gamma(\overline{D^0} \to K^+\pi^-, t)} \propto 1 - \tau \Tilde{y}_{\text{CP}}, 
\label{Eq:CP_over_CF}
\ee
where we have neglected the quadratic terms in the charm mixing and CP-violating parameters that come with higher powers in $\tau$. 
Then, the slope $\Tilde{y}_{\text{CP}}$ is extracted from Eq.~(\ref{Eq:CP_over_CF}) through a linear fit  \cite{E791:1999bzz,LHCb:2022gnc, FOCUS:2000kxx,CLEO:2001lgl,Belle:2009xzl,BaBar:2012bho,BESIII:2015ado,Belle:2015etc,LHCb:2018zpj,Belle:2019xha}, and we have:
\be
\begin{aligned}
2\Tilde{y}_{\text{CP}} =   y \cos \phi_2 &  \bigg(  \left \vert \frac{q}{p}\right \vert + \left \vert \frac{p}{q} \right \vert  \bigg) -  \ x \sin \phi_2 \bigg( \left \vert \frac{q}{p}\right \vert -  \left \vert \frac{p}{q} \right \vert  \bigg)  \\  - &  r_D^{K \pi}\bigg[  \cos \phi_2 ( y \cos \delta_D^{K \pi} + x \sin \delta_D^{K\pi} )\bigg( \left \vert \frac{q}{p} \right \vert + \left \vert \frac{p}{q} \right \vert  \bigg) - \sin \phi_2 (  x \cos \delta_D^{K\pi}   - y \sin \delta_D^{K\pi}  )  \bigg( \left \vert \frac{q}{p} \right \vert - \left \vert \frac{p}{q} \right \vert  \bigg) \bigg].
\end{aligned}
\label{Eq:ytilde_CP}
\ee
CP-violating ratios can be defined through the time-dependent asymmetries between the rates of  $D^0$ and $\overline{D^0}$ decays to CP-even final states as follows:
\be
A^{\text{CP}}_D(f, t) = \frac{\Gamma(D^0 \to f, t) - \Gamma(\overline{D^0} \to f, t)}{\Gamma(D^0 \to f, t) + \Gamma(\overline{D^0} \to f, t)} = a_D^f +  \tau   \Delta Y^f.
\label{Eq:time_dependent_cp_asymmetry}
\ee
Experiments measure the slopes $\Delta Y^f$ of  Eq.~(\ref{Eq:time_dependent_cp_asymmetry}) for the modes $D^0 \to K^+K^-$ and $D^0 \to \pi^+\pi^-$\cite{CDF:2014wyb, LHCb:2015xyd, LHCb:2017ejh, LHCb:2019dom, LHCb:2021vmn, LHCb:2021vmn}, and we can define the following combinations
\begin{equation}
    \begin{aligned}
        \Delta Y^{KK} - & \  \Delta Y^{\pi\pi}, \\ 
        \Delta Y =  \frac{1}{2} (  \Delta Y^{KK} \ &   + \Delta Y^{\pi\pi}).
    \end{aligned}
    \label{Eq:DYf}
\end{equation}
We use the relations in Eq.~(\ref{Eq:DYf}) to constrain the parameter $\Delta Y^{KK}$, which enters the observables coming from the WS/RS ratios of $D^0 \to K^{\pm} \pi^{\mp}$ in Eq.~(\ref{Eq:Dctilde}). Moreover, since in the U-spin symmetric limit the final state-dependent contributions to $\Delta Y^f$ are equal and opposite in sign for $K^+K^-$ and $\pi^+\pi^-$ \cite{Kagan:2020vri}, averaging the two modes allows $\Delta Y$ to be expressed in the framework of approximate universality as 
\be
\Delta Y = \frac{1}{2} \bigg[  x \sin \phi_2 \bigg(   \left \vert \frac{q}{p} \right \vert + \left \vert \frac{p}{q} \right \vert \bigg)  - y \cos \phi_2 \bigg( \left \vert \frac{q}{p} \right \vert - \left \vert \frac{p}{q} \right \vert \bigg ) \bigg],
\label{Eq:DeltaY}
\ee
up to quadratic corrections in the charm mixing and CPV parameters. \\ 
Notice that we do not include the latest measurement of $\Delta Y^f$ from $D^0 \to \pi^+\pi^-\pi^0$ decays by LHCb~\cite{LHCb:2024jpt} in the combination, as it is less precise than the observables in Eq.~(\ref{Eq:DYf}). Moreover, the application of approximate universality in this case is not as good as for $\Delta Y$ since there is no U-spin cancellation of  final-state dependent contributions. \\ 
We consider in the combination also measurements of the time-integrated CP asymmetry for the mode $K^+K^-$~\cite{LHCb:2014kcb, LHCb:2016nxk, LHCb:2022lry, BaBar:2007tfw, Belle:2008ddg, DiCanto:2012ufu, CDF:2011ejf} which, from Eq.~(\ref{Eq:time_dependent_cp_asymmetry}), can be written as
\be
A_D^{\text{CP}}(f) = a_D^f  + \MEANMOD{\tau}^f_{E} \Delta Y^f,
\label{Eq:integrated_asymmetry}
\ee
where we introduced the so-called  average decay time $\MEANMOD{\tau}^f_E$, which depends on the final state $f$ and on the experimental environment $E$. We assume $\MEANMOD{\tau}^{\pi\pi}_{B\text{facts}} = $ $\MEANMOD{\tau}^{KK}_{B\text{facts}} = 1$ for measurements performed at the $B$ factories.  \\ 
The first signal of direct CP violation in the charm system \cite{LHCb:2019hro} was observed in the difference between the asymmetries in Eq.~(\ref{Eq:integrated_asymmetry}) 
 for the modes $K^+K^-$ and $\pi^+\pi^-$, which we fit as
\be
\Delta A^{\text{CP}}_D = a_D^{KK} - a_D^{\pi \pi} +    \MEANMOD{\tau}^{KK}_{E}  \Delta Y^{KK}   -  \MEANMOD{ \tau }_{E}^{\pi \pi} \Delta Y^{\pi\pi}. 
\label{Eq:difference_integrated_cp_asymmetry}
\ee

\subsection{Multi-body final states}
\label{Sec:Multibody}
The most precise observables for multi-body final states constraining together the charm mixing and CP-violating parameters  come from the Dalitz plot analyses of $D^0 \to K^0_S \pi^+\pi^-$ decays \cite{LHCb:2019mxy, LHCb:2022cak}.
The reconstruction of the two-dimensional phase space and the decay time is performed, in a model-independent way, by partitioning them into bins, which we indicate as $\pm i$ and $j$, respectively, with $i$ and $-i$ connected by a CP transformation. For each of the bins, ratios analogous to WS/RS in Eq.~(\ref{Eq:WS/RS})  are measured: 
\be
\frac{\Gamma_{-ij}(D^0 \to K^0_S \pi^+\pi^-)}{\Gamma_{ij}(D^0 \to K^0_S \pi^+ \pi^-)}, \qquad \qquad
 \frac{\Gamma_{ij}(\overline{D^0} \to K^0_S \pi^+\pi^-)}{\Gamma_{-ij}(\overline{D^0} \to K^0_S \pi^+ \pi^-)}.
\label{Eq:ratios_Kpipi}
\ee
Fitting the quantities in Eq.~(\ref{Eq:ratios_Kpipi}) with a quadratic expansion in $\tau$ of the decay rates, the following observables are determined:
\be
\begin{aligned}
    2x_{\text{CP}} = x \cos \phi_2 \bigg( \left \vert \frac{q}{p}\right \vert + \left \vert \frac{p}{q}\right \vert  \bigg) + y \sin \phi_2 \bigg( \left \vert \frac{q}{p}\right \vert - \left \vert \frac{p}{q}\right \vert  \bigg), \ \ \ \    2y_{\text{CP}} = y \cos \phi_2 \bigg( \left \vert \frac{q}{p}\right \vert + \left \vert \frac{p}{q}\right \vert  \bigg) - x \sin \phi_2 \bigg( \left \vert \frac{q}{p}\right \vert - \left \vert \frac{p}{q}\right \vert  \bigg), \\ 
     2 \Delta x = x \cos \phi_2 \bigg( \left \vert \frac{q}{p}\right \vert - \left \vert \frac{p}{q}\right \vert  \bigg) + y \sin \phi_2 \bigg( \left \vert \frac{q}{p}\right \vert + \left \vert \frac{p}{q}\right \vert  \bigg), \ \ \ \    2 \Delta y = y \cos \phi_2 \bigg( \left \vert \frac{q}{p}\right \vert - \left \vert \frac{p}{q}\right \vert  \bigg) - x \sin \phi_2 \bigg( \left \vert \frac{q}{p}\right \vert + \left \vert \frac{p}{q}\right \vert  \bigg). 
\end{aligned}
\label{Eq:xcp,ycp,dx,dy}
\ee
Notice that for $D\to K^0_S \pi\pi$ measurements performed at $B$ factories we take into account $K-\overline{K}$ mixing following Ref.~\cite{Kagan:2020vri} (see Eq.~(144)).
Besides the Dalitz plot analysis, the phase space dependence of the $D$ mesons decay amplitudes to a multi-body final state $f$, such as $K^0_SK^{-} \pi^{+}$,  can be parametrized through the following two quantities: 
\be
\begin{aligned}  
(r_D^{f})^2 =  \   \frac{\int \text{d}\Phi_D \left \vert \mathcal{A}_{\overline{D}}^{f} \right \vert^2}{\int \text{d}\Phi_D \left \vert \mathcal{A}_D^f  \right \vert^2} = & \frac{\int \text{d}\Phi_D \left \vert \mathcal{A}_{D}^{\overline{f}} \right \vert^2}{\int \text{d}\Phi_D \left \vert \mathcal{A}_{\overline{D}}^{\overline{f}}  \right \vert^2},  \\ 
\kappa_D^f e^{- i \delta_D^f} =   \frac{\int \text{d} \Phi_D  \mathcal{A}_{\overline{D}}^f \mathcal{A}_{D}^{f*}}{\sqrt{\int \text{d}\Phi_D \left \vert \mathcal{A}_D^{f} \right \vert^2} \sqrt{\int \text{d}\Phi_D \left \vert \mathcal{A}_{\overline{D}}^{f} \right \vert^2}} =&   \frac{\int \text{d} \Phi_D \mathcal{A}_{D}^{\overline{f}}\mathcal{A}_{\overline{D}}^{\overline{f}*}}{\sqrt{\int \text{d}\Phi_D \left \vert \mathcal{A}_{\overline{D}}^{\overline{f}} \right \vert^2} \sqrt{\int \text{d}\Phi_D \left \vert \mathcal{A}_{D}^{\overline{f}} \right \vert^2}},
\label{Eq:multi-body_parameters}
\end{aligned}
\ee
where direct CP violation has been neglected.
The integrals in Eq.~(\ref{Eq:multi-body_parameters}) are performed over the $D$ phase space coordinates $\Phi_D$. The parameter in the upper row is a straightforward generalization for multi-body final states of the ratios introduced in Eqs.~(\ref{Eq:rf}) and~(\ref{Eq:rf_CP}) for two-body decays. 
\noindent The interference integral in the bottom row is described simply through its magnitude $\kappa_D^f$, also known as the coherence factor, and its integrated strong phase $\delta_D^f$.  
For quasi-CP eigenstates, such as $\pi^+\pi^-\pi^0$, the imaginary part of the interference integral over the full phase space in Eq.~(\ref{Eq:multi-body_parameters}) is vanishing. Then, instead of the coherence factor and the integrated strong phase, it is customary to use one single real parameter: the CP-even fraction $F_{+}^f$, defined as  
\be
F^f_{+} = \frac{1}{2} \bigg[ 1 + \kappa_D^f \cos \delta^f_D \bigg],
\label{Eq:CP_even_fraction}
\ee
which has already been introduced in Eq.~(\ref{Eq:asymmetry_quantum_correlated_charm}). From its definition in Eq.~(\ref{Eq:CP_even_fraction}), the CP-even fraction is non-negative and it is unity for two-body CP-even states. 
\\ The decay parameters for $K^0_SK^{\pm} \pi^{\mp}, \  K^{\pm}\pi^{\mp},$~and $K^{\pm} \pi^{\mp}\pi^0$ final states are measured from the ratio between the branching fractions of the DCS and CF decays \cite{CLEO:2012obf, LHCb:2015lnk, BESIII:2025vbt}, given by 
\be
\frac{\mathcal{B}(D^0 \to \overline{f})}{\mathcal{B}(D^0 \to f)} = \frac{(r_D^{f})^2 - \kappa_D^{f} r_D^{f} ( y \cos \delta_D^{f} - x \sin \delta_D^{f})}{1 - \kappa_D^{f} r_D^{f} ( y \cos \delta_D^{f} + x \sin \delta_D^{f})}, 
\label{Eq:ratio_BF_DCS/CF}
\ee
where we have neglected CP violation and the non-linear terms in the mixing parameters. 

\section{Beauty observables}
\label{Sec:Beauty_Obs}
Several charm mixing and decay parameters also enter time-integrated measurements of the so-called $B$ meson cascade decays to  $D^0-\overline{D^0}$   mixed states \cite{Rama:2013voa,  LHCb:2021dcr, LHCb-CONF-2024-004}. 
We show how the CKM angle $\gamma$ appears in the equations describing the most used and precise observables available to date. 
We also discuss the extraction of $\gamma$ from the CP-violating phase of the interference between neutral $B$ decays to charmed mesons with and without mixing.

\subsection{Cascade decays}
\label{Sec:Cascade_Decays}
\begin{figure}[]
    \centering
\includegraphics[width=0.45\columnwidth]{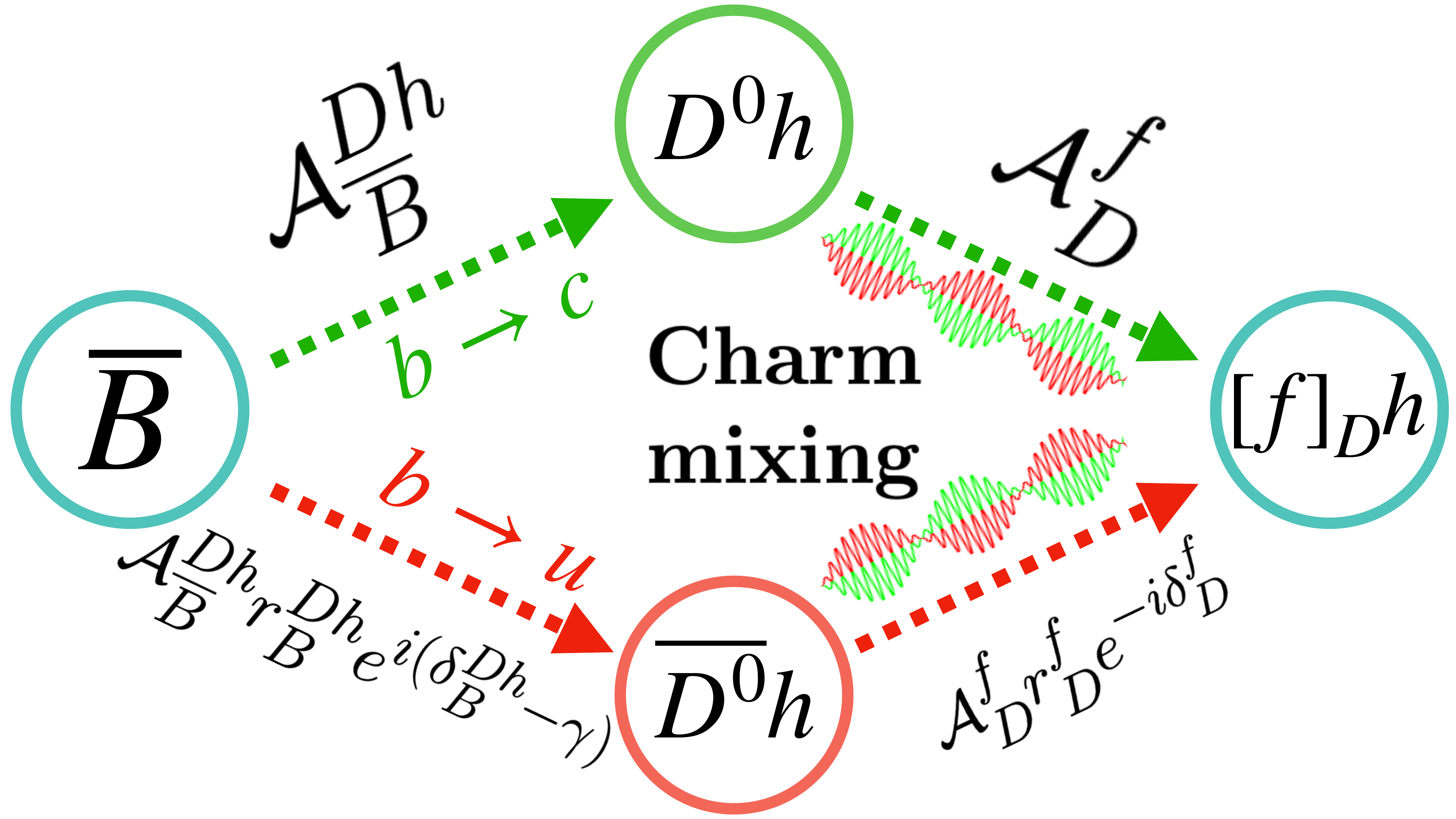}
\caption{Schematic representation of $B \to D$ cascade decays to two-body final states. The CKM angle $\gamma$ is extracted from the interference between the green and the red paths, whose amplitudes must be summed coherently. See the text for the definitions of the parameters and a  generalization for multi-body final states. }
\label{Fig:cascade_scheme}
\end{figure}

Beauty cascade decays \cite{Belle:2006cuz, Belle:2011ac, Belle:2013dtr,  LHCb:2015dlc, LHCb:2017egy, LHCb:2016bsl, LHCb:2020hdx, LHCb:2020vut, LHCb:2021mmv, Belle:2023lha, Belle:2023yoe, LHCb:2024oco,LHCb:2024ett, LHCb:2022nng, Belle:2010xyn, Belle:2019uav, LHCb:2020yot, Belle:2021efh, LHCb:2023yjo, LHCb:2023lib, LHCb:2023kpr, LHCb:2023ayf, BaBar:2007dro, BaBar:2008qcq, BaBar:2009dzx, BaBar:2010hvw, BaBar:2010otv, BaBar:2011rud, CDF:2009wnr, CDF:2011xrp, Belle:2006lys, BaBar:2010uep, Belle:2015roy} are processes where a $B$ meson goes to an hadron state $h$ and a $D^0$ or $\overline{D^0}$ meson, which subsequently oscillates and decays to one of the final states $f$ introduced in Sec.~\ref{Sec:CHARM_OBS}. In the following, we indicate with $B$ a positively charged $B^+$ or a neutral  meson $B^0_{(s)}$.  \\
Since the final states $f$ are accessible to both the neutral $D$ mesons, the total amplitude for the cascade decays is given by the coherent sum of the two paths in Fig.~\ref{Fig:cascade_scheme}. 
\noindent Then, the CKM angle $\gamma$ can be measured from the interference between the favoured $b \to c$ and suppressed $b \to u$ quark transitions and the corresponding decay amplitudes can be arranged as 
\be
(r_B^{Dh})^2 = \frac{\int \text{d} \Phi_B \left \vert \mathcal{A}_{B}^{D h} \right \vert^2}{\int \text{d} \Phi_B \left \vert \mathcal{A}_{B}^{\overline{D} h} \right \vert^2} , \qquad \qquad \kappa_{B}^{Dh} e^{-i(\delta_B^{Dh} + \gamma)} = \frac{\int \text{d} \Phi_B \mathcal{A}_{B}^{\overline{D}h} (\mathcal{A}_{B}^{Dh})^* }{\sqrt{\int \text{d} \Phi_B  \left \vert \mathcal{A}_{B}^{Dh} \right \vert^2} \sqrt{\int \text{d} \Phi_B \left \vert \mathcal{A}_{B}^{\overline{D} h }\right \vert^2}},
\label{Eq:ratio_B}
\ee
with the integrals performed over the $B$ meson phase space $\Phi_B$. 
For two-body final states, the expressions in Eq.~(\ref{Eq:ratio_B}) simplify since there is no phase space dependence and the coherence factor is unity. 
\\ Then, the time-integrated rate of the cascade decay $\overline{B} \to [f]_Dh$ is found to be
\be
\Gamma(\overline{B} \to [f]_Dh) \propto 1 + (r^f_D r_B^{Dh})^2 + 2 \kappa_D^f \kappa_B^{Dh} r_D^f r_B^{Dh} \cos(\delta_B^{Dh} - \delta_D^f - \gamma) + \Gamma_\mathrm{mix}(x,y,\gamma, f,h), 
\label{Eq:B_time_integrated_rate}
\ee
where $\Gamma_\mathrm{mix}(x,y,\gamma,f,h)$ takes into account $D^0-\overline{D^0}$ mixing and it is given by  
\be
\begin{aligned}
\Gamma_\mathrm{mix}(x,y,\gamma, f,h) = & \  - \alpha y \bigg[ \kappa_D^f r_D^f \cos \delta_D^f (1 + (r_B^{Dh})^2) + \kappa_B^{Dh} r_B^{Dh} \cos(\delta_B^{Dh} - \gamma) (1 + (r_D^f)^2)   \bigg] \\ 
& \ + \alpha x  \bigg[ \kappa_D^f r_D^f \sin \delta_D^f ( (r_B^{Dh})^2 - 1) + \kappa_{B}^{Dh} r_{B}^{Dh} \sin(\delta_{B}^{Dh}  - \gamma) (1 - (r_D^{f})^2)  \bigg],
\end{aligned}
\label{Eq:MT_gamma}
\ee
up to quadratic corrections in the charm mixing and CP-violating parameters and neglecting direct CP violation. 
\noindent The $\alpha$ coefficient in Eq.~(\ref{Eq:MT_gamma}) describes the impact on $\Gamma_\mathrm{mix}(x,y,\gamma,f,h)$ of the non-trivial dependence of the signal selection efficiency on time and it is one when the latter is constant \cite{Rama:2013voa}. 
\\ Since at this order in the charm mixing and CP-violating parameters the decay rate of the CP conjugated process is obtained simply by replacing $\gamma$ with $- \gamma$ in Eq.~(\ref{Eq:B_time_integrated_rate}), the difference between $\Gamma (\overline{B}  \to  [f]_{D}h)$ and $\Gamma (B \to [\overline{f}]_{D} \overline{h})$ gives access to $\sin \gamma$. This is commonly referred to as the time-integrated CP asymmetry, which is an example of the so-called Gronau-London-Wyler (GLW) \cite{Gronau:1991dp, Gronau:1990ra} and Atwood-Dunietz-Soni (ADS) \cite{Atwood:1996ci, Atwood:2000ck} observables that we use in our combination. A complete list of the GLW and ADS modes considered in this work is reported in Tab.~\ref{Tab:ADSGLW}, while their parametrizations are derived in Apps.~\ref{Sec:GLW} and \ref{Sec:ADS}. We have taken the coherence factors for the modes $B \to DK^{*}$ from~\cite{LHCb:2017egy} and~\cite{LHCb:2016bsl} for charged and neutral $B$ mesons, respectively. 
\noindent For cascade decays with multi-body final states of the charm~\cite{Giri:2003ty, Belle:2004bbr, BPGGSZ_Belle}, as $K^0_SK^+K^-$, $K^0_S\pi^+\pi^-$, $\pi^+ \pi^- K^+K^-$, and $\pi^+\pi^-K^+K^-$, analyses of the variation of the decay rates across the phase space are available \cite{Belle:2019uav, LHCb:2020yot, Belle:2021efh, LHCb:2023yjo, LHCb:2023lib, LHCb:2023kpr, LHCb:2023ayf, Belle:2010xyn, LHCb:2024ett}. 
The Dalitz plots of these events are fitted using a model for the $D$ meson decay amplitudes or by binning the phase space and solving a system of linear equations in a model-independent way. The first method is more precise statistically with respect to the latter but introduces an additional systematic uncertainty due to the model. 
\begin{table}[]
\centering
\resizebox{\linewidth}{!}{
\begin{tabular}{c|c|c|c|c}
    \hline \hline & & & & \\ [-2.75 ex] 
   \textbf{GLW Observables} &    \textbf{ADS Observables} &  $\mathbf{B} \to \mathbf{D^0}$ \textbf{decays} & $\mathbf{D^0}$ \textbf{decays} & \textbf{Ref.} \\ \hline \hline & & &  \\ [-2.75 ex]
   $A_B^{\text{CP}}(XX\pi^0, h)$, &   $A^{\text{fav,sup}}(K\pi\pi^0, h)$, & $B^{\pm} \to D h^{\pm}$ & $D \to K \pi \pi^0$, &  \cite{LHCb:2021mmv} \\ [1. ex]
      $R^{\text{CP}}(XX\pi^0, K\pi\pi^0, K, \pi)$  &    $R_{\pm}(K\pi\pi^0,h)$  &  &  $D \to XX \pi^0$ &  \\ [1. ex]
   \hline & & &  \\ [-2.75 ex]  
  $A_B^{\text{CP}}(XX, h\pi\pi)$,  &     $A^{\text{fav}}(K\pi, h\pi\pi)$,    & $B^{\pm} \to D h^{\pm} \pi \pi$ & $D \to K \pi$, & \cite{LHCb:2015dlc}\\ [1. ex]
  $R^{\text{CP}}(XX, K\pi, K \pi \pi, \pi \pi \pi)$   &     $R_{\pm}(K\pi, h\pi\pi)$     & & $D \to XX$ & \\ [1. ex]
    \hline  & & &  \\  [-2.5 ex]
  $A_B^{\text{CP}}(XX, K^{*\pm})$, $R(XX, K\pi, K^{*\pm})$   &  $A^{\text{fav,sup}}(K\pi, K^{*\pm})$, $R^{\text{ADS}}(K\pi, K^{*\pm})$  & $B^{\pm} \to D K^{  * \pm}$  & $D \to K \pi$, $D \to XX$,  &  \cite{LHCb:2024ett}  \\ [1. ex]
   $A_B^{\text{CP}}(4\pi, K^{*\pm})$, $R(4\pi, K3\pi, K^{*\pm})$   &  $A^{\text{fav,sup}}(K3\pi, K^{*\pm})$, $R^{\text{ADS}}(K3\pi, K^{*\pm})$   &  & $D \to K3\pi$, $D \to 4 \pi$ & \\  [1.25 ex] \hline & & &  \\ [-2.5 ex]
   $A_B^{\text{CP}}(XX, K^{*0})$, $R(XX, K\pi, K^{*0})$  & $A^{\text{fav}}(K\pi, K^{*0})$, $R_{\pm}(K\pi, K^{*0})$ & $B^0 \to D K^{*0}$   & $D \to K \pi$, $D \to XX$, & \cite{LHCb:2024oco} \\ [1. ex]
   $A_B^{\text{CP}}(4\pi, K^{*0})$, $R(4\pi, K3\pi, K^{*0})$   & $A^{\text{fav}}(K3\pi, K^{*0})$, $R_{\pm}(K3\pi, K^{*0})$  & $B^0_s \to D\overline{K^{*0}} $  & $D\to K3\pi$, $D \to 4\pi$ & \\   [1.25 ex] \hline & & &  \\ [-2.5 ex]
   $A_B^{\text{CP}}(XX,h)$, $R^{\text{CP}}(XX, K\pi, K, \pi)$    & $A^{\text{fav}}(K\pi, K)$, $R_{\pm}(K\pi, h)$  & $B^{\pm} \to D h^{\pm}$ & $D \to K \pi$, $D \to XX$ & \cite{LHCb:2020hdx} \\ [1. ex] \hline & & &  &  \\ [-2.75 ex]
$A_B^{\text{CP}}(XX,h(\gamma,\pi^0))$,    & $A^{\text{fav}}(K\pi, K(\gamma, \pi^0))$,  & $B^{\pm} \to [D\pi^0]_{D^*} h^{\pm}$ & $D \to K \pi$, & \cite{LHCb:2020hdx} \\[1. ex]
$R^{\text{CP}}(XX, K\pi, K (\gamma, \pi^0), \pi(\gamma, \pi^0))$    &  $R_{\pm}(K\pi, h(\gamma,\pi^0))$  & $B^{\pm} \to  [D\gamma]_{D^*} h^{\pm}$ &  $D \to XX$  &  \\ [1. ex] \hline & & &  &  \\ [-2.75 ex]   
  & $A^{\text{fav,sup}}(K^0_SK\pi, h)$, &  & &  \\ [1.25 ex]
  ------ & $R^{\text{fav,sup}}(K^0_SK\pi, K, \pi)$,   & $B^{\pm} \to D h^{\pm}$  &  $D \to K^0_S K \pi$  & \cite{LHCb:2020vut},  \\ [1. ex]
  & $R^{\text{ADS}}(K^0_SK\pi, \pi)$ &  &  & \cite{Belle:2023lha} \\ [1. ex]\hline & & &  \\  [-2.5 ex]
  $A_B^{\text{CP}}(XX\pi\pi, h)$, $R^{\text{CP}}(XX\pi\pi, K3\pi, K, \pi)$ & ------  & $B^{\pm} \to D h^{ \pm}$ & $D \to X X \pi \pi$  & \cite{LHCb:2023yjo} \\ [1. ex] \hline & & &  \\  [-2.5 ex]
  ------ & $R^{\text{ADS}}(K\pi, h)$, $A^{\text{sup}}(K\pi, h)$ & $B^{\pm} \to D h^{\pm}$ & $D \to K\pi$  & \cite{Belle:2011ac} \\ [1. ex] \hline & & &  \\  [-2.5 ex] 
   ------ & $R^{\text{ADS}}(K\pi \pi^0, h)$, $A^{\text{sup}}(K\pi \pi^0, h)$ & $B^{\pm} \to D h^{\pm}$ & $D \to K\pi\pi^0$  & \cite{Belle:2013dtr}  \\ [1. ex] \hline & & &  \\  [-2.5 ex] 
   $A_B^{\text{CP}}(K^+K^-,K)$, $R^{\text{CP}}(K^+K^-,K\pi,K,\pi)$    &  & $B^{\pm} \to Dh^{\pm}$ & $D \to K^+K^-$, $D \to K\pi$, & \cite{Belle:2023yoe}  \\ [1. ex] 
   $A_B^{\text{CP}}(K^0_S\pi^0,K)$ , $R^{\text{CP}}(K^0_S\pi^0,K\pi, K,\pi)$    &  ------   &  & $D \to K^0_S \pi^0$ & \\ [1. ex] \hline & & &  \\  [-2.5 ex] 
     $A_B^{\text{CP}}(f^{\pm},K(\pi^0))$,  &  & $B^{\pm} \to [D\pi^0]_{D^*}h^{\pm}$ & $D \to f^{\pm}$, & \cite{Belle:2006cuz} \\ [1. ex] 
     $R^{\text{CP}}(f^{\pm},K\pi,K(\pi^0),\pi(\pi^0))$   & ------ &  & $D \to K\pi$ &  
     \\ [1. ex]\hline & & &  \\  [-2.5 ex]
  $A_B^{\text{CP}}(f^{\pm}, K)$, $R^{\text{CP}}(f^\pm, K\pi, K, \pi)$ & ------  & $B^{\pm} \to D K^{ \pm}$ & $D \to f^\pm$, $D \to K\pi$  & \cite{BaBar:2010hvw}
   \\ [1. ex]\hline & & &  \\  [-2.5 ex]
  $A_B^{\text{CP}}(f^{\pm}, K)$, $R^{\text{CP}}(f^\pm, K\pi, K, \pi)$ & ------  & $B^{\pm} \to D^* K^{ \pm}$ & $D \to f^\pm$, $D \to K\pi$  & \cite{BaBar:2008qcq}
  \\ [1. ex]\hline & & &  \\  [-2.5 ex]
  $A_B^{\text{CP}}(f^{\pm}, K^{*\pm})$, $R(f^\pm, K\pi, K^{*\pm})$ & $A^{\mathrm{sup}}(K\pi, K^{*\pm})$, $R^{\mathrm{ADS}}(K\pi, K^{*\pm})$  & $B^{\pm} \to D K^{* \pm}$ & $D \to f^\pm$, $D \to K\pi$  & \cite{BaBar:2009dzx}
  \\ [1. ex]\hline & & &  \\  [-2.5 ex]
  $A_B^{\text{CP}}(\pi^+\pi^-\pi^0, K)$ & ------  & $B^{\pm} \to D K^{ \pm}$ & $D \to \pi^+\pi^-\pi^0$  & \cite{BaBar:2007dro}
  \\ [1. ex] \hline & & &  &  \\ [-2.75 ex]
   & $[R_-(K\pi, h(\gamma, \pi^0)) + R_+(K\pi, h(\gamma, \pi^0))]/2$  & $B^{\pm} \to [D\pi^0]_{D^*} h^{\pm}$ & $D \to K \pi$ & \cite{BaBar:2010otv} \\[1. ex]
   ------ &  $\frac{R_-(K\pi, h(\gamma, \pi^0)) - R_+(K\pi, h(\gamma, \pi^0))}{R_-(K\pi, h(\gamma, \pi^0)) + R_+(K\pi, h(\gamma, \pi^0))}$  & $B^{\pm} \to  [D\gamma]_{D^*} h^{\pm}$ &   &
     \\ [1. ex]\hline & & &  \\  [-2.5 ex]
   & $[R_-(K\pi, h) + R_+(K\pi, h)]/2$ & $B^{\pm} \to D h^{ \pm}$ & $D \to K\pi$  & \cite{BaBar:2010otv}
   \\[1. ex]
  ------  &  $\frac{R_-(K\pi, h) - R_+(K\pi, h)}{R_-(K\pi, h) + R_+(K\pi, h)}$  &  &   &
    \\ [1. ex]\hline & & &  \\  [-2.5 ex]
  ------ & $R_\pm(K\pi\pi^0, K)$  & $B^{\pm} \to D K^{\pm}$ & $D \to K\pi\pi^0$  & \cite{BaBar:2011rud}
  \\ [1. ex]\hline & & &  \\  [-2.5 ex]
  $A_B^{\text{CP}}(XX, K)$, $R^{\text{CP}}(XX, K\pi, K, \pi)$ & ------  & $B^{\pm} \to D K^{ \pm}$ & $D \to XX$, $D \to K\pi$  & \cite{CDF:2009wnr}
  \\ [1. ex]\hline & & &  \\  [-2.5 ex]
  ------ & $A^{\mathrm{sup}}(K\pi, h)$, $R^{\mathrm{ADS}}(K\pi, h)$  & $B^{\pm} \to D h^{ \pm}$ &  $D \to K\pi$  & \cite{CDF:2011xrp}
  \\ [1. ex]  \hline \hline 
    \end{tabular} }
    \caption{GLW and ADS observables coming from $B$ mesons cascade decays. The $XX$ notation denotes both $K^+K^-$ and $\pi^+\pi^-$ states, while we use $K\pi$  for $K^{\mp} \pi^{\pm}$. $h$ stands for both $K$ and $\pi$ mesons. We refer to $\pi^+\pi^-\pi^+\pi^-$ and $K^{\mp} \pi^{\pm} \pi^+\pi^-$ final states as $4\pi$ and $K3\pi$, respectively. We use $f^{\pm}$ as a shorthand notation for CP-even and CP-odd eigenstates, respectively. The observables are described in details in Apps.~\ref{Sec:GLW} and~\ref{Sec:ADS}.}
    \label{Tab:ADSGLW}
    \end{table}

\noindent In both cases, the following cartesian observables are obtained:
\be
x_{\pm}^{Dh} = r_B^{Dh} \cos(\delta_B^{Dh} \pm \gamma), \qquad  \qquad y_{\pm}^{Dh} = r_B^{Dh} \sin(\delta_B^{Dh} \pm \gamma).
\label{Eq:GGSZ}
\ee   
Measuring the quantities in Eq.~(\ref{Eq:GGSZ}) for beauty decays with a pion in the final state is particularly hard due to the small value of $r_B^{D\pi} \sim \mathcal{O}(10^{-3})$. 
\begin{table}[]
        \centering
    \resizebox{11.5cm}{!}{
    \begin{tabular}{c|c|c|c}
     \hline \hline & &  & \\ [-2.75 ex] 
   \textbf{BPGGSZ cartesian observables}  &  $\mathbf{B} \to \mathbf{D^0}$ \textbf{decays} & $\mathbf{D^0}$ \textbf{decays} & \textbf{Ref.} \\ \hline \hline & & &  \\ [-2.5 ex]
     $x_{\pm}^{DK^{*0}}$, $y_{\pm}^{DK^{*0}}$  & $B^0 \to D K^{*0}$ & $D \to K^0_S X X$ & \cite{LHCb:2023ayf}   \\ [1. ex]   \hline & & &  \\ [-2.5 ex] 
      $x^{DK}_{\pm}$, $x^{D\pi}_{\xi}$  & $B^{\pm} \to DK^{\pm}$ & $D \to K^0_S \pi^+\pi^-\pi^0$ & \cite{Belle:2019uav} \\ [1. ex] 
     $y_{\pm}^{DK}$, $y^{D\pi}_{\xi}$  & $B^{\pm} \to D\pi^{\pm}$ & & \\ [1. ex] \hline & & &  \\ [-2.5 ex] 
    $x^{DK}_{\pm}$, $x^{D\pi}_{\xi}$  & $B^{\pm} \to DK^{\pm}$ & $D \to K^0_S X X$ & \cite{LHCb:2020yot, Belle:2021efh} \\ [1. ex] 
     $y_{\pm}^{DK}$, $y^{D\pi}_{\xi}$  & $B^{\pm} \to D\pi^{\pm}$ & & \\ [1. ex] \hline & & &  \\ [-2.5 ex] 
     $x^{DK}_{\pm}$, $x^{D\pi}_{\xi}$  & $B^{\pm} \to DK^{\pm}$ & $D \to K^+K^- \pi^+ \pi^-$ & \cite{LHCb:2023yjo} \\ [1. ex] 
     $y_{\pm}^{DK}$, $y^{D\pi}_{\xi}$  & $B^{\pm} \to D\pi^{\pm}$ & &
     \\ [1. ex] \hline & & &  \\ [-2.5 ex] 
     $x^{[D\pi^0]_{D^*}K}_{\pm}$, $x^{[D\gamma]_{D^*}K}_{\pm}$  & $B^{\pm} \to [D\pi^0]_{D^*}K^{\pm}$ & $D \to K^0_S \pi^+ \pi^-$ & \cite{Belle:2010xyn} \\ [1. ex] 
     $y_{\pm}^{[D\pi^0]_{D^*}K}$, $y^{[D\gamma]_{D^*}K}_{\pm}$  & $B^{\pm} \to [D\gamma]_{D^*}K^{\pm}$& & \\ [1. ex] \hline & & &  \\ [-2.5 ex] 
     $x^{D^*K}_{\pm}$, $x^{D^*\pi}_{\xi}$  & $B^{\pm} \to D^*K^{\pm}$ & $D \to K^0_S X X$ & \cite{LHCb:2023lib, LHCb:2023kpr} \\ [1. ex] 
     $y_{\pm}^{D^*K}$, $y^{D^*\pi}_{\xi}$  & $B^{\pm} \to D^*\pi^{\pm}$ & & \\ [1. ex]  \hline & & &  \\ [-2.5 ex] 
     $x_{\pm}^{DK^{*\pm}}$, $y_{\pm}^{DK^{*\pm}}$  & $B^\pm \to D K^{*\pm}$ & $D \to K^0_S X X$ & \cite{LHCb:2024ett} \\ [1. ex]  \hline & & &  \\ [-2.5 ex] 
    $x^{DK}_{\pm}$, $x^{D\pi}_{\xi}$  & $B^{\pm} \to DK^{\pm}$ & $D \to K^{\mp}\pi^{\pm}\pi^+\pi^-$ & \cite{LHCb:2022nng}\tablefootnote{We thank Matthew W. Kenzie for providing us with the results of the analysis in terms of cartesian observables.} \\ [1. ex] 
     $y_{\pm}^{DK}$, $y^{D\pi}_{\xi}$  & $B^{\pm} \to D\pi^{\pm}$ & &
     \\ [1. ex]  \hline & & &  \\ [-2.5 ex] 
     $\sqrt{(x_\pm^{DK} - (2F_+^{\pi\pi\pi^0}-1))^2+(y_{\pm}^{DK})^2}$  & $B^\pm \to D K^{\pm}$ & $D \to \pi^+\pi^-\pi^0$ & \cite{BaBar:2007dro} 
     \\ [1. ex] 
     $\arctan \bigg[ y_{\pm}^{DK}/(x_{\pm}^{DK} - (2F_+^{\pi\pi\pi^0}-1) ) \bigg]$  &  & & 
     \\ [1.2 ex] \hline & & &  \\ [-2.5 ex] 
     $x^{DK}_{\pm}$, $y_{\pm}^{DK}$, $x^{D^*K}_{\pm}$, $y_{\pm}^{D^*K}$  &  $B^{\pm} \to D^{(*)}K^{\pm}$ & $D \to K^0_S X X$ & \cite{BaBar:2010uep} \\ [1.2 ex] \hline & & &  \\ [-2.5 ex] 
     $x_{\pm}^{DK^{*\pm}}$,  $y_{\pm}^{DK^{*\pm}}$  & $B^{\pm} \to DK^{*\pm}$ & $D \to K^0_S \pi\pi$ & \cite{BaBar:2010uep}
     \\ [1. ex]  \hline & & &  \\ [-2.5 ex] 
     $x_{\pm}^{DK^{*\pm}}$, $y_{\pm}^{DK^{*\pm}}$  & $B^\pm \to D K^{*\pm}$ & $D \to K^0_S \pi\pi$ & \cite{Belle:2006lys}
     \\ [1. ex]  \hline & & &  \\ [-2.5 ex] 
     $x_{\pm}^{DK^{*0}}$, $y_{\pm}^{DK^{*0}}$  & $B^0 \to D K^{*0}$ & $D \to K^0_S \pi\pi$ & \cite{Belle:2015roy}
     \\ [1. ex]  
     \hline \hline
    \end{tabular}}
    \caption{BPGGSZ cartesian observables used in the combination. Here, $XX$ stands for both the $K^+K^-$ and $\pi^+\pi^-$ states. The dependence of the observables on the fit parameters are reported in Eqs.~(\ref{Eq:GGSZ}) and~(\ref{Eq:matrix_elements}).}
        \label{Tab:GGSZ}
    \end{table}

\noindent In this case, the cartesian observables for $\overline{B} \to [f]_D \pi$ are written 
as rotated with respect to the ones entering $\overline{B} \to [f]_D K$ through the transformation 
\be
\begin{pmatrix}
    x_{\pm}^{D\pi} \\ 
    y_{\pm}^{D\pi}
\end{pmatrix}  = \begin{pmatrix}
    x_{\xi}^{D\pi} & - y_{\xi}^{D\pi} \\ 
    y_{\xi}^{D\pi} & x_{\xi}^{D\pi}
\end{pmatrix} \begin{pmatrix}
    x_{\pm}^{DK} \\ 
    y_{\pm}^{DK}
\end{pmatrix},
\label{Eq:GGSZ_pi}
\ee
where $x_{\xi}^{D\pi}$ and $y_{\xi}^{D\pi}$ are given by 
\be
x_{\xi}^{D\pi} = \frac{r_B^{D\pi}}{r_B^{DK}} \cos(\delta_B^{D\pi} - \delta_B^{DK}), \qquad \qquad   y_{\xi}^{D\pi} = \frac{r_B^{D\pi}}{r_B^{DK}} \sin(\delta_B^{D\pi} - \delta_B^{DK}).
\label{Eq:matrix_elements}
\ee
Then, the coefficients in Eq.~(\ref{Eq:matrix_elements}) are extracted simultaneously with $x_{\pm}^{DK}$ and $y_{\pm}^{DK}$. The cartesian observables used in our combination are reported in Tab.~\ref{Tab:GGSZ}. \\  
It is worth noting that the expressions in Eqs.~(\ref{Eq:B_time_integrated_rate}), (\ref{Eq:MT_gamma}), (\ref{Eq:GGSZ}) and~(\ref{Eq:matrix_elements})  are still valid for cascade decays with the $D^0$ or $\overline{D^0}$ meson produced together with a $\pi^0$ meson or a photon $\gamma$ by the decay of a $D^*$ intermediate state, simply replacing the strong phase $\delta_D^f$ with $\delta_{D^*}^f$ or $\delta_{D^*}^f + \pi$, respectively. 
\begin{figure}[]
    \centering
\includegraphics[width=0.45\columnwidth]{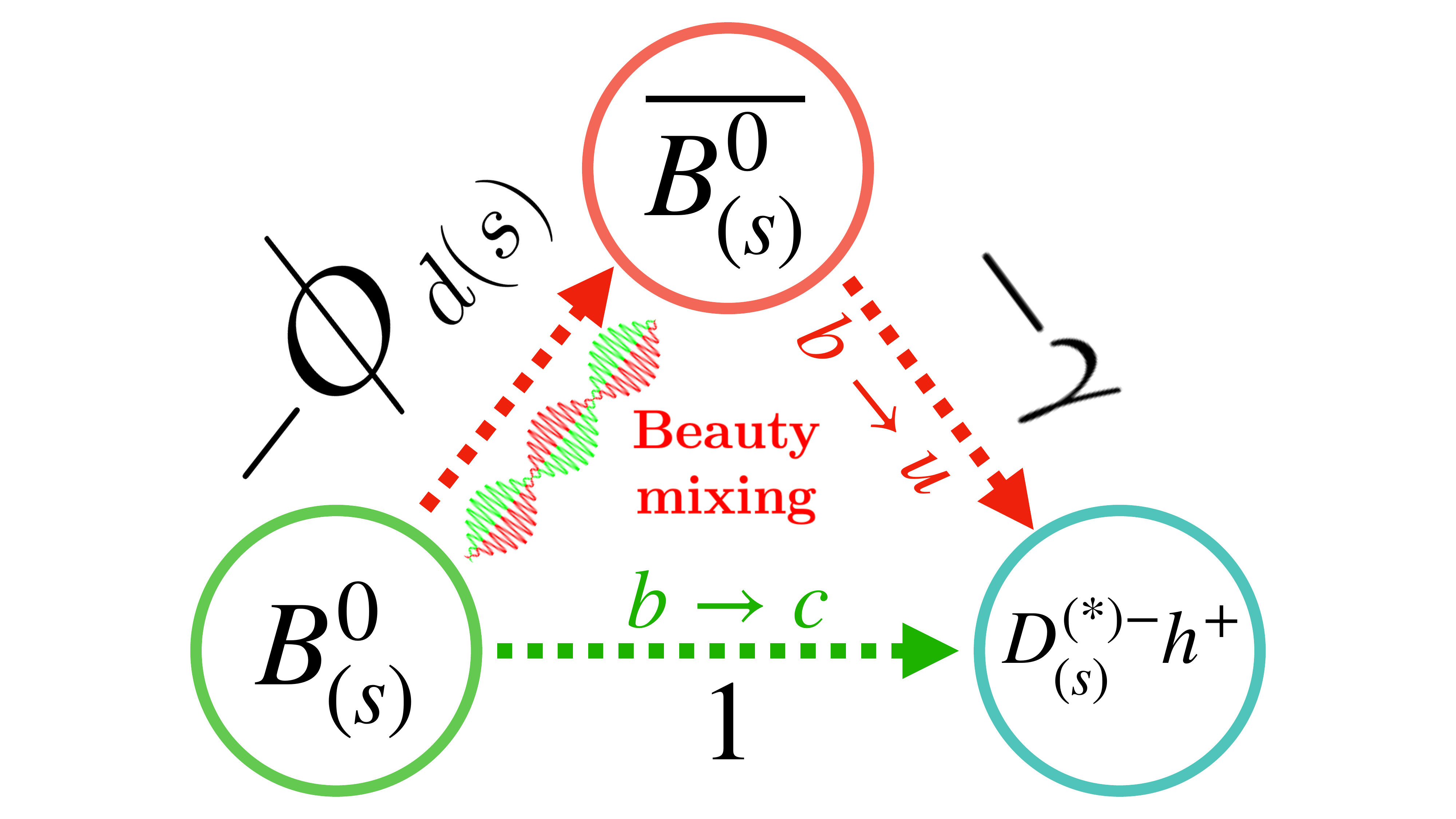}
\caption{Schematic representation of the extraction of the CKM angle $\gamma$ from the CP-violating phase of the interference between neutral $B$ decays to charmed mesons with and without mixing. Here, $h^+$ stands for $\pi^+$ and $\rho^+$($K^+$ and $K^+\pi^+\pi^-$) for $B^{0}_{(s)}$ decays.}
\label{Fig:neutralB_scheme}
\end{figure}

\subsection{Neutral \texorpdfstring{$B$}{B} decays to charmed mesons}
The CKM angle $\gamma$ can be obtained from the CP-violating phase of the interference between neutral $B$ decays to charmed mesons $f$ (\emph{e.g.} $f = D^{(*)-}\pi^+, D^-\rho^+$, $ D^-_s K^+$, and $D_s^-K^+\pi^+\pi^-$) with and without mixing, as shown schematically in Fig.~\ref{Fig:neutralB_scheme}.
The decay amplitudes for these processes can be organized into ratios of magnitudes, coherence factors and strong phases through the following integrals over the phase space  $\Phi_{B^0_{(s)}}$: 
\be
(r_{B^0_{(s)}}^f)^2 = \frac{\int \text{d} \Phi_{B^0_{(s)}} \left \vert \mathcal{A}_{\overline{B}^{0}_{(s) }}^{f} \right \vert^2}{  \int \text{d} \Phi_{B^0_{(s)}} \left \vert \mathcal{A}_{B^0_{(s)}}^f \right \vert^2}, \qquad \qquad \kappa_{B^0_{(s)}}^f e^{i (\delta_{B^0_{(s)}}^f + \varphi_{(s)})} =  \frac{\int \text{d} \Phi_{B^0_{(s)}} \mathcal{A}_{\overline{B}^0_{(s)}}^f \  \mathcal{A}_{B^0_{(s)}}^{f*} }{\sqrt{\int \text{d} \Phi_{B^0_{(s)}} \left \vert \mathcal{A}_{\overline{B}^{0}_{(s) }}^{f} \right \vert^2} \sqrt{\int \text{d} \Phi_{B^0_{(s)}}  \left \vert \mathcal{A}_{B^0_{(s)}}^f \right \vert^2}  },
\label{Eq:B0_decay_amplitudes}
\ee
where $\varphi_{(s)}$ are the weak phases entering the decay, which are given in the SM by a combination of the relevant CKM matrix elements. 
Then, the time-dependent rates of neutral $B^0_{(s)}$ meson decays can be written as \cite{Dunietz:2000cr, Artuso:2015swg}
\be
\Gamma(B^0_{(s)} \to f, t) \propto \cosh( t \Delta \Gamma_{(s)} /2) - G_f \sinh( t \Delta \Gamma_{(s)} /2) + C_f \cos(t \Delta m_{(s)}) - S_f \sin(t \Delta m_{(s)}),
\label{Eq:B0_rates}
\ee
\noindent where $\Delta m_{(s)}$ and $\Delta \Gamma_{(s)}$ are the differences in masses and lifetimes between the heavier and lighter eigenstates of the $B^0_{(s)}$ mixing Hamiltonian. 
The observables $C_f$, $G_f$ and $S_f$ are fitted from the data \cite{LHCb:2018zap, LHCb:2020qag, LHCB-PAPER-2024-020, BaBar:2005jis, BaBar:2006slj, Belle:2006lts, Belle:2011dhx} and are given by
\be
C_f = \frac{1 - (r_{B^0_{(s)}}^f)^2}{1 + (r_{B^0_{(s)}}^f)^2},   \ \ \   G_f = \frac{- 2 \eta_{(s)} \kappa_{B^0_{(s)}}^f r_{B^0_{(s)}}^f }{1 + (r_{B^0_{(s)}}^f)^2} \cos(\delta_{B^0_{(s)}}^f - (  \phi_{d(s)} + \gamma ) ),  \ \ \     S_f = \frac{ 2 \eta_{(s)}\kappa_{B^0_{(s)}}^f r_{B^0_{(s)}}^f }{1 + (r_{B^0_{(s)}}^f)^2} \sin(\delta_{B^0_{(s)}}^f - (  \phi_{d(s)} + \gamma ) ),
\label{Eq:B0_obs}
\ee
with $\eta=-1$ and $\eta_{s} =1$. 
Here, we have neglected CP violation in pure $B^0_{(s)} - \overline{B}^0_{(s)}$ mixing, as well as direct CP violation. 
\noindent The phases $\phi_{(d,s)}$ are obtained in the SM from the CKM angles $\beta = \arg( -V_{td}^* V_{cd} V_{cb}^* V_{tb} )$ and  $\beta_s = \arg( -V_{cs}^* V_{ts} V_{tb}^* V_{cb} )$ as $\phi_d =  2 \beta $ and $\phi_s = - 2 \beta_s$, up to corrections to the fourth power of the sine of the Cabibbo angle. Similar observables to the ones in Eq.~(\ref{Eq:B0_obs}) can be obtained for the CP conjugate final states by changing the sign of $\eta_{(s)}$ and of the CP-violating phases between mixing and decay $-(\phi_{d(s)} + \gamma)$. 
Beyond the SM, the relations between $\phi_{d(s)}$ and $\beta_{(s)}$ could be altered, but one can still use the experimental values of $\phi_{(d,s)}$ to take into account $B^0_{(s)}$ mixing effects in the extraction of $\gamma$. 
The measurements used in the combination are reported in Tab.~\ref{Tab:B0_obs}, while inputs for $\phi_d$ and $\phi_s$ are taken from the UTfit Summer 2024 update~\cite{ICHEP24-Bona}  (including also the theoretical uncertainty~\cite{Ciuchini:2005mg}) and from the
HFLAV Summer 2024 average of $B^0_s \to J/\psi \phi$ modes~\cite{HFLAV:2022pwe}, respectively. 
    \begin{table}[]
    \centering
    \begin{tabular}{c|c|c}
    \hline \hline & & \\  [-2.5 ex]
    \textbf{Observables} & $\mathbf{B^0_{(s)}} \to \mathbf{D_{(s)}^{\mp} } \mathbf{h^{\pm}}$ \textbf{decays}  & \textbf{Ref.} \\ [ 1. ex] \hline \hline & &  \\ [-2. ex]
  $C_{D_s^{\mp} K^{\pm}}$, \ \ \ \ \ \ \ \ \  $G_{D_s^{\mp} K^{\pm}}$,  \ \ \ \ \ \ \ \ \  $S_{D_s^{\mp} K^{\pm}}$ & $B^0_s \to D_s K$ & \cite{LHCB-PAPER-2024-020}\\ [1. ex] \hline & &  \\ [-2. ex]
     $S_{D^{\mp} \pi^{\pm}}$ & $B^0 \to D \pi$ & \cite{LHCb:2018zap}\\ [1. ex] \hline & &  \\ [-2. ex] 
    $C_{D_s^{\mp} K^{\pm} \pi \pi}$, \ \ \ \ \ \ \ \ \ $G_{D_s^{\mp} K^{\pm} \pi \pi}$, \ \ \ \ \ \ \ \ \ $S_{D_s^{\mp} K^{\pm} \pi \pi}$ & $B^0_s \to D_s K \pi \pi$ & \cite{LHCb:2020qag}
    \\ [1. ex] \hline & &  \\ [-2. ex]
     $-(S_{D^{-}\pi^{+}}+S_{D^+\pi^-})/2$, \ $(S_{D^{-}\pi^{+}}-S_{D^+\pi^-})/2$  & $B^0 \to D \pi$ & \cite{BaBar:2006slj, Belle:2006lts}
     \\ [1. ex] \hline & &  \\ [-2. ex]
     $-(S_{D^{*-}\pi^{+}}+S_{D^{*+}\pi^-})/2$, \ $(S_{D^{*-}\pi^{+}}-S_{D^{*+}\pi^-})/2$  & $B^0 \to D^* \pi$ & \cite{BaBar:2005jis, BaBar:2006slj, Belle:2006lts, Belle:2011dhx}
      \\ [1. ex] \hline & &  \\ [-2. ex]
     $-(S_{D^{-}\rho^{+}}+S_{D^{+}\rho^-})/2$, \ $(S_{D^{-}\rho^{+}}-S_{D^{+}\rho^-})/2$  & $B^0 \to D \rho$ & \cite{BaBar:2006slj}
    \\ [1. ex] \hline \hline
    \end{tabular}
    \caption{ Observables coming from time-dependent analyses of neutral $B$ mesons decays to charmed mesons. The observables are defined in Eq.~(\ref{Eq:B0_obs}).}
    \label{Tab:B0_obs}
    \end{table}

\noindent For the purpose of using $\gamma$ in the Unitarity Triangle analysis, the correlation between the results for $\gamma$ and the inputs used for $\phi_{(d,s)}$ should be taken into account. 
However, in practice, for current experimental uncertainties such correlation is lost in the global fit, although it might move the central value of $\gamma$ at the degree level when only $B_{s}$ decays are considered.

\section{Results}
\label{Sec:Results}
\begin{table}[] 
\centering 
\begin{tabular}{c|ccc|ccc} 
\hline \hline & & & & & &    \\ [-2.5 ex] 
& \multicolumn{3}{|c|}{\textbf{All modes}} & \multicolumn{3}{|c}{\textbf{Charm modes only}} \\ \cline{2-7}   & & & & & &  \\ [-2.5 ex] 
 \textbf{Par.} &  \textbf{Value} & \textbf{Unc.} & \textbf{95.4\% Prob.} & \textbf{Value} & \textbf{Unc.} & \textbf{95.4\% Prob.}  \\ [0.75 ex] \hline  &                          & & & & &  \\ [-2.5 ex] 
$\phi_2^M[\degree]$ & $ 0.13 $ & $ \pm 0.70 $ & $[ -1.27 , 1.54 ]$ & $ 0.07 $ & $ \pm 0.73 $ & $[ -1.42 , 1.53 ]$ \\ [0.75 ex]  
$\phi_2^\Gamma[\degree]$ & $ 2.1 $ & $ \pm 1.6 $ & $[ -1.0 , 5.2 ]$ & $ 1.9 $ & $ \pm 1.6 $ & $[ -1.1 , 4.9 ]$ \\ [0.75 ex]  
$\phi_2[\degree]$ & $ -1.5 $ & $ \pm 1.1 $ & $[ -3.7 , 0.6 ]$ & $ -1.3 $ & $ \pm 1.1 $ & $[ -3.5 , 0.7 ]$ \\ [0.75 ex]  
$\vert q/p \vert -1[\%]$ & $ -1.6 $ & $ \pm 1.5 $ & $[ -4.5 , 1.4 ]$ & $ -1.4 $ & $ \pm 1.5 $ & $[ -4.3 , 1.5 ]$ \\ [0.75 ex]  
$x_{12} \simeq x[\permil]$ & $ 4.01 $ & $ \pm 0.43 $ & $[ 3.16 , 4.83 ]$ & $ 3.96 $ & $ \pm 0.43 $ & $[ 3.12 , 4.80 ]$ \\ [0.75 ex]  
$y_{12} \simeq y[\permil]$ & $ 6.10 $ & $ \pm 0.17 $ & $[ 5.77 , 6.45 ]$ & $ 6.28 $ & $ \pm 0.23 $ & $[ 5.84 , 6.74 ]$ \\ [0.75 ex]  
$a_D^{KK}[\permil]$ & $ 0.40 $ & $ \pm 0.53 $ & $[ -0.63 , 1.43 ]$ & $ 0.40 $ & $ \pm 0.53 $ & $[ -0.64 , 1.44 ]$ \\ [0.75 ex]  
$a_D^{\pi\pi}[\permil]$ & $ 2.34 $ & $ \pm 0.60 $ & $[ 1.15 , 3.53 ]$ & $ 2.34 $ & $ \pm 0.61 $ & $[ 1.15 , 3.52 ]$ \\ [0.75 ex]  
$\phi_{12}[\degree]$ & $ -2.0 $ & $ \pm 1.8 $ & $[ -5.6 , 1.7 ]$ & $ -1.8 $ & $ \pm 1.9 $ & $[ -5.5 , 1.9 ]$ \\ [0.75 ex]  
$r_D^{K\pi}[\%]$ & $ 5.8573 $ & $ \pm 0.0094 $ & $[ 5.8387 , 5.8758 ]$ & $ 5.8601 $ & $ \pm 0.0099 $ & $[ 5.8406 , 5.8796 ]$ \\ [0.75 ex]  
$\delta_D^{K\pi}[\degree]$ & $ 191.4 $ & $ \pm 2.4 $ & $[ 186.6 , 196.0 ]$ & $ 194.1 $ & $ \pm 3.2 $ & $[ 187.6 , 200.3 ]$  \\ [.75 ex] \hline \hline 
\end{tabular} 
\caption{Results for the charm part of the combination when using all the observables (``All modes'') or exluding the beauty measurements (``Charm only''). The half-width of the smallest interval containing at least $68 \%$ probability (Unc.) and its center (Value) are reported for each quantity of interest. The smallest interval containing at least $95 \%$ probability is reported as well.  }
\label{Tab:results}
\end{table}

We combine the observables in Tabs.~\ref{Tab:D_obs}, \ref{Tab:ADSGLW}, \ref{Tab:GGSZ} and~\ref{Tab:B0_obs} in a Bayesian framework, employing as likelihood the product of Gaussian distributions for each set of correlated measurements. The fit parameters are assumed to follow uniform distributions in a sufficiently large range. It is worth noting that the beauty observables described in Sec.~\ref{Sec:Beauty_Obs} and Apps.~\ref{Sec:GLW} and~\ref{Sec:ADS} are symmetric under the simultaneous transformation $\gamma \to \gamma \pm \pi$ and $\delta_B^{Dh} \to \delta_{B}^{Dh} \pm \pi$, for each of the $B$ modes considered, and we have limited our study to just one of these solutions. Other solutions can be found by adding $\pm \pi$ to our results. The posterior pdfs of the parameters are determined through a Markov Chain Monte Carlo (MCMC) algorithm\footnote{The fit code is available at \url{https://github.com/silvest/GammaDDbar} \cite{palma_silvestgammaddbar_2024}.} implemented using the Bayesian Analysis Toolkit (BAT) software package \cite{Caldwell:2008fw}. We perform four combinations using different subsets of the beauty observables, divided by the species of the $B$ mesons: all the modes together, only charged $B$, only neutral $B$ and $B_s$ modes.  Furthermore, we perform a combination of the charm observables only, to compare the results with the ones obtained by HFLAV and to assess the impact of the combined analysis.
\begin{figure}[]
    \centering
\includegraphics[width=0.3\columnwidth]{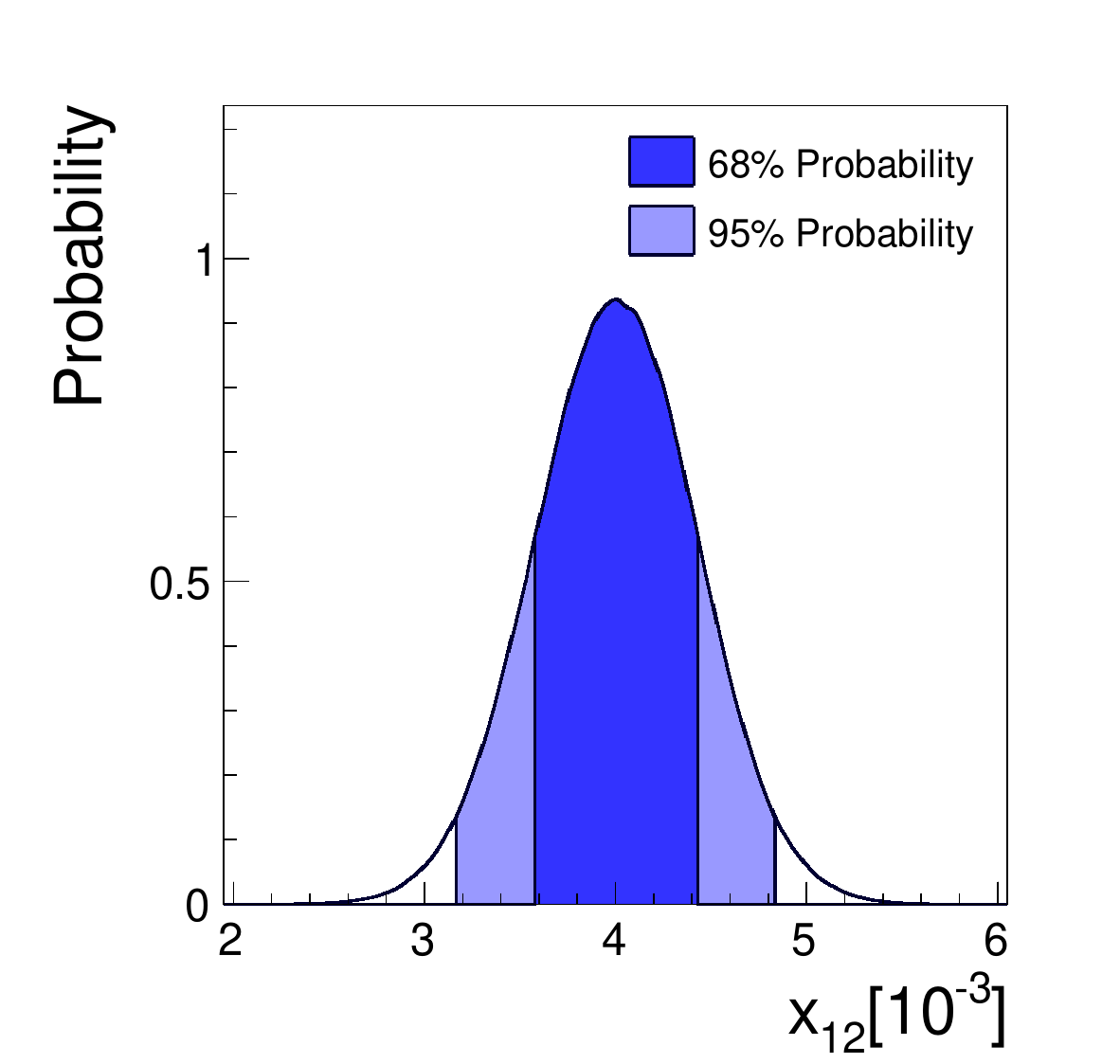}
\includegraphics[width=0.3\columnwidth]{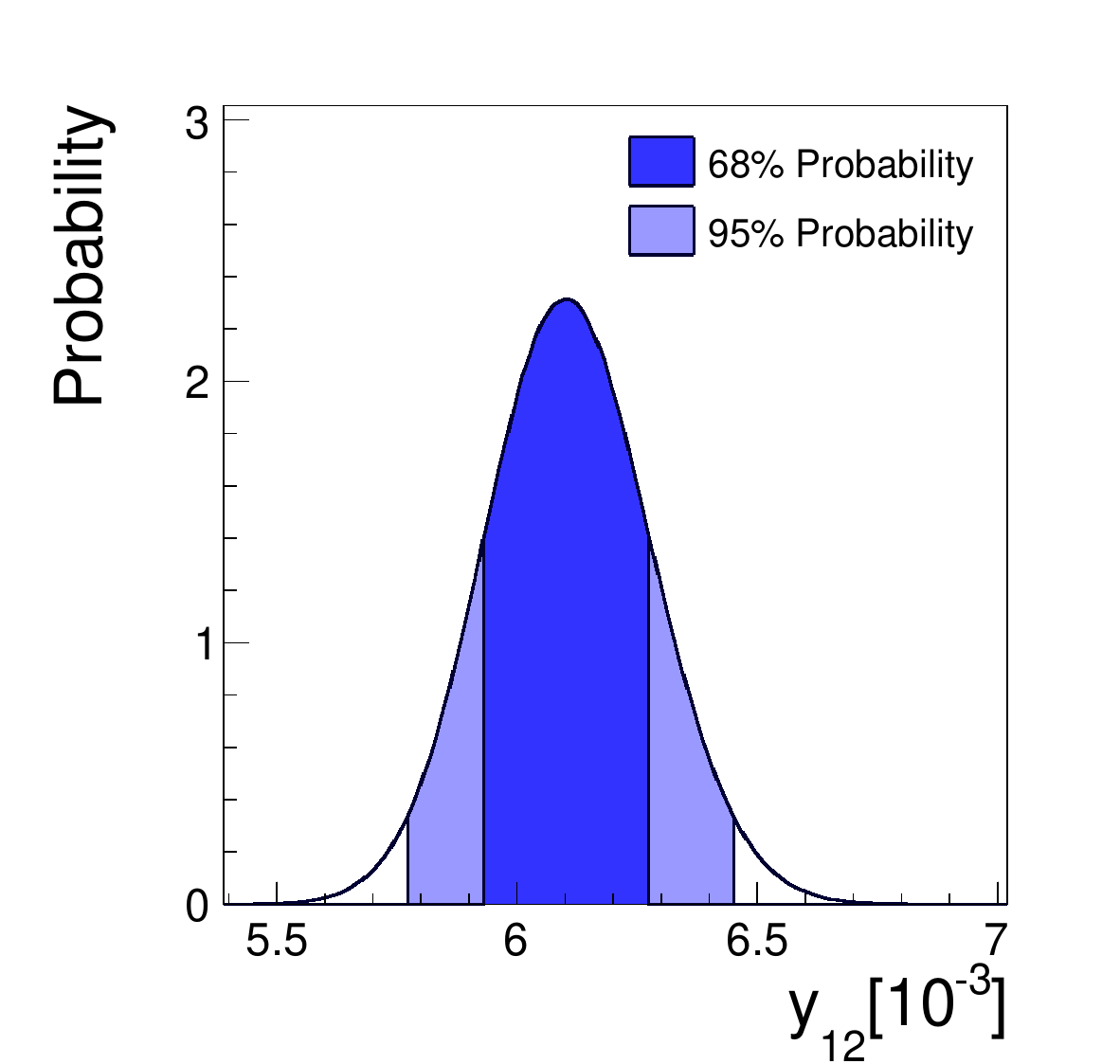}
\includegraphics[width=0.3\columnwidth]{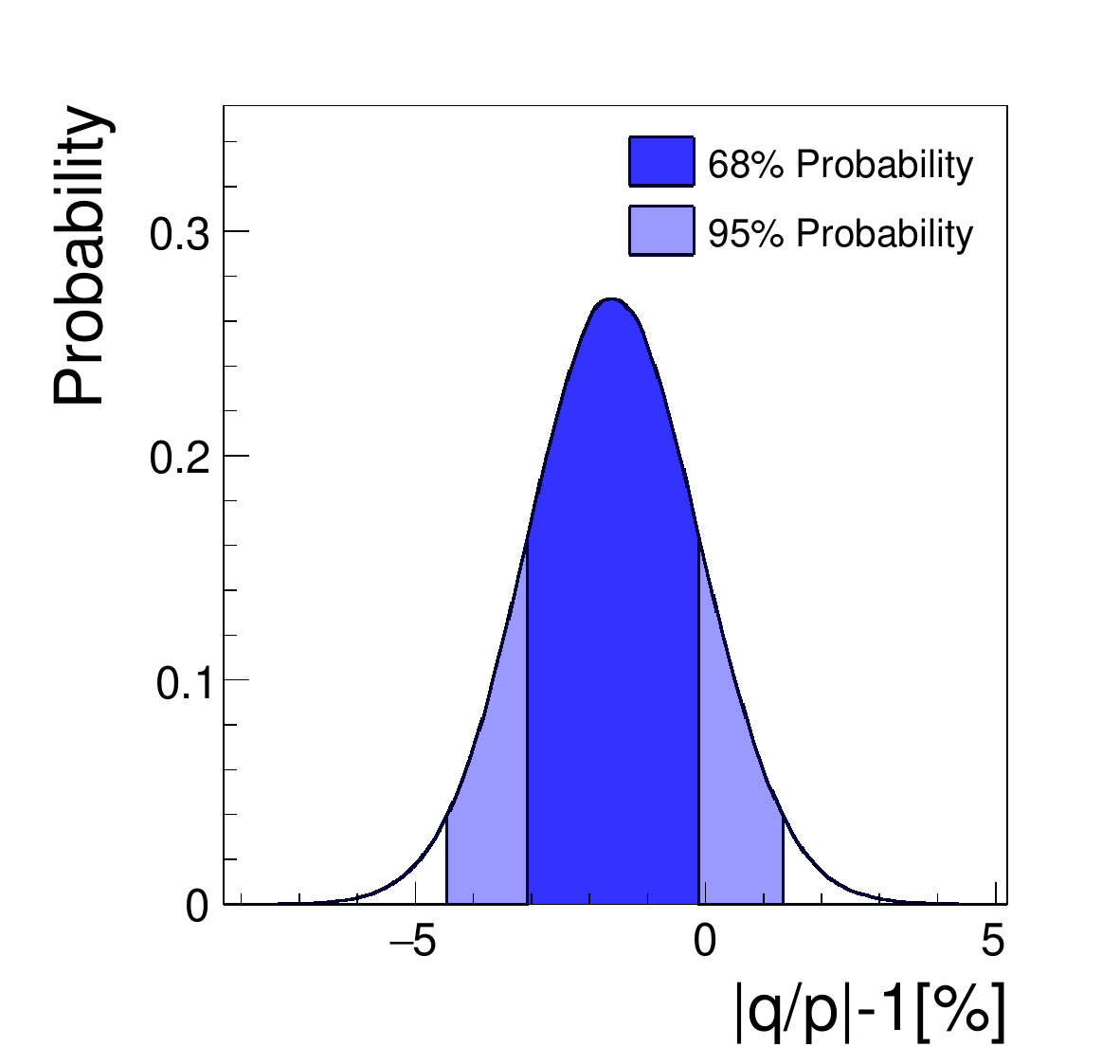}
\includegraphics[width=0.3\columnwidth]{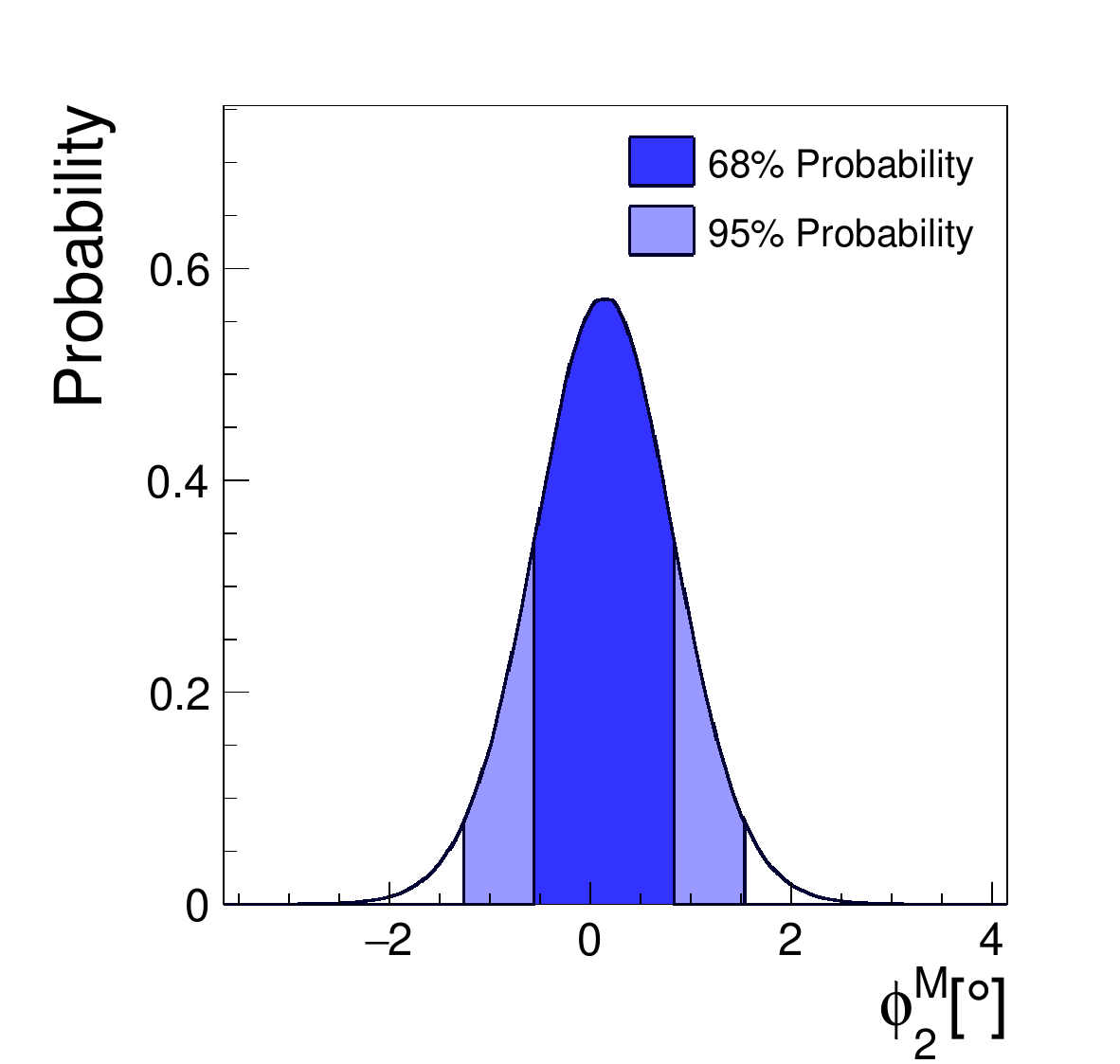}
\includegraphics[width=0.3\columnwidth]{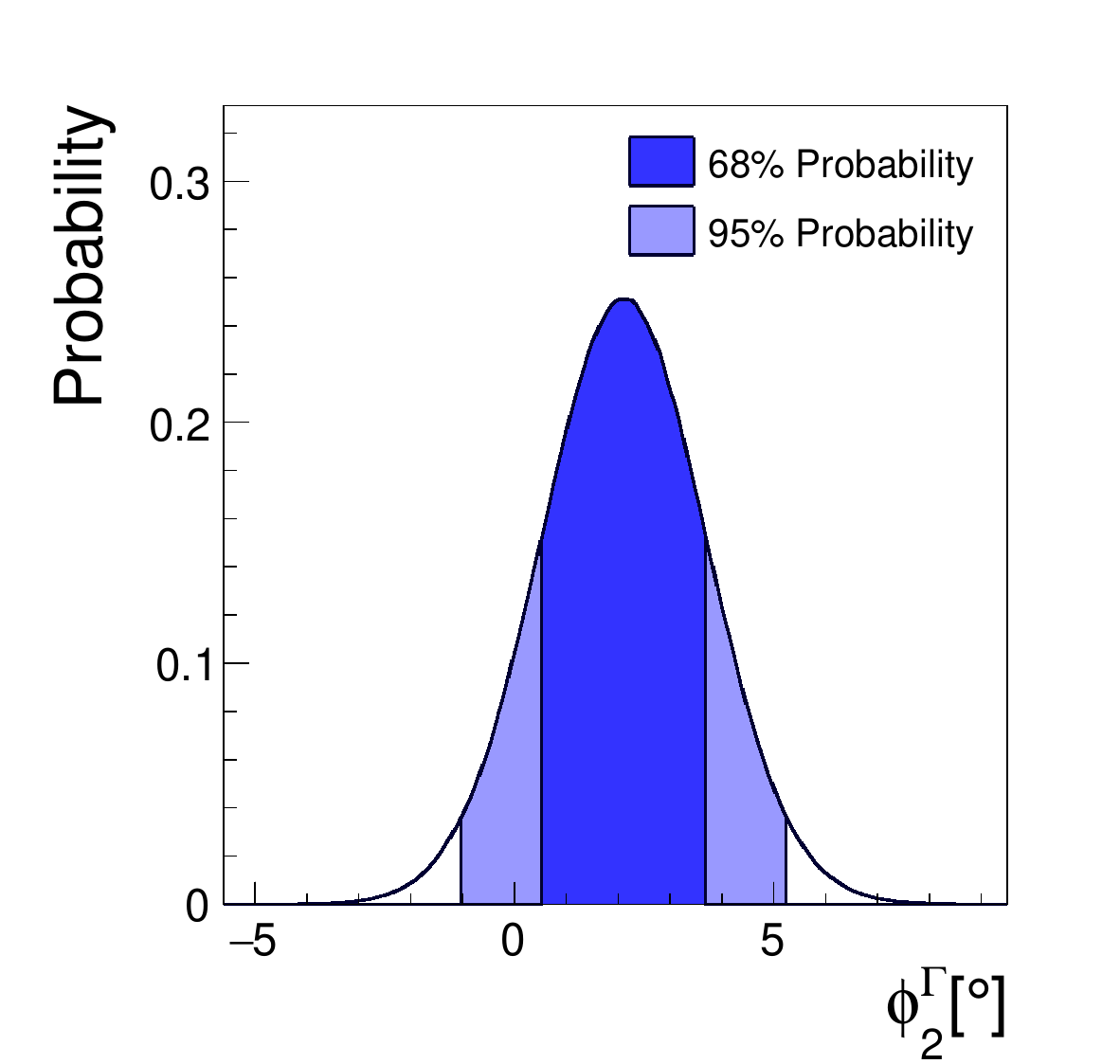}
\includegraphics[width=0.3\columnwidth]{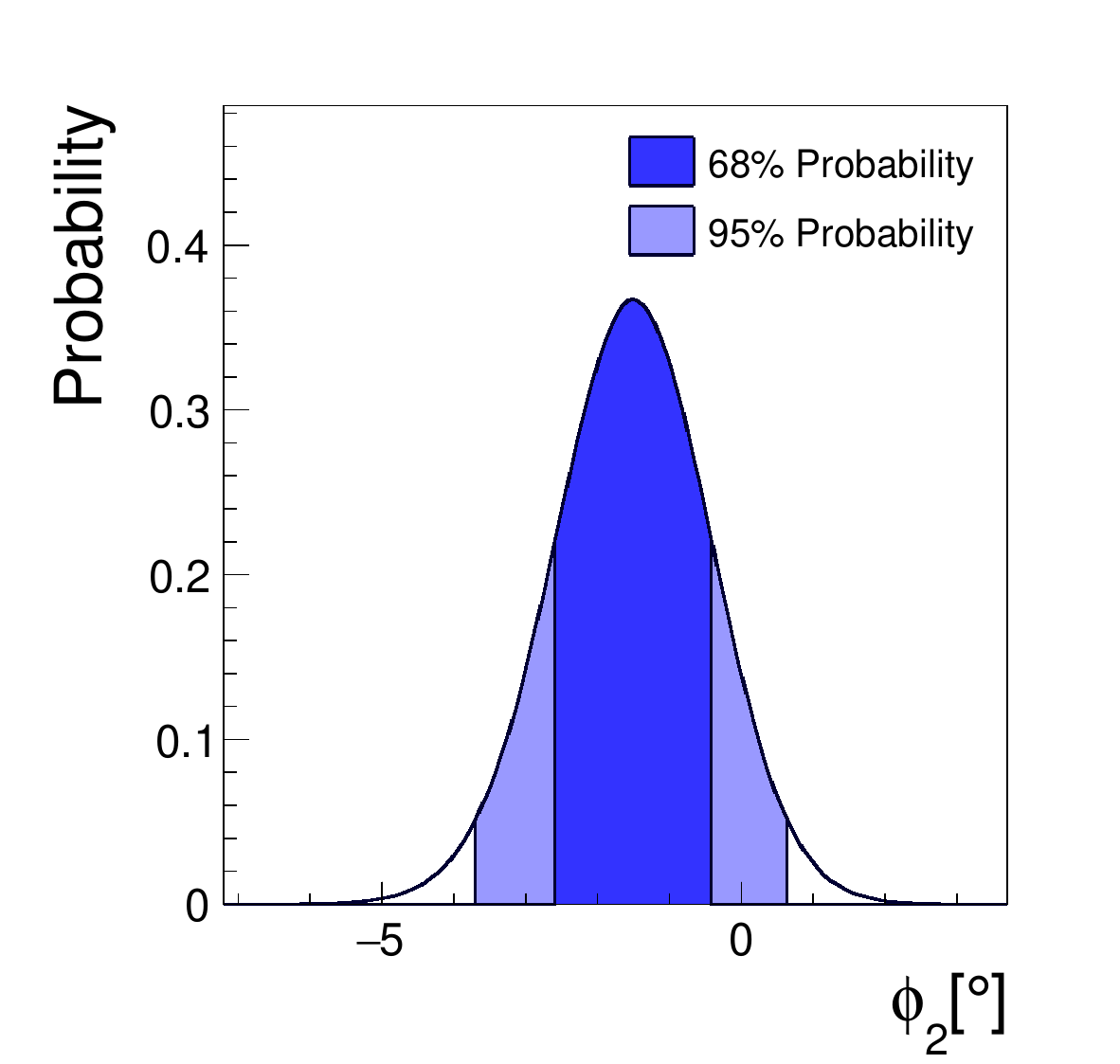}
\caption{Pdfs of the charm mixing and CP-violating parameters obtained using all the observables.} 
\label{Fig:all_1D}
\end{figure}

\begin{figure}[]
    \centering
\includegraphics[width=0.32\columnwidth]{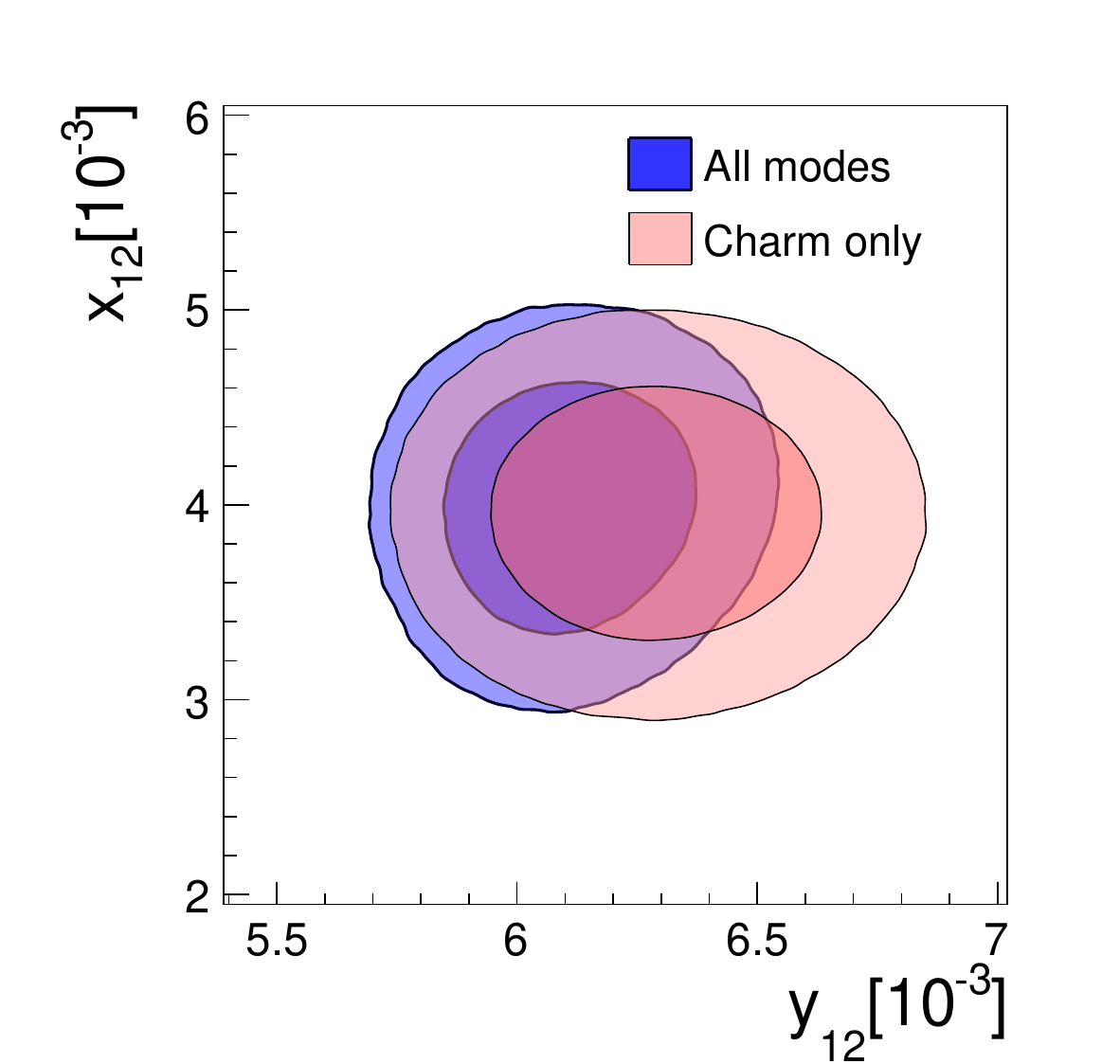}
\includegraphics[width=0.32\columnwidth]{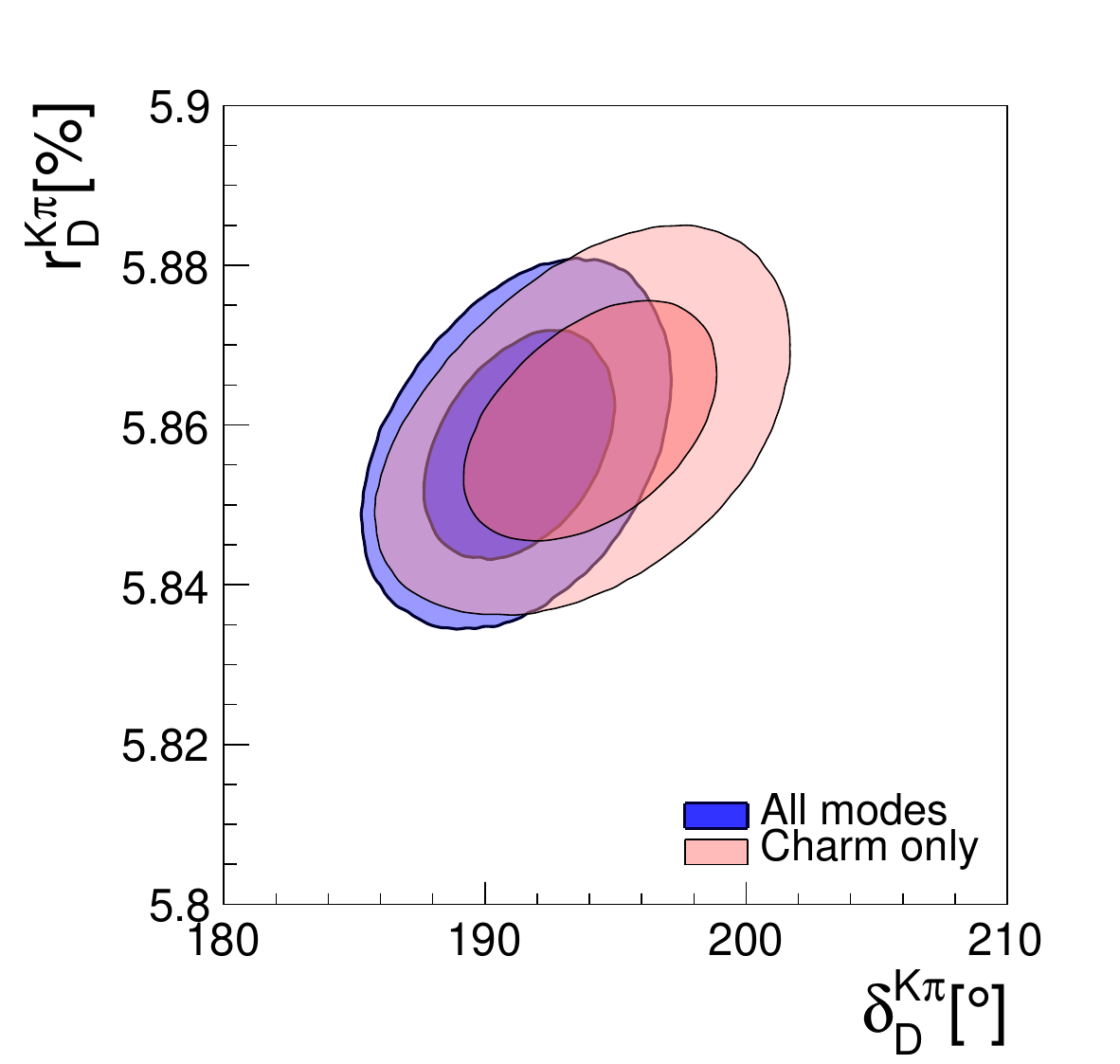}
\includegraphics[width=0.32\columnwidth]{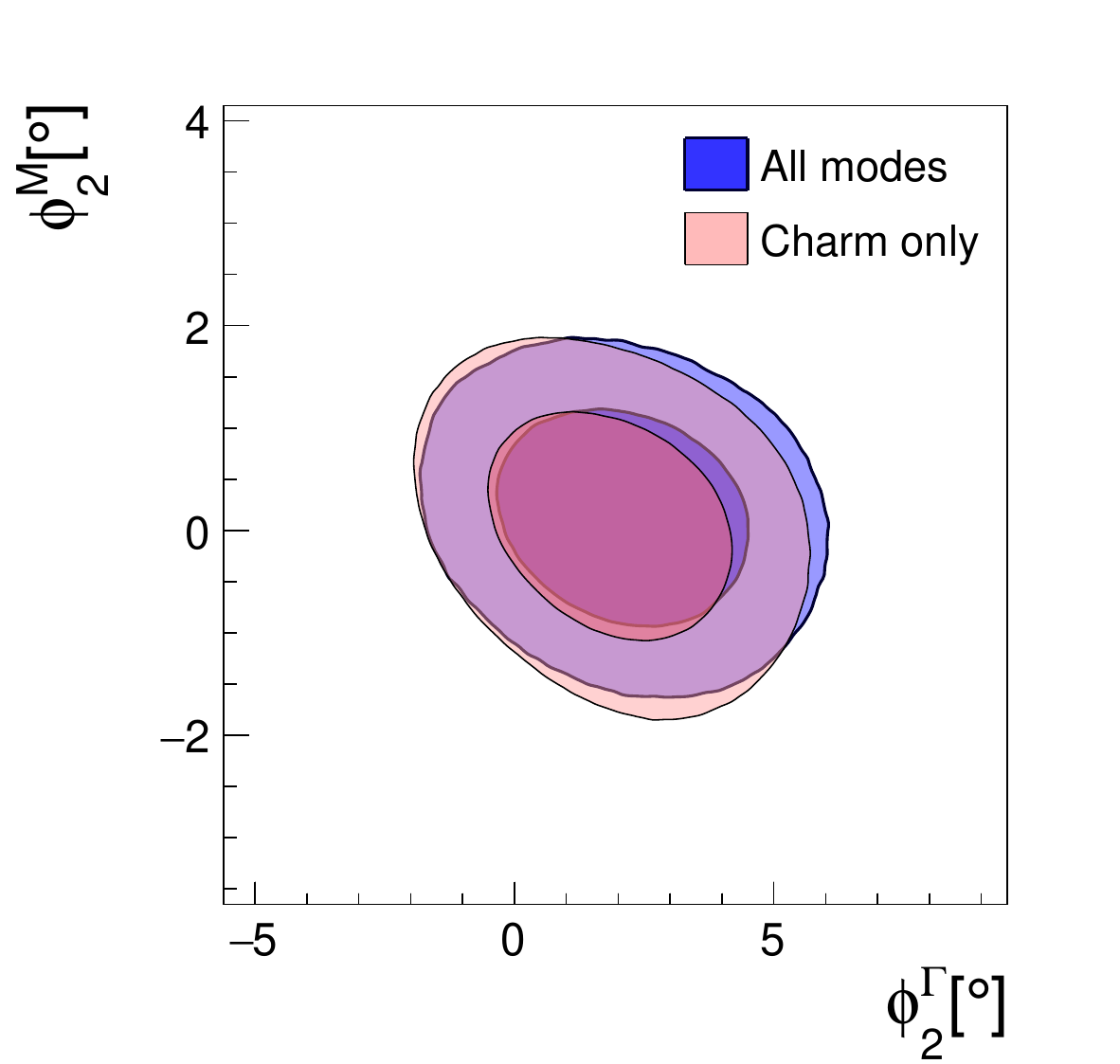}
\caption{Pdfs of the charm mixing and CP-violating parameters obtained using all the observables (``All modes'') or using only the charm modes (``Charm only''). 
Darker (lighter) regions corresponds to $ 68 \%$ ($ 95\%$) probability.} 
\label{Fig:charm_pars_2D}
\end{figure}

\begin{figure}[]
    \centering
\includegraphics[width=0.4\columnwidth]{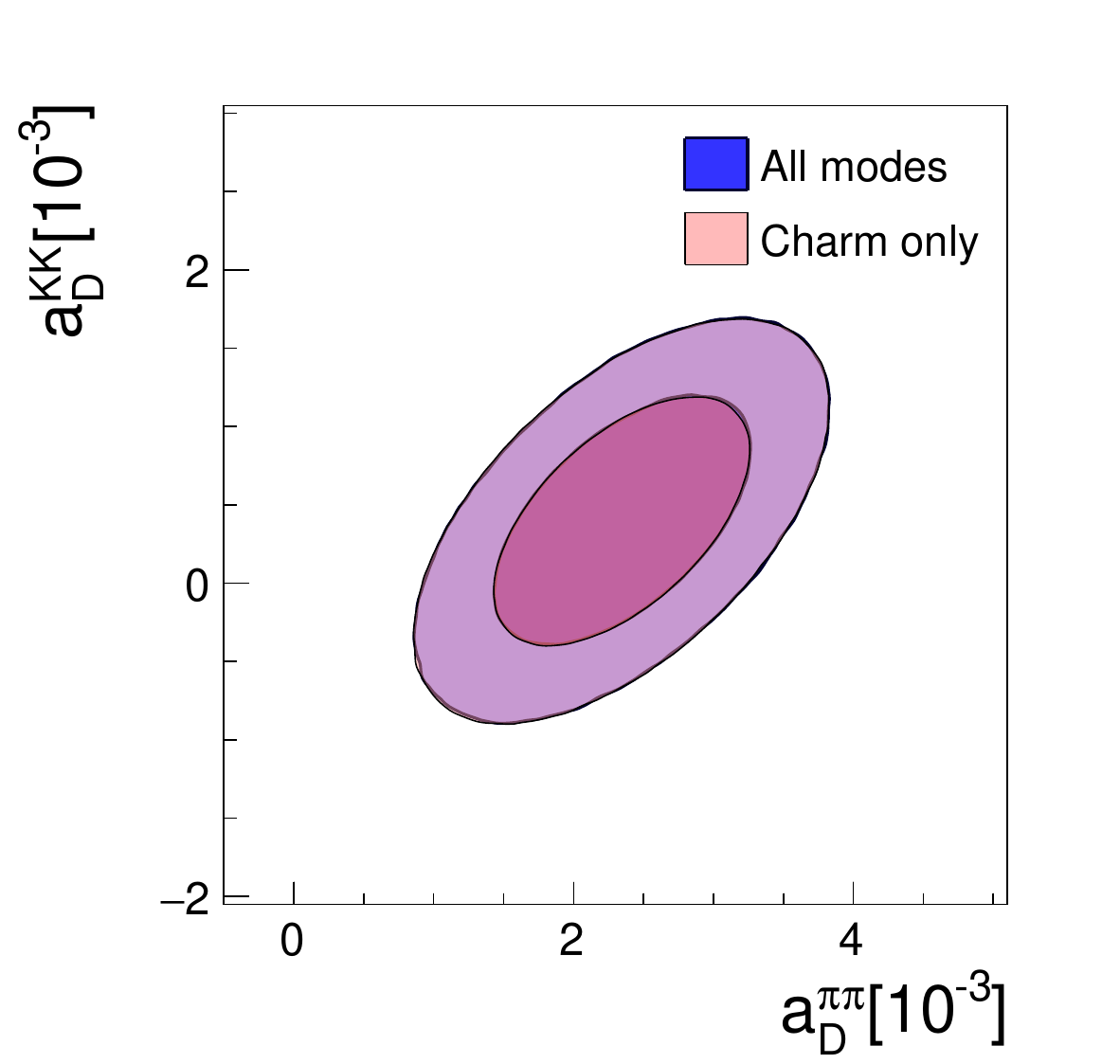}
\caption{Pdf of the parameters describing the direct CP violation in $D^0 \to K^+K^-$ ($a_D^{KK}$) and $D^0 \to \pi^+\pi^-$ ($a_D^{\pi\pi}$) decays. Darker (lighter) regions corresponds to $ 68 \%$ ($ 95\%$) probability.}
\label{Fig:adKK_vs_aDpipi}
\end{figure}

\noindent We report in Tab.~\ref{Tab:results} the charm mixing and CP-violating parameters obtained using all the inputs or excluding the beauty modes, while the corresponding pdfs are shown in Fig.~\ref{Fig:all_1D}. Two-dimensional contours are depicted in Fig.~\ref{Fig:charm_pars_2D}.  
The charm  parameters are compatible with the estimates found by the 2023 analysis by HFLAV \footnote{ See the ``No subleading ampl.~for CF/DCS decays" fit at \scriptsize \url{https://hflav-eos.web.cern.ch/hflav-eos/charm/CKM23/results_mix_cpv.html}.} \cite{HFLAV:2022pwe}. An interesting byproduct of our fit is the value of the direct CP asymmetries in $D^0\to K^+K^-$ and $D^0 \to \pi^+ \pi^-$ decays within the framework of approximate universality:
\begin{equation}
    a_D^{KK} = (4.0 \pm 5.3)\cdot 10^{-4}\,,\qquad a_D^{\pi\pi} = (23.4 \pm 6.0)\cdot 10^{-4}\,,\qquad \rho = 57\%\,.
    \label{Eq:values_direct_CPV}
\end{equation}
We report their two-dimensional pdf in Fig.~\ref{Fig:adKK_vs_aDpipi}. The values in Eq.~(\ref{Eq:values_direct_CPV}) provide evidence for direct CP violation in $D^0 \to \pi^+\pi^-$ at approximately $4\sigma$, together with a sizable deviation from the U-spin expectation of $a_D^{KK} \sim - a_D^{\pi\pi}$. 
Notice that HFLAV and LHCb fit $a_D^{KK}$ and $a_D^{\pi\pi}$ neglecting final-state dependent contributions to the linear parts entering the time-integrated CP asymmetries in Eqs.~(\ref{Eq:integrated_asymmetry}) and~(\ref{Eq:difference_integrated_cp_asymmetry}) (i.e. $\Delta Y^f =\Delta Y$ ). These terms are not known at present and could be non-negligible since they have a relative size of the order of the U-spin breaking parameter with respect to the final-state independent part~\cite{Kagan:2020vri}. On the other hand, the relations we adopted exploit the average between $D^0 \to K^+K^-$ and $D^0 \to \pi^+\pi^-$ decays to get an additional suppression of $\mathcal{O}(\epsilon)$ of the final-state dependent contributions. At this level of precision, the probability intervals found for $a_D^{KK}$ and $a_D^{\pi\pi}$ are compatible between the two fitting strategies, except for small shifts of their central values. As the precision of $A_D^{\text{CP}}(f)$ measurements improves, differences between the fits might become relevant. \\ 
Adding beauty observables to the combination mainly improves the determination of the strong phase $\delta_{D}^{K\pi}$, which in turn improves the determination of $y_{12}$ from  $D \to K \pi$ decays.\footnote{We are indebted to Tommaso Pajero for pointing this out to us.}
\begin{table}[]
\centering
\begin{tabular}{c|cc}
\hline \hline &  &      \\ [-2. ex]
$\mathbf{B}$ \textbf{meson types} & \textbf{Value} & \textbf{Unc.}    \\  [1. ex] \hline &  &      \\ [-2. ex]
$\textbf{All modes}: \ \gamma [\degree]$ & $65.7$ & $\pm 2.5$  \\ [1. ex] \hline &    &    \\ [-2. ex]
$\mathbf{B^{\pm}:} \ \gamma_{B^{\pm}} [\degree]$ & $65.5$ & $\pm 2.7$  \\ [1. ex] \hline &    &    \\ [-2. ex]
 $\mathbf{B^0:} \ \gamma_{B^0} [\degree]$ & $57$ & $\pm 13$ \\ [1. ex ] \hline &    &    \\ [-2. ex]
$\mathbf{B^0_s:} \ \gamma_{B^0_s} [\degree]$ & $77.4$ & $\pm 9.6$  \\ [1. ex]  \hline \hline
\end{tabular}
\caption{Probability intervals for $\gamma$ when using all the measurements and when splitting the inputs of the combination: only charged $B$ measurements 
($\gamma_{B^{\pm}}$), only neutral $B$ ($\gamma_{B^0}$) and $B_{s}$ modes ($\gamma_{B^0_s}$). The central value (Value) and the half-width of the smallest interval containing at least
$68 \%$ probability (Unc.) are reported.}
\label{Tab:gamma_comparison}
\end{table}

\begin{figure}[]
    \centering
\includegraphics[width=0.45\columnwidth]{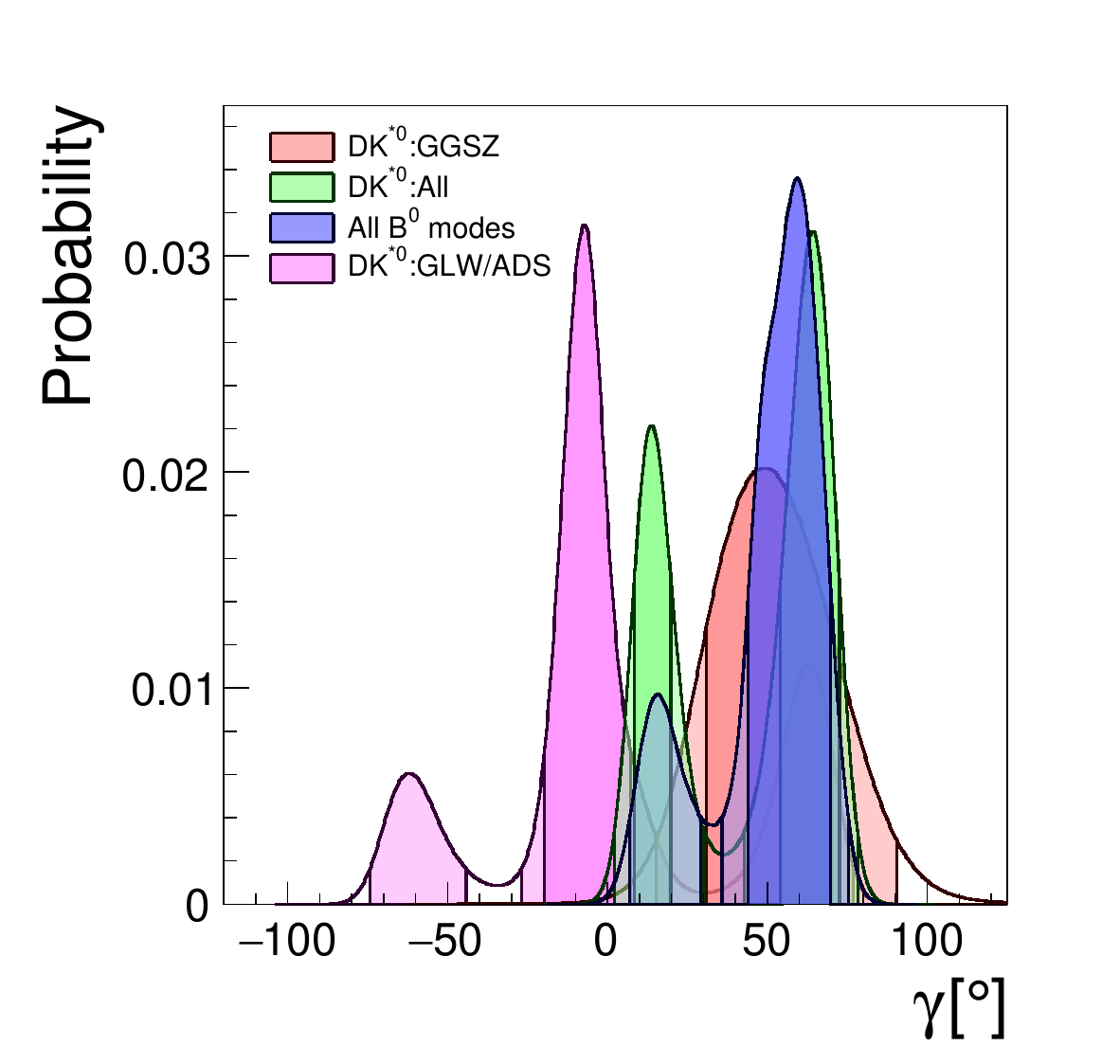}
\includegraphics[width=0.45\columnwidth]{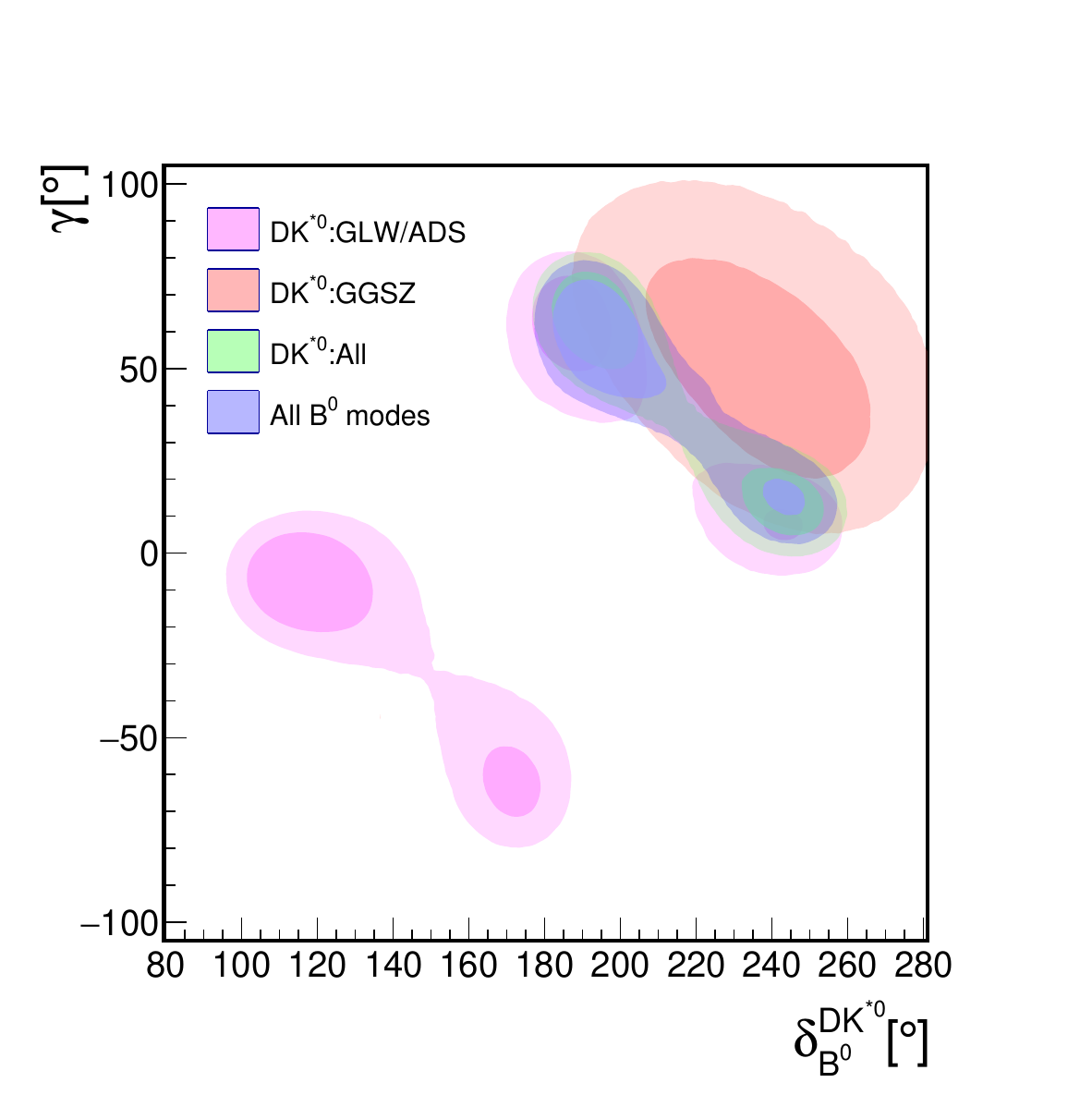}
\caption{Left figure: pdfs of $\gamma$ obtained using different types of $B^0$ observables. Right figure: correlation between $\gamma$ and the strong phase entering $B \to DK^{*0}$ decays. }
\label{Fig:gamma_B0}
\end{figure}

Let us now discuss the results for the CKM angle $\gamma$. 
\begin{figure}[]
    \centering
\includegraphics[width=0.45\columnwidth]{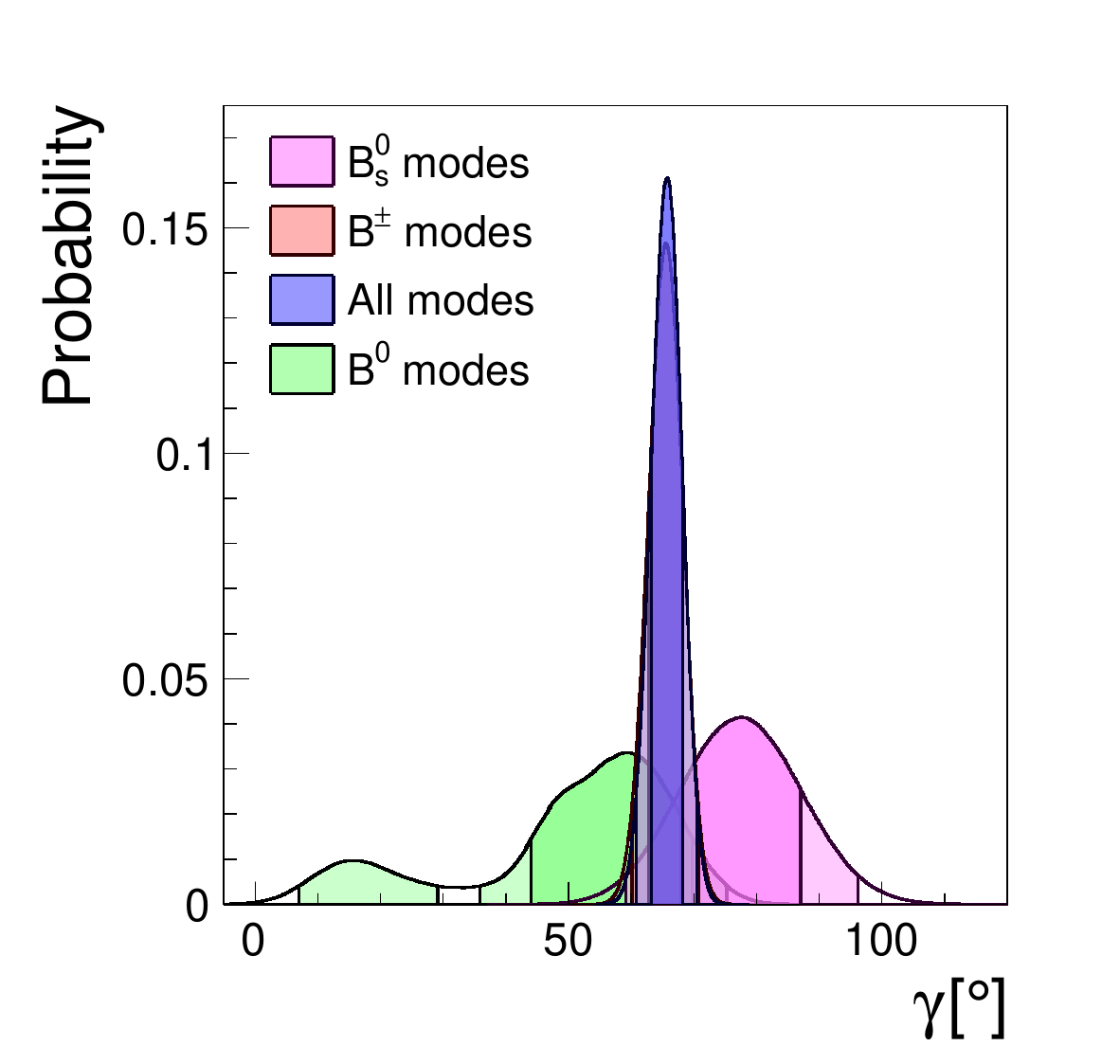}
\includegraphics[width=0.45\columnwidth]{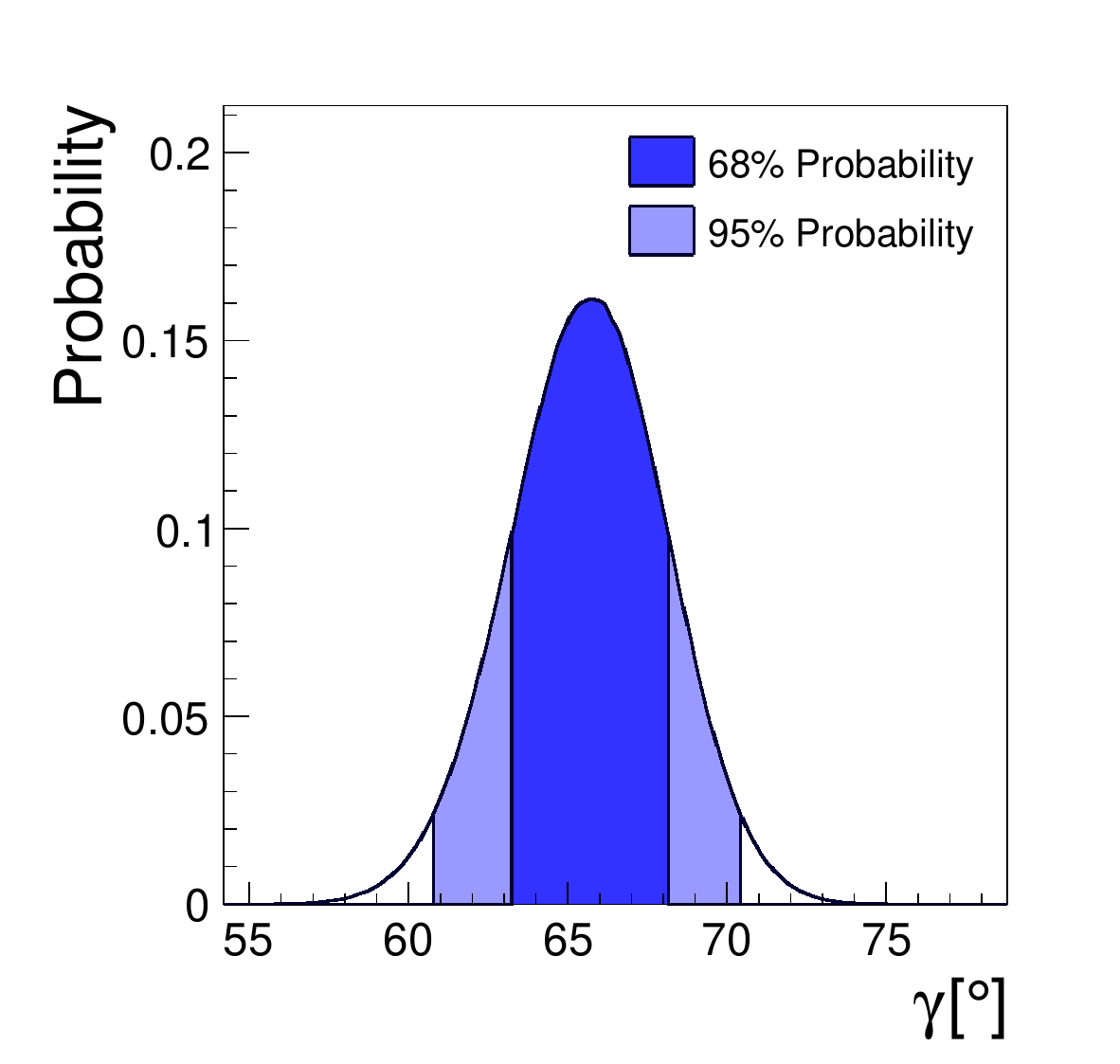}
\caption{Left figure: pdfs of the estimates of $\gamma$ obtained combining all the beauty modes and using separately measurements involving only charged $B$, only $B^0$ or $B^0_s$ mesons. Right figure: pdf of $\gamma$ obtained by combining all the modes.}
\label{Fig:gamma_comparison}
\end{figure}
The estimates extracted using separately measurements of beauty decays with different types of initial  $B$ mesons are reported in Tab.~\ref{Tab:gamma_comparison} and their pdfs are shown in Fig.~\ref{Fig:gamma_comparison};  we also report in Fig.~\ref{Fig:gamma_B0} the pdfs obtained using the different categories of $B^0$ observables.  
 The CKM angle $\gamma$ obtained using all the inputs  falls within the confidence intervals found by the latest LHCb combination \cite{LHCb-CONF-2024-004}, as the other beauty and charm parameters.  
The $\gamma$ estimates extracted from different subsets of beauty observables are all compatible with each other, substantially reducing the moderate tension of $2.2 \sigma$ between $\gamma_{B^{\pm}}$ and $\gamma_{B^0}$ observed in previous combinations \cite{LHCb-CONF-2022-003,DiPalma:2024bzk}. 
This change is driven by the most precise measurement 
of $B^0 \to DK^{*0}$ cascade decays recently published  by LHCb \cite{LHCb:2024oco}, which already in their work provided a $\gamma$ estimate compatible with our result. The full results of the fit are reported in Tabs.~\ref{Tab:pars_charm_res} and~\ref{Tab:B_pars_res} and Fig.~\ref{Fig:other_2D}.
Correlation between $D$ meson mixing parameters and the CKM angle $\gamma$ is below $5\%$, as reported in Tab.~\ref{Tab:corr}. Correlations between $\gamma$ and $\phi_{d(s)}$ is $1\%$ ($-2\%$) when combining all the observables, while is $-5\%$ ($-8\%$) when considering only neutral $B^0_{d(s)}$ measurements.

\section{Conclusions}
\label{Sec:Conclusions}
We presented a Bayesian combination of charm and beauty observables to determine the CKM angle $\gamma$ and the neutral $D$ mixing and CP-violating parameters simultaneously in the framework of the SM and employing approximate universality for $D^0-\overline{D^0}$ mixing. \\ 
The charm parameters are given by
\begin{gather*}
   x_{12} \simeq x = (0.401 \pm 0.043) \%, \qquad  y_{12} \simeq  y = (0.610 \pm 0.017) \% , \qquad   \qquad   \phi_2^M = (0.13 \pm 0.70) \degree , \qquad \phi_2^{\Gamma} = (2.1 \pm 1.6) \degree, \\ \\  
 \phi_2 = (-1.5 \pm 1.1) \degree,  \qquad    \qquad  \left \vert q/p \right \vert -1 = (-1.6 \pm 1.5) \%,
\end{gather*}
while the CKM angle $\gamma$ is found to be 
\begin{equation*}
\gamma = (65.7 \pm 2.5) \degree,
\label{Eq:summary_gamma}
\end{equation*}
in perfect agreement with the indirect determination from the Unitary triangle analysis $\gamma_{\mathrm{UT}} = (65.6 \pm 1.4)\degree$ \cite{ICHEP24-Bona}.
The uncertainties on the two CP-violating phases $\phi_2^{M, \Gamma}$ are still one order of magnitude larger than what is needed to test the predictions $\phi_2^M \sim \phi_2^{\Gamma} \sim 0.1 \degree$ obtained from U-spin arguments in Eq.~(\ref{Eq:SM_estimate_phi2MG}), and a factor of seven larger than the upper bound of $\left \vert \phi_2^\Gamma \right \vert < 0.3 \degree$ in Eq.~(\ref{Eq:alter_phi2G}).
The direct CP asymmetries in $D^0\to K^+K^-$ and $D^0 \to \pi^+ \pi^-$ decays within the framework of approximate universality are given by
\begin{equation}
    a_D^{KK} = (4.0 \pm 5.3)\cdot 10^{-4}\,,\qquad a_D^{\pi\pi} = (23.4 \pm 6.0)\cdot 10^{-4}\,.
\end{equation}
To check the consistency of our combination of measurements sensitive to the angle $\gamma$, we extracted the value of $\gamma$ separately from charged $B$, $B^0$ and $B^0_s$ meson decays.
We found three compatible estimates, given by  
\begin{equation*}
\gamma_{B^{\pm}} = (65.5 \pm 2.7) \degree ,  \qquad \qquad \gamma_{B^0} = (57 \pm 13)\degree, \qquad \qquad \gamma_{B^0_s} = (77.4 \pm 9.6) \degree.
\end{equation*}

\vspace{10pt}
\begin{acknowledgments}
We warmly thank F.~Blanc, A.~Gilman, M.~Morello and S.~Vecchi for correspondence on LHCb results. We are indebted to T.~Pajero, M.W.~Kenzie, A.~Gilman, J.~Butter and M.P.~Whitehead for pointing out to us several problems in the first version of this manuscript. This work was supported in part
by the European Union - Next Generation EU under italian MUR grant PRIN-2022-RXEZCJ and by the HSE basic research fund. 
\end{acknowledgments}
\vspace{20pt}

\appendix

\section{GLW Observables}
\label{Sec:GLW}
The GLW observables consist of CP-conserving and CP-violating ratios of time-integrated rates of $B$ cascade decays when the final states $f$ are (quasi-)CP eigenstates. \\ 
Here, we report the definitions of the observables in the leftmost column of Tab.~\ref{Tab:ADSGLW} and their expressions up to quadratic corrections in charm mixing and CP-violating parameters, neglecting direct CP violation.  \\ 
The CP asymmetry $A_B^{\text{CP}}(f,h)$ is defined as 
\be
A_B^{\text{CP}}(f,h) = \frac{\Gamma(\overline{B} \to [f]_D h) - \Gamma(B \to [f]_D \overline{h})}{\Gamma(\overline{B} \to [f]_D h) + \Gamma(B \to [f]_D \overline{h})} = \frac{2 r_B^{Dh} \kappa_{B}^{Dh} \sin \gamma \sin \delta_B^{Dh}}{(1+(r_B^{Dh})^2) \frac{1 - (2F_{+}^f - 1) \alpha y}{2F_{+}^{f} -1 - \alpha y} + 2 r_B^{Dh} \kappa_B^{Dh} \cos \gamma \cos \delta_B^{Dh}}.
\label{Eq:B_CP_Asymmetry}
\ee
The ratio of rates for decays in which the $D$ meson is reconstructed as a CP eigenstate and decays with CF final states $f^*$, as $K^-\pi^+ (\pi^0)$ or $K^-\pi^+\pi^+\pi^-$, is given by 
\be
R(f,f^*,h) = \frac{\mathcal{B}(D^0 \to f^*)}{\mathcal{B}(D^0 \to f)} \frac{\Gamma(\overline{B} \to [f]_D h) + \Gamma(B \to [f]_D \overline{h})}{\Gamma(\overline{B} \to [f^*]_D h) + \Gamma(B \to [\overline{f^*}]_D \overline{h}) }.
\label{Eq:B_charge_average}
\ee
In terms of charm parameters, Eq.~(\ref{Eq:B_charge_average}) can be written as 
\be
R(f,f^*,h) = \frac{(1+(r_B^{Dh})^2) (1 - (2F_+^{f}-1) \alpha y) + 2 r_B^{Dh} \kappa_B^{Dh} \cos \gamma \cos \delta_B^{Dh} ( (2 F_+^{f} -1 ) - \alpha y )}{1 + (r_D^{f^*} r_B^{Dh})^2 + 2 r_B^{Dh} \kappa_B^{Dh} r_D^{f^*} \kappa_D^{f^*}  \cos \gamma \cos( \delta_B^{Dh} - \delta_D^{f^*}) + \Gamma_\mathrm{mix}^+(x,y,  \gamma,f^*,h)},
\label{Eq:B_charge_average_pars}
\ee
with $\Gamma^+_{\text{mix}}(x,y,\gamma,f^*,h)$ arising due to the mixing terms in Eq.~(\ref{Eq:MT_gamma}), given by 
\be
\begin{aligned}
\Gamma_\mathrm{mix}^+(x,y,\gamma,f^*,h) =& \  - \alpha y \bigg[ r_D^{f^*} \kappa_D^{f^*} (1 + (r_B^{Dh})^2) \cos \delta_D^{f^*}  +  r_B^{Dh} \kappa_B^{Dh} (1 + (r_D^{f^*})^2 ) \cos \gamma \cos \delta_B^{Dh}  \bigg]  \\ 
        & \   + \alpha x \bigg[ r_B^{Dh} \kappa_B^{Dh} (1 - (r_D^{f^*})^2) \cos \gamma \sin \delta_B^{Dh}  -  r_D^{f^*} \kappa_D^{f^*} (1 - (r_{B}^{Dh})^2) \sin \delta_D^{f^*}    \bigg]. 
\end{aligned}
\label{Eq:B_charge_average_MT}
\ee
Sometimes, it is useful to avoid the dependence on the charm branching fractions in Eq.~(\ref{Eq:B_charge_average}) by measuring directly the so-called double ratio, which is related  to $R(f,f^*,h)$ for two different meson states $h$ and $h'$ as
\be
R^{\text{CP}}(f,f^*,h,h') = \frac{R(f,f^*,h)}{R(f,f^*,h')}.
\label{Eq:RCP}
\ee

\section{ADS Observables}
\label{Sec:ADS}
In this section, we give an overview of the ADS observables considered in the combination in Tab.~\ref{Tab:ADSGLW}, expanding them in terms of the charm parameters and neglecting the non-linear terms, as well as direct CP violation. The ADS observables are defined through time-integrated rates of $B$ cascade decays, involving CF/DCS decays of the $D$ mesons.   As in Sec.~\ref{Sec:2bodyCF/DCS}, we refer to the CF mode of the $D^0$ meson as $f$. \\ 
The simplest observables that can be measured are the CP conjugate suppressed  over favoured   ratios
\be
R^{-}(f,h) = \frac{\Gamma(\overline{B} \to [\overline{f}]_D h)}{\Gamma(\overline{B} \to [f]_D h)}, \qquad \qquad R^{+}(f,h) = \frac{\Gamma(B \to [f]_D \overline{h})}{\Gamma(B \to [\overline{f}]_D \overline{h})}.
\label{Eq:Rpm}
\ee
The rates for the favoured processes are reported in Eq.~(\ref{Eq:B_time_integrated_rate}), while for the suppressed modes, we have 
\be
\Gamma(\overline{B} \to [\overline{f}]_D h) \propto   (r_D^f)^2 + (r_B^{Dh})^2 + 2 r_B^{Dh} r_D^f \kappa_B^{Dh} \kappa_D^f \cos(\delta_B^{Dh} + \delta_D^f - \gamma) + \Gamma_\mathrm{mix}(-x,y,\gamma,f,h),
\label{Eq:B_time_integrated_sup}
\ee
with $\Gamma_\mathrm{mix}(x,y,\gamma,f,h)$ defined in Eq.~(\ref{Eq:MT_gamma}). The rate for the CP conjugate mode can be obtained simply replacing $\gamma$ with $-\gamma$ in Eq.~(\ref{Eq:B_time_integrated_sup}). \\ 
Suppressed (sup) and favoured (fav) decays can be used also to define the following  asymmetries 
\be
A^{\text{sup}}(f,h) = \frac{\Gamma(\overline{B} \to [\overline{f}]_D h) - \Gamma(B \to [f]_D \overline{h})}{\Gamma(\overline{B} \to [\overline{f}]_D h) + \Gamma(B \to [f]_D \overline{h})}, \qquad \qquad  A^{\text{fav}}(f,h) = \frac{\Gamma(\overline{B} \to [f]_D h) - \Gamma(B \to [\overline{f}]_D \overline{h})}{\Gamma(\overline{B} \to [f]_D h) + \Gamma(B \to [\overline{f}]_D \overline{h})}.
\label{Eq:Afav_sup}
\ee
In terms of charm parameters, $A^{\text{sup}}(f,h)$ reads 
\be
A^{\text{sup}}(f,h) = \frac{2 r_D^f r_B^{Dh} \kappa_B^{Dh} \kappa_D^f \sin \gamma \sin(\delta_B^{Dh} + \delta_D^f) + \Gamma_\mathrm{mix}^{-}(x,y,\gamma,f,h)}{(r_B^{Dh})^2 + (r_D^f)^2 + 2 r_D^f r_B^{Dh} \kappa_D^f \kappa_B^{Dh} \cos \gamma \cos(\delta_B^{Dh} + \delta_D^f) + \Gamma_\mathrm{mix}^+(-x,y,\gamma,f,h)},
\label{Eq:Asup_pars}
\ee
where $\Gamma_\mathrm{mix}^+(x,y,\gamma,f,h)$ has been already defined in Eq.~(\ref{Eq:B_charge_average_MT}) for the GLW modes, while $\Gamma_\mathrm{mix}^-(x,y,\gamma,f,h)$ is given by 
\be
\begin{aligned}
\Gamma_\mathrm{mix}^-(x,y,\gamma,f,h) =& - \alpha r_{B}^{Dh} \kappa_B^{Dh} \sin \gamma \bigg[ y \sin \delta_B^{Dh} (1+(r_D^f)^2) - x \cos \delta_B^{Dh} (1 - (r_D^f)^2 ) \bigg]. 
\end{aligned}
\label{Eq:MT'}
\ee
In the same way, we find for the favoured asymmetry:
\be
A^{\text{fav}}(f,h) = \frac{2 r_D^f r_B^{Dh} \kappa_B^{Dh} \kappa_D^f \sin \gamma \sin(\delta_B^{Dh} - \delta_D^f) + \Gamma_\mathrm{mix}^-(-x,y,\gamma,f,h)}{1 + (r_D^{f} r_B^{Dh})^2 + 2 r_B^{Dh} \kappa_B^{Dh} r_D^{f} \kappa_D^{f}  \cos \gamma \cos( \delta_B^{Dh} - \delta_D^{f}) + \Gamma_\mathrm{mix}^+(x,y,\gamma,f,h)}.
\label{Eq:Afav_pars}
\ee
Favoured and suppressed charm modes can be arranged into CP-conserving ratios as 
\be
R^{\text{ADS}}(f,h) = \frac{\Gamma(\overline{B} \to [\overline{f}]_D h) + \Gamma(B \to [f]_D \overline{h})}{\Gamma(\overline{B} \to [f]_D h) + \Gamma(B \to [\overline{f}]_D \overline{h})},
\label{Eq:RADS} 
\ee
or using two different channels for beauty decays, $\overline{B} \to [f]_Dh$ and $\overline{B} \to [f]_Dh'$, through the following observables
\be
R^{\text{sup}}(f,h, h') = \frac{\Gamma(\overline{B} \to [\overline{f}]_D h) + \Gamma(B \to [f]_D \overline{h})}{\Gamma(\overline{B} \to [\overline{f}]_D h') + \Gamma(B \to [f]_D \overline{h'})}, \qquad \qquad R^{\text{fav}}(f,h,h') = \frac{\Gamma(\overline{B} \to [f]_D h) + \Gamma(B \to [\overline{f}]_D \overline{h})}{\Gamma(\overline{B} \to [f]_D h') + \Gamma(B \to [\overline{f}]_D \overline{h'})}.
\label{Eq:Rfavsup}
\ee
The ADS ratio in Eq.~(\ref{Eq:RADS}) can be expressed in terms of charm parameters as 
\be
R^{\text{ADS}}(f,h) = \frac{(r_B^{Dh})^2 + (r_D^f)^2 + 2 r_D^f r_B^{Dh} \kappa_D^f \kappa_B^{Dh} \cos \gamma \cos(\delta_B^{Dh} + \delta_D^f) + \Gamma_\mathrm{mix}^+(-x,y,\gamma,f,h)}{1 + (r_D^{f} r_B^{Dh})^2 + 2 r_B^{Dh} \kappa_B^{Dh} r_D^{f} \kappa_D^{f}  \cos \gamma \cos( \delta_B^{Dh} - \delta_D^{f}) + \Gamma_\mathrm{mix}^+(x,y,\gamma,f,h)}.
\label{Eq:RADS_pars}
\ee
The fit equation for $R^{\text{sup}}(f,h,h')$ is given by 
\be
R^{\text{sup}}(f,h,h') = \frac{\mathcal{B}(\overline{B} \to D^0 h)}{\mathcal{B}(\overline{B} \to D^0 h')} \frac{(r_B^{Dh})^2 + (r_D^f)^2 + 2 r_D^f r_B^{Dh} \kappa_D^f \kappa_B^{Dh} \cos \gamma \cos(\delta_B^{Dh} + \delta_D^f) + \Gamma_\mathrm{mix}^+(-x,y,\gamma,f,h)}{(r_B^{Dh'})^2 + (r_D^f)^2 + 2 r_D^f r_B^{Dh'} \kappa_D^f \kappa_B^{Dh'} \cos \gamma \cos(\delta_B^{Dh'} + \delta_D^f) + \Gamma_\mathrm{mix}^+(-x,y,\gamma,f,h')},
\label{Eq:Rsup_pars}
\ee
while for $R^{\text{fav}}(f,h,h')$ we have 
\be
R^{\text{fav}}(f,h,h') =  \frac{\mathcal{B}(\overline{B} \to D^0 h)}{\mathcal{B}(\overline{B} \to D^0 h')} \frac{1 + (r_D^{f} r_B^{Dh})^2 + 2 r_B^{Dh} \kappa_B^{Dh} r_D^{f} \kappa_D^{f}  \cos \gamma \cos( \delta_B^{Dh} - \delta_D^{f}) + \Gamma_\mathrm{mix}^+(x,y,\gamma,f,h)}{1 + (r_D^{f} r_B^{Dh'})^2 + 2 r_B^{Dh'} \kappa_B^{Dh'} r_D^{f} \kappa_D^{f}  \cos \gamma \cos( \delta_B^{Dh'} - \delta_D^{f}) + \Gamma_\mathrm{mix}^+(x,y,\gamma,f,h')}.
\label{Eq:Rfav_pars}
\ee

\section{Fit results}
\label{Sec:all_results}
\begin{figure}[]
    \centering
\includegraphics[width=0.33\columnwidth]{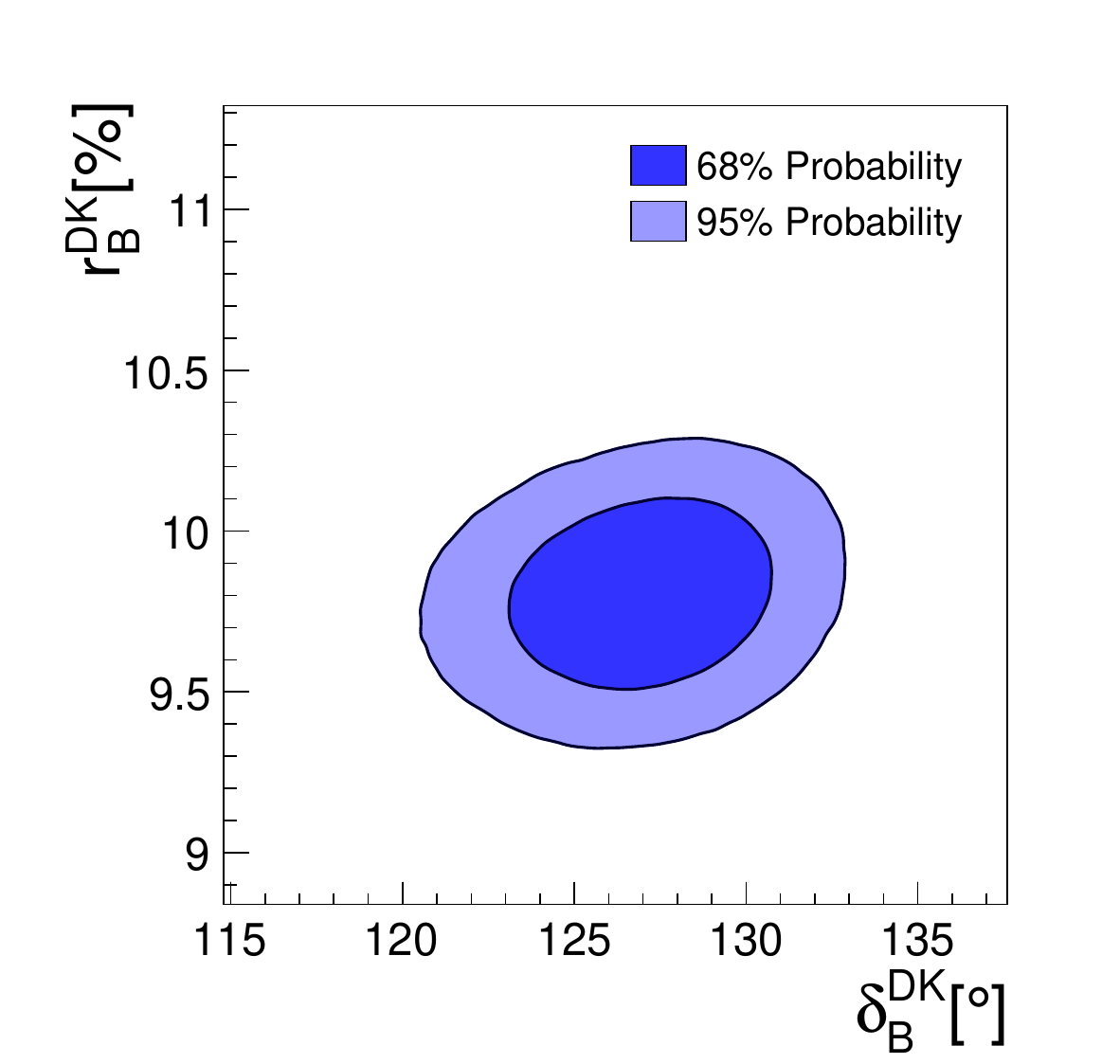}
\includegraphics[width=0.33\columnwidth]{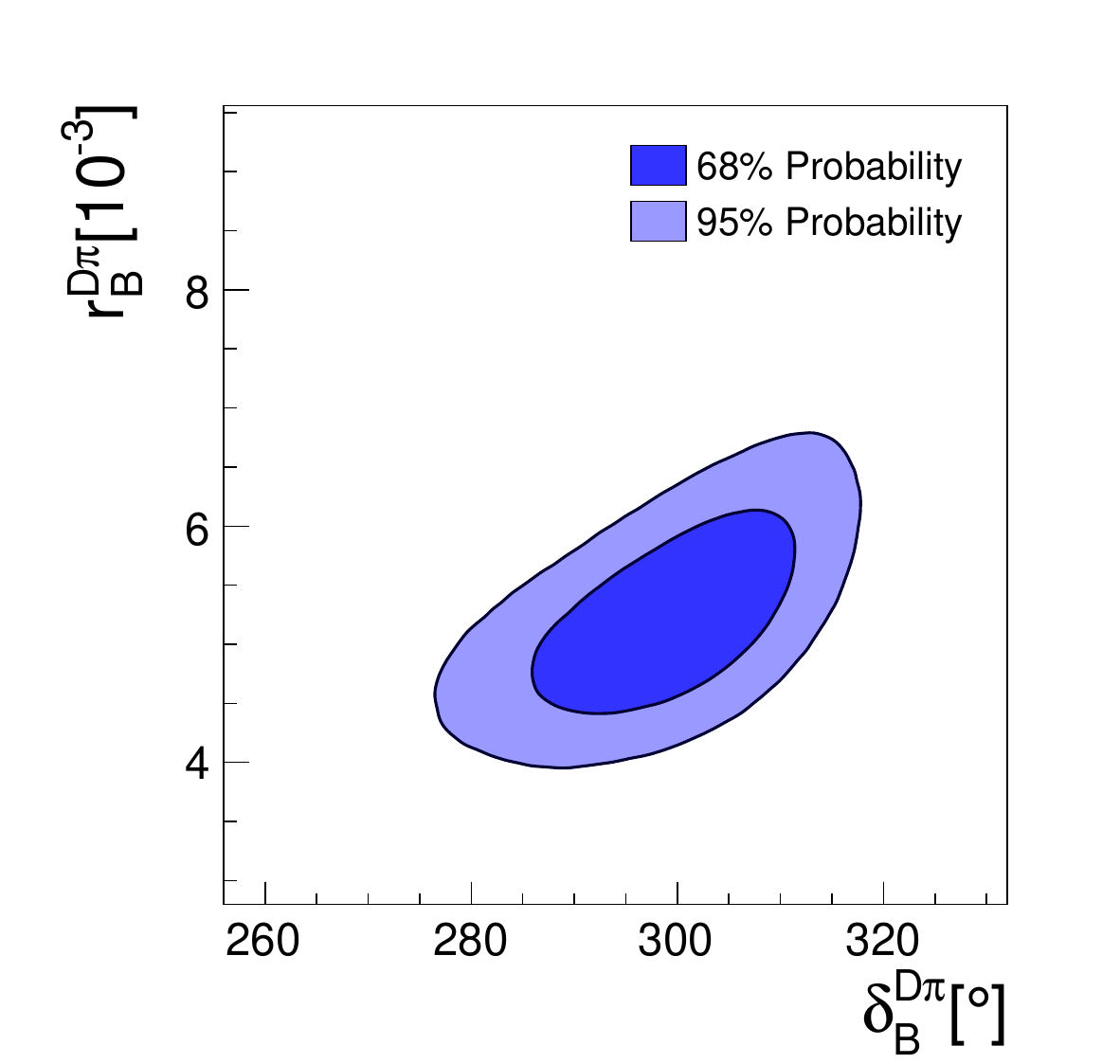}
\caption{Two-dimensional contours of the beauty decay parameters describing the most precise observables of the combination obtained by combining all the modes.}
\label{Fig:other_2D}
\end{figure}
We report the results of the fit obtained by combining all the observables. 
The correlation matrix of the most physically relevant parameters is reported in Tab.~\ref{Tab:corr}. 
Tabs.~\ref{Tab:pars_charm_res}, ~\ref{Tab:B_pars_res}, and~\ref{Tab:taus} present the probability intervals found for the charm and beauty parameters. In Fig.~\ref{Fig:other_2D}, we report two-dimensional contours of the ratio of decay amplitudes and strong phases describing the most precise modes used in the combination: $B^{\pm} \to D \pi^{\pm} (K^{\pm})$.
\begin{table}[] 
\centering 
\resizebox{13cm}{!}{ 
\begin{tabular}{c|ccccccccccccccc} 
\hline \hline &  &  &  &  & &  &       \\ [-2.5 ex] 
 & $\gamma$ & $r_{B}^{DK}$ & $\delta_{B}^{DK}$ & $r_{B}^{D\pi}$ & $\delta_{B}^{D\pi}$ & $r_D^{K\pi}$ & $\delta_D^{K\pi}$ & $x_{12}$ & $y_{12}$ & $\phi_2^M$ & $\phi_2^\Gamma$ & $\phi_2$ & $\vert q/p \vert -1$ & $a_D^{KK}$ & $a_D^{\pi\pi}$  \\ [0.75 ex] \hline \hline &  &  &  &  &  &  & &     \\ [-2.5 ex] 
$\gamma$ & $100$ & $35$ & $58$ & $3$ & $18$ & $1$ & $5$ & $-$  & $5$ & $-1$ & $-1$ & $1$ & $-$  & $-$  & $-$   \\  [0.75 ex] \hline &  &  &  &  &  &  &      \\ [-2.5 ex] 
$r_{B}^{DK}$ &  & $100$ & $17$ & $-6$ & $2$ & $-4$ & $-9$ & $1$ & $-8$ & $1$ & $1$ & $-1$ & $-1$ & $-$  & $-$   \\  [0.75 ex] \hline &  &  &  &  &  &  &      \\ [-2.5 ex] 
$\delta_{B}^{DK}$ &  &  & $100$ & $-11$ & $12$ & $-14$ & $-38$ & $2$ & $-33$ & $4$ & $5$ & $-5$ & $-3$ & $-$  & $-$   \\  [0.75 ex] \hline &  &  &  &  &  &  &      \\ [-2.5 ex] 
$r_{B}^{D\pi}$ &  &  &  & $100$ & $60$ & $7$ & $10$ & $-1$ & $8$ & $-1$ & $-1$ & $1$ & $1$ & $-$  & $-$   \\  [0.75 ex] \hline &  &  &  &  &  &  &      \\ [-2.5 ex] 
$\delta_{B}^{D\pi}$ &  &  &  &  & $100$ & $1$ & $-9$ & $-1$ & $-10$ & $1$ & $2$ & $-2$ & $-1$ & $-$  & $-$   \\  [0.75 ex] \hline &  &  &  &  &  &  &      \\ [-2.5 ex] 
$r_D^{K\pi}$ &  &  &  &  &  & $100$ & $38$ & $36$ & $3$ & $-5$ & $-3$ & $7$ & $-1$ & $-1$ & $-1$  \\  [0.75 ex] \hline &  &  &  &  &  &  &      \\ [-2.5 ex] 
$\delta_D^{K\pi}$ &  &  &  &  &  &  & $100$ & $6$ & $88$ & $-11$ & $-13$ & $14$ & $8$ & $-1$ & $-1$  \\  [0.75 ex] \hline &  &  &  &  &  &  &      \\ [-2.5 ex] 
$x_{12}$ &  &  &  &  &  &  &  & $100$ & $8$ & $1$ & $-1$ & $8$ & $-4$ & $1$ & $1$  \\  [0.75 ex] \hline &  &  &  &  &  &  &      \\ [-2.5 ex] 
$y_{12}$ &  &  &  &  &  &  &  &  & $100$ & $-10$ & $-10$ & $11$ & $6$ & $-1$ & $-$   \\  [0.75 ex] \hline &  &  &  &  &  &  &      \\ [-2.5 ex] 
$\phi_2^M$ &  &  &  &  &  &  &  &  &  & $100$ & $-18$ & $1$ & $53$ & $43$ & $42$  \\  [0.75 ex] \hline &  &  &  &  &  &  &      \\ [-2.5 ex] 
$\phi_2^\Gamma$ &  &  &  &  &  &  &  &  &  &  & $100$ & $-98$ & $-93$ & $3$ & $1$  \\  [0.75 ex] \hline &  &  &  &  &  &  &      \\ [-2.5 ex] 
$\phi_2$ &  &  &  &  &  &  &  &  &  &  &  & $100$ & $84$ & $-11$ & $-9$  \\  [0.75 ex] \hline &  &  &  &  &  &  &      \\ [-2.5 ex] 
$\vert q/p \vert -1$ &  &  &  &  &  &  &  &  &  &  &  &  & $100$ & $14$ & $15$  \\  [0.75 ex] \hline &  &  &  &  &  &  &      \\ [-2.5 ex] 
$a_D^{KK}$ &  &  &  &  &  &  &  &  &  &  &  &  &  & $100$ & $57$  \\  [0.75 ex] \hline &  &  &  &  &  &  &      \\ [-2.5 ex] 
$a_D^{\pi\pi}$ &  &  &  &  &  &  &  &  &  &  &  &  &  &  & $100$  \\  [0.75 ex] \hline \hline \end{tabular} }  
\caption{Correlation matrix for the most significant quantities in percent. Values smaller than $1\%$ are not reported. Large correlations between $\phi_2^{M, \Gamma}$, $|q/p|-1$ and $\phi_2$ are due to the fact that they are not independent parameters, as shown in Eqs.~(\ref{Eq:connection_qop,x,y}) and~(\ref{Eq:phi2}). }
\label{Tab:corr}
\end{table}
\begin{table}[] 
\centering 
\begin{tabular}{cccc|cccc} 
\hline \hline & & & & & & & \\ [-2.75 ex] 
\textbf{Parameter} & \textbf{Value} & \textbf{Unc.} & $\mathbf{95.4 \%}$ \textbf{Prob.} &    \textbf{Parameter} & \textbf{Value} & \textbf{Unc.} &         $\mathbf{95.4 \%}$ \textbf{Prob.} \\ [0.75 ex]                       \hline \hline                    & & & & & & &   \\ [-2.5 ex] 
$\phi_2^M[\degree]$ & $ 0.13 $ & $ \pm 0.70 $ & $[ -1.27 , 1.54 ]$  & $\phi_2^\Gamma[\degree]$ & $ 2.1 $ & $ \pm 1.6 $ & $[ -1.0 , 5.2 ]$  \\  [0.75 ex] \hline & & & & & \\ [-2.75 ex] 
$\phi_2[\degree]$ & $ -1.5 $ & $ \pm 1.1 $ & $[ -3.7 , 0.6 ]$  & $\vert q/p \vert -1[\%]$ & $ -1.6 $ & $ \pm 1.5 $ & $[ -4.5 , 1.4 ]$  \\  [0.75 ex] \hline & & & & & \\ [-2.75 ex] 
$x_{12} \simeq x[\permil]$ & $ 4.01 $ & $ \pm 0.43 $ & $[ 3.16 , 4.83 ]$  & $y_{12} \simeq y[\permil]$ & $ 6.10 $ & $ \pm 0.17 $ & $[ 5.77 , 6.45 ]$  \\  [0.75 ex] \hline & & & & & \\ [-2.75 ex] 
$a_D^{KK}[\permil]$ & $ 0.40 $ & $ \pm 0.53 $ & $[ -0.63 , 1.43 ]$  & $a_D^{\pi\pi}[\permil]$ & $ 2.34 $ & $ \pm 0.60 $ & $[ 1.15 , 3.53 ]$  \\  [0.75 ex] \hline & & & & & \\ [-2.75 ex] 
$\phi_{12}[\degree]$ & $ -2.0 $ & $ \pm 1.8 $ & $[ -5.6 , 1.7 ]$  & $\kappa_D^{K3\pi}$ & $ 0.444 $ & $ \pm 0.092 $ & $[ 0.261 , 0.617 ]$  \\  [0.75 ex] \hline & & & & & \\ [-2.75 ex] 
$r_D^{K3\pi}$ & $ 0.05502 $ & $ \pm 0.00070 $ & $[ 0.05364 , 0.05636 ]$  & $\delta_D^{K3\pi}[\degree]$ & $ 164 $ & $ \pm 21 $ & $[ 126 , 208 ]$  \\  [0.75 ex] \hline & & & & & \\ [-2.75 ex] 
$r_D^{K\pi}[\%]$ & $ 5.8573 $ & $ \pm 0.0094 $ & $[ 5.8387 , 5.8758 ]$  & $\delta_D^{K\pi}[\degree]$ & $ 191.4 $ & $ \pm 2.4 $ & $[ 186.6 , 196.0 ]$  \\  [0.75 ex] \hline & & & & & \\ [-2.75 ex] 
$\kappa_D^{K\pi\pi^0}$ & $ 0.791 $ & $ \pm 0.039 $ & $[ 0.713 , 0.868 ]$  & $r_D^{K\pi\pi^0}$ & $ 0.04433 $ & $ \pm 0.00099 $ & $[ 0.04239 , 0.04626 ]$  \\  [0.75 ex] \hline & & & & & \\ [-2.75 ex] 
$\delta_D^{K\pi\pi^0}[\degree]$ & $ 201.0 $ & $ \pm 7.9 $ & $[ 185.2 , 216.3 ]$  & $\kappa_D^{K^0_SK\pi}$ & $ 0.824 $ & $ \pm 0.094 $ & $[ 0.652 , 0.995 ]$  \\  [0.75 ex] \hline & & & & & \\ [-2.75 ex] 
$r_D^{K^0_SK\pi}$ & $ 0.6192 $ & $ \pm 0.0052 $ & $[ 0.6090 , 0.6297 ]$  & $\delta_D^{K^0_SK\pi}[\degree]$ & $ 13 $ & $ \pm 14 $ & $[ -15 , 39 ]$  \\  [0.75 ex] \hline & & & & & \\ [-2.75 ex] 
$F^{4\pi}_+$ & $ 0.7444 $ & $ \pm 0.0097 $ & $[ 0.7252 , 0.7635 ]$  & $F^{KK\pi\pi}_+$ & $ 0.756 $ & $ \pm 0.013 $ & $[ 0.732 , 0.781 ]$  \\  [0.75 ex] \hline & & & & & \\ [-2.75 ex] 
$F^{\pi\pi\pi^0}_+$ & $ 0.9417 $ & $ \pm 0.0040 $ & $[ 0.9337 , 0.9496 ]$  & $F^{KK\pi^0}_+$ & $ 0.641 $ & $ \pm 0.017 $ & $[ 0.608 , 0.674 ]$  \\  [0.75 ex] \hline & & & & & \\ [-2.75 ex] 
$(\Delta Y^{KK} - \Delta Y^{\pi\pi})[\permil]$ & $ 0.24 $ & $ \pm 0.26 $ & $[ -0.27 , 0.75 ]$  \\ [0.75 ex]  \hline \hline 
\end{tabular} 
\caption{Results for the charm part of the combination. For each of the parameters, we have reported the central value (``Value") and the smallest intervals containing at least $ 68\%$ (``Unc.") and $ 95\%$ probabilities.}
\label{Tab:pars_charm_res}
\end{table} 
\begin{table}[] 
\centering 
\begin{tabular}{c|cccc|cccc}  
& \textbf{Parameter}     & \textbf{Value} & \textbf{Unc.} &  $\mathbf{95.4 \%}$ \textbf{Prob.}  & \textbf{Parameter}     & \textbf{Value} & \textbf{Unc.} &  $\mathbf{95.4 \%}$ \textbf{Prob.}  \\                      [0.5 ex]                         \hline \hline  & & & & & & & &  \\[-2.ex]   
\multirow{2}{*}{ \textbf{CKM Angles}  }  & $\gamma[\degree]$ & $ 65.7 $ & $ \pm 2.5 $ & $[ 60.8 , 70.4 ]$  & $\phi_d/2[\degree]$ & $ 22.13 $ & $ \pm 0.60 $ & $[ 20.96 , 23.32 ]$  \\  & & & & &  & \\[-2.ex]  
 & $\phi_s[\degree]$ & $ -3.36 $ & $ \pm 0.80 $ & $[ -4.93 , -1.79 ]$ \\  [1.ex] \hline & & & & &  & \\[-2.ex]\multirow{6}{*}{$\mathbf{B^{\pm}}$ } & $\kappa_{B}^{DK^{*\pm}}$ & $ 0.950 $ & $ \pm 0.043 $ & $[ 0.852 , 1.001 ]$  & $r_{B}^{DK^{*\pm}}$ & $ 0.0994 $ & $ \pm 0.0090 $ & $[ 0.0806 , 0.1162 ]$  \\  & & & & &  & \\[-2.ex]  
\multirow{6}{*}{ \textbf{Parameters} } & $\delta_{B}^{DK^{*\pm}}[\degree]$ & $ 51 $ & $ \pm 12 $ & $[ 30 , 80 ]$  & $r_{B}^{DK}[\%]$ & $ 9.80 $ & $ \pm 0.20 $ & $[ 9.42 , 10.19 ]$  \\  & & & & &  & \\[-2.ex]  
 & $\delta_{B}^{DK}[\degree]$ & $ 127.0 $ & $ \pm 2.5 $ & $[ 121.9 , 131.8 ]$  & $r_{B}^{D\pi}[\permil]$ & $ 5.18 $ & $ \pm 0.56 $ & $[ 4.17 , 6.38 ]$  \\  & & & & &  & \\[-2.ex]  
 & $\delta_{B}^{D\pi}[\degree]$ & $ 299.8 $ & $ \pm 8.5 $ & $[ 282.0 , 315.1 ]$  & $r_{B}^{D^*K}$ & $ 0.1078 $ & $ \pm 0.0088 $ & $[ 0.0905 , 0.1251 ]$  \\  & & & & &  & \\[-2.ex]  
 & $\delta_{B}^{D^*K}[\degree]$ & $ -47.6 $ & $ \pm 5.9 $ & $[ -60.0 , -36.7 ]$  & $r_{B}^{D^*\pi}$ & $ 0.0058 $ & $ \pm 0.0030 $ & $[ 0.0001 , 0.0123 ]$  \\  & & & & &  & \\[-2.ex]  
 & $\delta_{B}^{D^*\pi}[\degree]$ & $ 39 $ & $ \pm 31 $ & $[ -4 , 133 ]$  & $\frac{\mathcal{B}(B\to DK)}{\mathcal{B}(B\to D\pi)}$ & $ 0.0782 $ & $ \pm 0.0038 $ & $[ 0.0707 , 0.0857 ]$  \\  & & & & &  & \\[-2.ex]  
 & $r_{B}^{DK\pi\pi}$ & $ 0.048 $ & $ \pm 0.031 $ & $[ 0.000 , 0.096 ]$  & $\delta_{B}^{DK\pi\pi}[\degree]$ & \multicolumn{3}{l}{$[-226, -154]\cup[-75, 69]~@~68.5\%$ prob.}   \\  & & & & & &  &  & \\[-2.ex]  
 & $\kappa_{B}^{DK\pi\pi}$ & $ 0.28 $ & $ \pm 0.28 $ & $[ 0.00 , 0.92 ]$  & $r_{B}^{D\pi\pi\pi}$ & $ 0.0099 $ & $ \pm 0.0099 $ & $[ 0.0000 , 0.0365 ]$  \\  & & & & &  & &  &\\[-2.ex]
 & $\delta_{B}^{D\pi\pi\pi}[\degree]$ & \multicolumn{2}{l}{$[-204, -100]\cup[-72, 11]$} & $@~68.9\%$ prob.  & $\kappa_{B}^{D\pi\pi\pi}$ & $ 0.23 $ & $ \pm 0.23 $ & $[ 0.00 , 0.88 ]$  \\  & & & & & & \\[-2.ex]  
\hline\multirow{3}{*}{$\mathbf{B^0}$ } & $r_{B^0}^{D\pi}$ & $ 0.0229 $ & $ \pm 0.0090 $ & $[ 0.0058 , 0.0439 ]$  & $\delta_{B^0}^{D\pi}[\degree]$ & $ 32 $ & $ \pm 31 $ & $[ -43 , 75 ]$  \\  & & & & &  & \\[-2.ex]  
\textbf{Parameters} & $\kappa_{B^0}^{DK^{*0}}$ & $ 0.933 $ & $ \pm 0.025 $ & $[ 0.885 , 0.982 ]$  & $r_{B^0}^{DK^{*0}}$ & $ 0.231 $ & $ \pm 0.015 $ & $[ 0.200 , 0.259 ]$  \\  & & & & &  & \\[-2.ex]  
 & $\delta_{B^0}^{DK^{*0}}[\degree]$ & $ 192.6 $ & $ \pm 6.0 $ & $[ 180.9 , 205.2 ]$  & $r_{B^0}^{D^*\pi}$ & $ 0.0244 $ & $ \pm 0.0083 $ & $[ 0.0099 , 0.0476 ]$  \\  & & & & &  & \\[-2.ex]  
 & $\delta_{B^0}^{D^*\pi}[\degree]$ & $ -34 $ & $ \pm 31 $ & $[ -74 , 41 ]$  & $r_{B^0}^{D\rho}$ & $ 0.060 $ & $ \pm 0.060 $ & $[ 0.000 , 0.261 ]$  \\  & & & & &  & \\[-2.ex]  
 & $\delta_{B^0}^{D\rho}[\degree]$ & $ -77 $ & $ \pm 27 $ & $[ -265 , 93 ]$ \\  [1.ex] \hline & & & & &  & \\[-2.ex]\multirow{4}{*}{} & $r_{B^0_s}^{D_sK}$ & $ 0.331 $ & $ \pm 0.035 $ & $[ 0.264 , 0.402 ]$  & $\delta_{B^0_s}^{D_sK}[\degree]$ & $ -11.0 $ & $ \pm 6.0 $ & $[ -22.8 , 0.8 ]$  \\  & & & & &  & \\[-2.ex]  
$\mathbf{B^0_s}$ & $\kappa_{B^0_s}^{D_sK\pi\pi}$ & $ 0.76 $ & $ \pm 0.14 $ & $[ 0.51 , 1.00 ]$  & $r_{B^0_s}^{D_sK\pi\pi}$ & $ 0.453 $ & $ \pm 0.077 $ & $[ 0.301 , 0.605 ]$  \\  & & & & &  & \\[-2.ex]  
\textbf{Parameters} & $\delta_{B^0_s}^{D_sK\pi\pi}[\degree]$ & $ -14 $ & $ \pm 12 $ & $[ -39 , 10 ]$  & $\kappa_{B^0_s}^{D\overline{K^{*0}}}$ & $ 0.40 $ & $ \pm 0.30 $ & $[ 0.00 , 0.92 ]$  \\  & & & & &  & \\[-2.ex]  
 & $r_{B^0_s}^{D\overline{K^{*0}}}$ & $ 0.057 $ & $ \pm 0.029 $ & $[ 0.000 , 0.098 ]$  & $\delta_{B^0_s}^{D\overline{K^{*0}}}[\degree]$ & $ 72 $ & $ \pm 58 $ & $[ -44 , 215 ]$ \\ [1.ex] \hline \hline 
\end{tabular} 
\caption{Results for the beauty part of the combination. For each of the parameters, we have reported the central value (``Value") and the smallest intervals containing at least $ 68\%$ (``Unc.") and $  95\%$ probabilities. For the phases $\delta_B^{DK\pi\pi}$ and $\delta_B^{D\pi\pi\pi\pi}$, we report the two disjoint intervals at $68.5\%$ and $68.9\%$ probabilities, respectively.}
\label{Tab:B_pars_res}
\end{table} 
\begin{table}[] 
\centering 
\resizebox{\linewidth}{!}{
\begin{tabular}{cccl|cccl} 
\hline \hline & & & & & & & \\ [-2.75 ex] 
\textbf{Parameter} & \textbf{Value} & \textbf{Unc.} & \textbf{Exp.} &    \textbf{Parameter} & \textbf{Value} & \textbf{Unc.} &         \textbf{Exp.} \\ [0.75 ex]                       \hline \hline                    & & & & & & &   \\ [-2.5 ex] 
$(\MEANMOD{\tau}^{KK}_{\Delta A_D^{\mathrm{CP}}} + \MEANMOD{\tau}^{\pi\pi}_{\Delta A_D^{\mathrm{CP}}})/2$ & $1.73$ & $ \pm 0.10 $ & LHCb Run2, $\pi$-tag \cite{LHCb:2019hro}  & $\MEANMOD{\tau}^{KK}_{\Delta A_D^{\mathrm{CP}}} - \MEANMOD{\tau}^{\pi\pi}_{\Delta A_D^{\mathrm{CP}}}$ & $ 0.1350 $ & $ \pm 0.0020 $ &  LHCb Run2, $\pi$-tag \cite{LHCb:2019hro}  \\  [0.75 ex] \hline & & & & & \\ [-2.75 ex] 
$(\MEANMOD{\tau}^{KK}_{\Delta A_D^{\mathrm{CP}}} + \MEANMOD{\tau}^{\pi\pi}_{\Delta A_D^{\mathrm{CP}}})/2$ & $1.210$ & $ \pm 0.010 $ & LHCb Run2, $\mu$-tag \cite{LHCb:2019hro}  & $\MEANMOD{\tau}^{KK}_{\Delta A_D^{\mathrm{CP}}} - \MEANMOD{\tau}^{\pi\pi}_{\Delta A_D^{\mathrm{CP}}}$ & $ -0.0030 $ & $ \pm 0.0010 $ & LHCb Run2, $\mu$-tag \cite{LHCb:2019hro}  \\  [0.75 ex] \hline & & & & & \\ [-2.75 ex] 
$\MEANMOD{t}^{KK}_{A_D^{\mathrm{CP}}}[\mathrm{ps}]$ & $0.7315$ & $ \pm 0.0020 $ & LHCb Run2, $D^{+}$-cal \cite{LHCb:2022lry}  & $\MEANMOD{t}^{KK}_{A_D^{\mathrm{CP}}}[\mathrm{ps}]$ & $0.6868$ & $ \pm 0.0014 $ & LHCb Run2, $D_s^{+}$-cal \cite{LHCb:2022lry}  \\  [0.75 ex] \hline & & & & & \\ [-2.75 ex] 
$\MEANMOD{\tau}^{KK}_{\Delta A_D^{\mathrm{CP}}}$ & $2.152$ & $ \pm 0.016 $ & LHCb Run1, $\pi$-tag \cite{LHCb:2016nxk}  & $\MEANMOD{\tau}^{\pi\pi}_{\Delta A_D^{\mathrm{CP}}}$ & $ 2.037 $ & $ \pm 0.015 $ & LHCb Run1, $\pi$-tag \cite{LHCb:2016nxk}  \\  [0.75 ex] \hline & & & & & \\ [-2.75 ex] 
$\MEANMOD{\tau}^{KK}_{\Delta A_D^{\mathrm{CP}}}$ & $1.0820$ & $ \pm 0.0041 $ & LHCb Run1, $\mu$-tag \cite{LHCb:2016nxk}  & $\MEANMOD{\tau}^{\pi\pi}_{\Delta A_D^{\mathrm{CP}}}$ & $ 1.0679 $ & $ \pm 0.0041 $ & LHCb Run1, $\mu$-tag \cite{LHCb:2016nxk}  \\  [0.75 ex] \hline & & & & & \\ [-2.75 ex] 
$\MEANMOD{\tau}^{KK}_{A_D^{\mathrm{CP}}}$ & $  2.239$ & $ \pm 0.019 $ & LHCb Run1, $\pi$-tag \cite{LHCb:2016nxk}  & $\MEANMOD{\tau}^{KK}_{ A_D^{\mathrm{CP}}}$ & $ 1.0510 $ & $ \pm 0.0041 $ & LHCb Run1, $\mu$-tag \cite{LHCb:2016nxk}  \\  [0.75 ex] \hline & & & & & \\ [-2.75 ex] 
$\MEANMOD{\tau}^{KK}_{A_D^{\mathrm{CP}}}$ & $2.65$ & $ \pm 0.03 $ & CDF \cite{CDF:2011ejf}  & $\MEANMOD{\tau}^{\pi\pi}_{A_D^{\mathrm{CP}}}$ & $ 2.40 $ & $ \pm 0.03 $ & CDF \cite{CDF:2011ejf}  
\\ [0.75 ex]  \hline \hline 
\end{tabular} }
\caption{Parameters corresponding to the average decay times entering the integrated CP asymmetries $A_D^{\mathrm{CP}}(f,t)$ in Eq.~(\ref{Eq:integrated_asymmetry}) and $\Delta A_D^{\mathrm{CP}}$ in Eq.~(\ref{Eq:difference_integrated_cp_asymmetry}). The column \textit{Exp.} lists the name of the experiment, the tagging (-tag) mode  employed, the calibration (-cal) method, and the reference describing the measurement.}
\label{Tab:taus}
\end{table} 

\newpage
\bibliographystyle{JHEP}
\bibliography{hepbiblio}

\providecommand{\href}[2]{#2}\begingroup\raggedright\begin{thebibliography}{100}

\bibitem{Glashow:1970gm}
S.~L. Glashow, J.~Iliopoulos, and L.~Maiani, {\it {Weak Interactions with Lepton-Hadron Symmetry}},  {\em Phys. Rev. D} {\bf 2} (1970) 1285--1292.

\bibitem{Cabibbo:1963yz}
N.~Cabibbo, {\it {Unitary Symmetry and Leptonic Decays}},  {\em Phys. Rev. Lett.} {\bf 10} (1963) 531--533.

\bibitem{Kobayashi:1973fv}
M.~Kobayashi and T.~Maskawa, {\it {CP Violation in the Renormalizable Theory of Weak Interaction}},  {\em Prog. Theor. Phys.} {\bf 49} (1973) 652--657.

\bibitem{Grossman:2006jg}
Y.~Grossman, A.~L. Kagan, and Y.~Nir, {\it {New physics and $CP$ violation in singly Cabibbo suppressed $D$ decays}},  {\em Phys. Rev.} {\bf D75} (2007) 036008, [\href{http://arxiv.org/abs/hep-ph/0609178}{{\tt hep-ph/0609178}}].

\bibitem{Ciuchini:2007cw}
M.~Ciuchini, E.~Franco, D.~Guadagnoli, V.~Lubicz, M.~Pierini, et~al., {\it {$D$ - $\bar{D}$ mixing and new physics: General considerations and constraints on the MSSM}},  {\em Phys.Lett.} {\bf B655} (2007) 162--166, [\href{http://arxiv.org/abs/hep-ph/0703204}{{\tt hep-ph/0703204}}].

\bibitem{Golowich:2007ka}
E.~Golowich, J.~Hewett, S.~Pakvasa, and A.~A. Petrov, {\it {Implications of $D^0$ - $\bar{D}^0$ Mixing for New Physics}},  {\em Phys.Rev.} {\bf D76} (2007) 095009, [\href{http://arxiv.org/abs/0705.3650}{{\tt arXiv:0705.3650}}].

\bibitem{Fajfer:2007dy}
S.~Fajfer, N.~Kosnik, and S.~Prelovsek, {\it {Updated constraints on new physics in rare charm decays}},  {\em Phys.Rev.} {\bf D76} (2007) 074010, [\href{http://arxiv.org/abs/0706.1133}{{\tt arXiv:0706.1133}}].

\bibitem{UTfit:2007eik}
{\bf UTfit} Collaboration, M.~Bona et~al., {\it {Model-independent constraints on $\Delta F=2$ operators and the scale of new physics}},  {\em JHEP} {\bf 03} (2008) 049, [\href{http://arxiv.org/abs/0707.0636}{{\tt arXiv:0707.0636}}].

\bibitem{Artuso:2008vf}
M.~Artuso, B.~Meadows, and A.~A. Petrov, {\it {Charm Meson Decays}},  {\em Ann. Rev. Nucl. Part. Sci.} {\bf 58} (2008) 249--291, [\href{http://arxiv.org/abs/0802.2934}{{\tt arXiv:0802.2934}}].

\bibitem{Brod:2011re}
J.~Brod, A.~L. Kagan, and J.~Zupan, {\it {Size of direct $CP$ violation in singly Cabibbo-suppressed $D$ decays}},  {\em Phys. Rev.} {\bf D86} (2012) 014023, [\href{http://arxiv.org/abs/1111.5000}{{\tt arXiv:1111.5000}}].

\bibitem{Pirtskhalava:2011va}
D.~Pirtskhalava and P.~Uttayarat, {\it {$CP$ Violation and Flavor $SU(3)$ Breaking in $D$-meson Decays}},  {\em Phys. Lett.} {\bf B712} (2012) 81--86, [\href{http://arxiv.org/abs/1112.5451}{{\tt arXiv:1112.5451}}].

\bibitem{Bhattacharya:2012ah}
B.~Bhattacharya, M.~Gronau, and J.~L. Rosner, {\it {CP asymmetries in singly-Cabibbo-suppressed $D$ decays to two pseudoscalar mesons}},  {\em Phys. Rev. D} {\bf 85} (2012) 054014, [\href{http://arxiv.org/abs/1201.2351}{{\tt arXiv:1201.2351}}].

\bibitem{Cheng:2012wr}
H.-Y. Cheng and C.-W. Chiang, {\it {Direct CP violation in two-body hadronic charmed meson decays}},  {\em Phys. Rev. D} {\bf 85} (2012) 034036, [\href{http://arxiv.org/abs/1201.0785}{{\tt arXiv:1201.0785}}]. [Erratum: Phys.Rev.D 85, 079903 (2012)].

\bibitem{Franco:2012ck}
E.~Franco, S.~Mishima, and L.~Silvestrini, {\it {The Standard Model confronts CP violation in $D^0 \to \pi^+\pi^-$ and $D^0 \to K^+K^-$}},  {\em JHEP} {\bf 05} (2012) 140, [\href{http://arxiv.org/abs/1203.3131}{{\tt arXiv:1203.3131}}].

\bibitem{Brod:2012ud}
J.~Brod, Y.~Grossman, A.~L. Kagan, and J.~Zupan, {\it {A Consistent Picture for Large Penguins in D -\ensuremath{>} pi+ pi-, K+ K-}},  {\em JHEP} {\bf 10} (2012) 161, [\href{http://arxiv.org/abs/1203.6659}{{\tt arXiv:1203.6659}}].

\bibitem{Giudice:2012qq}
G.~F. Giudice, G.~Isidori, and P.~Paradisi, {\it {Direct CP violation in charm and flavor mixing beyond the SM}},  {\em JHEP} {\bf 04} (2012) 060, [\href{http://arxiv.org/abs/1201.6204}{{\tt arXiv:1201.6204}}].

\bibitem{Feldmann:2012js}
T.~Feldmann, S.~Nandi, and A.~Soni, {\it {Repercussions of Flavour Symmetry Breaking on CP Violation in D-Meson Decays}},  {\em JHEP} {\bf 06} (2012) 007, [\href{http://arxiv.org/abs/1202.3795}{{\tt arXiv:1202.3795}}].

\bibitem{Keren-Zur:2012buf}
B.~Keren-Zur, P.~Lodone, M.~Nardecchia, D.~Pappadopulo, R.~Rattazzi, and L.~Vecchi, {\it {On Partial Compositeness and the CP asymmetry in charm decays}},  {\em Nucl. Phys. B} {\bf 867} (2013) 394--428, [\href{http://arxiv.org/abs/1205.5803}{{\tt arXiv:1205.5803}}].

\bibitem{Grossman:2019xcj}
Y.~Grossman and S.~Schacht, {\it {The emergence of the $\Delta U=0$ rule in charm physics}},  {\em JHEP} {\bf 07} (2019) 020, [\href{http://arxiv.org/abs/1903.10952}{{\tt arXiv:1903.10952}}].

\bibitem{Dery:2019ysp}
A.~Dery and Y.~Nir, {\it {Implications of the LHCb discovery of CP violation in charm decays}},  {\em JHEP} {\bf 12} (2019) 104, [\href{http://arxiv.org/abs/1909.11242}{{\tt arXiv:1909.11242}}].

\bibitem{Buras:2021rdg}
A.~J. Buras, P.~Colangelo, F.~De~Fazio, and F.~Loparco, {\it {The charm of 331}},  {\em JHEP} {\bf 10} (2021) 021, [\href{http://arxiv.org/abs/2107.10866}{{\tt arXiv:2107.10866}}].

\bibitem{Schacht:2021jaz}
S.~Schacht and A.~Soni, {\it {Enhancement of charm CP violation due to nearby resonances}},  {\em Phys. Lett. B} {\bf 825} (2022) 136855, [\href{http://arxiv.org/abs/2110.07619}{{\tt arXiv:2110.07619}}].

\bibitem{Schacht:2022kuj}
S.~Schacht, {\it {A U-spin anomaly in charm CP violation}},  {\em JHEP} {\bf 03} (2023) 205, [\href{http://arxiv.org/abs/2207.08539}{{\tt arXiv:2207.08539}}].

\bibitem{E791:1996klq}
{\bf E791} Collaboration, E.~M. Aitala et~al., {\it {Search for D0 - anti-D0 mixing in semileptonic decay modes}},  {\em Phys. Rev. Lett.} {\bf 77} (1996) 2384--2387, [\href{http://arxiv.org/abs/hep-ex/9606016}{{\tt hep-ex/9606016}}].

\bibitem{CLEO:2005oam}
{\bf CLEO} Collaboration, C.~Cawlfield et~al., {\it {Limits on neutral D mixing in semileptonic decays}},  {\em Phys. Rev. D} {\bf 71} (2005) 077101, [\href{http://arxiv.org/abs/hep-ex/0502012}{{\tt hep-ex/0502012}}].

\bibitem{BaBar:2004grg}
{\bf BaBar} Collaboration, B.~Aubert et~al., {\it {Search for $D^0 - \bar{D}^0$ mixing using semileptonic decay modes}},  {\em Phys. Rev. D} {\bf 70} (2004) 091102, [\href{http://arxiv.org/abs/hep-ex/0408066}{{\tt hep-ex/0408066}}].

\bibitem{BaBar:2007fup}
{\bf BaBar} Collaboration, B.~Aubert et~al., {\it {Search for $D^0 - \overline{D^0}$ mixing using doubly flavor tagged semileptonic decay modes}},  {\em Phys. Rev. D} {\bf 76} (2007) 014018, [\href{http://arxiv.org/abs/0705.0704}{{\tt arXiv:0705.0704}}].

\bibitem{Belle:2008qhk}
{\bf Belle} Collaboration, U.~Bitenc et~al., {\it {Improved search for $D^0 - \overline{D^0}$ mixing using semileptonic decays at Belle}},  {\em Phys. Rev. D} {\bf 77} (2008) 112003, [\href{http://arxiv.org/abs/0802.2952}{{\tt arXiv:0802.2952}}].

\bibitem{BaBar:2008xkf}
{\bf BaBar} Collaboration, B.~Aubert et~al., {\it {Measurement of $D^0 - \bar{D}^0$ mixing from a time-dependent amplitude analysis of $D^0 \to K^{+} \pi^{-} \pi^0$ decays}},  {\em Phys. Rev. Lett.} {\bf 103} (2009) 211801, [\href{http://arxiv.org/abs/0807.4544}{{\tt arXiv:0807.4544}}].

\bibitem{BaBar:2010nhz}
{\bf BaBar} Collaboration, P.~del Amo~Sanchez et~al., {\it {Measurement of $D^0 - \overline{D^0}$ mixing parameters using $D^0 \to K_S \pi^+\pi^-$ and $D^0 \to K_S K^+K^-$ decays}},  {\em Phys. Rev. Lett.} {\bf 105} (2010) 081803, [\href{http://arxiv.org/abs/1004.5053}{{\tt arXiv:1004.5053}}].

\bibitem{LHCb:2015lgi}
{\bf LHCb} Collaboration, R.~Aaij et~al., {\it {Model-independent measurement of mixing parameters in $D^{0} \to K_{S}^{0} \pi^{+} \pi^{-}$ decays}},  {\em JHEP} {\bf 04} (2016) 033, [\href{http://arxiv.org/abs/1510.01664}{{\tt arXiv:1510.01664}}].

\bibitem{BaBar:2016kvp}
{\bf BaBar} Collaboration, J.~P. Lees et~al., {\it {Measurement of the neutral $D$ meson mixing parameters in a time-dependent amplitude analysis of the $D^0\to\pi^+\pi^-\pi^0$ decay}},  {\em Phys. Rev. D} {\bf 93} (2016), no.~11 112014, [\href{http://arxiv.org/abs/1604.00857}{{\tt arXiv:1604.00857}}].

\bibitem{CLEO:2012fel}
{\bf CLEO} Collaboration, D.~M. Asner et~al., {\it {Updated Measurement of the Strong Phase in $D^0 \to K^+\pi^-$ Decay Using Quantum Correlations in $e^+e^- \to D^0 \bar{D}^0$ at CLEO}},  {\em Phys. Rev. D} {\bf 86} (2012) 112001, [\href{http://arxiv.org/abs/1210.0939}{{\tt arXiv:1210.0939}}].

\bibitem{CLEO:2012obf}
{\bf CLEO} Collaboration, J.~Insler et~al., {\it {Studies of the decays $D^0 \rightarrow K_S^0K^-\pi^+$ and $D^0 \rightarrow K_S^0K^+\pi^-$}},  {\em Phys. Rev. D} {\bf 85} (2012) 092016, [\href{http://arxiv.org/abs/1203.3804}{{\tt arXiv:1203.3804}}]. [Erratum: Phys.Rev.D 94, 099905 (2016)].

\bibitem{CDF:2013gvz}
{\bf CDF} Collaboration, T.~A. Aaltonen et~al., {\it {Observation of $D^0-\overline{D}^0$ Mixing Using the CDF II Detector}},  {\em Phys. Rev. Lett.} {\bf 111} (2013), no.~23 231802, [\href{http://arxiv.org/abs/1309.4078}{{\tt arXiv:1309.4078}}].

\bibitem{Belle:2014ydf}
{\bf Belle} Collaboration, T.~Peng et~al., {\it {Measurement of $D^0-\bar{D}^0$ mixing and search for indirect CP violation using $D^0\to K_S^0\pi^+\pi^-$ decays}},  {\em Phys. Rev. D} {\bf 89} (2014), no.~9 091103, [\href{http://arxiv.org/abs/1404.2412}{{\tt arXiv:1404.2412}}].

\bibitem{LHCb:2019mxy}
{\bf LHCb} Collaboration, R.~Aaij et~al., {\it {Measurement of the mass difference between neutral charm-meson eigenstates}},  {\em Phys. Rev. Lett.} {\bf 122} (2019), no.~23 231802, [\href{http://arxiv.org/abs/1903.03074}{{\tt arXiv:1903.03074}}].

\bibitem{LHCb:2022cak}
{\bf LHCb} Collaboration, R.~Aaij et~al., {\it {Model-independent measurement of charm mixing parameters in $B^- \to D^0 (\to K_S \pi^+\pi^-)\mu^-\overline{\nu}_{\mu}X$ decays}},  {\em Phys. Rev. D} {\bf 108} (2023), no.~5 052005, [\href{http://arxiv.org/abs/2208.06512}{{\tt arXiv:2208.06512}}].

\bibitem{Malde:2015mha}
S.~Malde, C.~Thomas, G.~Wilkinson, P.~Naik, C.~Prouve, J.~Rademacker, J.~Libby, M.~Nayak, T.~Gershon, and R.~A. Briere, {\it {First determination of the $CP$ content of $D \to \pi^+\pi^-\pi^+\pi^-$ and updated determination of the $CP$ contents of $D \to \pi^+\pi^-\pi^0$ and $D \to K^+K^-\pi^0$}},  {\em Phys. Lett. B} {\bf 747} (2015) 9--17, [\href{http://arxiv.org/abs/1504.05878}{{\tt arXiv:1504.05878}}].

\bibitem{BESIII:2024nnf}
{\bf BESIII} Collaboration, M.~Ablikim et~al., {\it {Measurements of the $CP$-even fractions of $D^0\to\pi^{+}\pi^{-}\pi^{0}$ and $D^0\to K^{+}K^{-}\pi^{0}$ at BESIII}},  \href{http://arxiv.org/abs/2409.07197}{{\tt arXiv:2409.07197}}.

\bibitem{BESIII:2024zco}
{\bf BESIII} Collaboration, M.~Ablikim et~al., {\it {Model-independent determination of the strong-phase difference between $D^0$ and $\bar{D}^0 \to \pi^+\pi^-\pi^+\pi^-$ decays}},  \href{http://arxiv.org/abs/2408.16279}{{\tt arXiv:2408.16279}}.

\bibitem{LHCb:2015lnk}
{\bf LHCb} Collaboration, R.~Aaij et~al., {\it {Studies of the resonance structure in $D^0\to K^0_S K^{\pm}\pi^{\mp}$ decays}},  {\em Phys. Rev. D} {\bf 93} (2016), no.~5 052018, [\href{http://arxiv.org/abs/1509.06628}{{\tt arXiv:1509.06628}}].

\bibitem{BESIII:2025vbt}
{\bf BESIII} Collaboration, M.~Ablikim et~al., {\it {Measurement of the branching fractions of doubly Cabibbo-suppressed $D$ decays}},  \href{http://arxiv.org/abs/2503.19542}{{\tt arXiv:2503.19542}}.

\bibitem{BESIII:2025ypr}
{\bf BESIII} Collaboration, M.~Ablikim et~al., {\it {Measurement of the strong-phase difference between $D^0$ and $\bar{D^0}\to K^+K^-\pi^+\pi^-$ in bins of phase space}},  \href{http://arxiv.org/abs/2502.12873}{{\tt arXiv:2502.12873}}.

\bibitem{LHCb:2016zmn}
{\bf LHCb} Collaboration, R.~Aaij et~al., {\it {First observation of $D^0-\bar D^0$ oscillations in $D^0\to K^+\pi^-\pi^+\pi^-$ decays and measurement of the associated coherence parameters}},  {\em Phys. Rev. Lett.} {\bf 116} (2016), no.~24 241801, [\href{http://arxiv.org/abs/1602.07224}{{\tt arXiv:1602.07224}}].

\bibitem{Betti:2021mpf}
F.~Betti, {\it {A Review of CP Violation Measurements in Charm at LHCb}},  {\em Symmetry} {\bf 13} (2021), no.~8 1482, [\href{http://arxiv.org/abs/2107.05305}{{\tt arXiv:2107.05305}}].

\bibitem{Libby:2014rea}
J.~Libby et~al., {\it {New determination of the $D^{0} \to K^{-} \pi^{+} \pi^{0}$ and $D^{0} \to K^{-} \pi^{+} \pi^{+} \pi^{-}$ coherence factors and average strong-phase differences}},  {\em Phys. Lett. B} {\bf 731} (2014) 197--203, [\href{http://arxiv.org/abs/1401.1904}{{\tt arXiv:1401.1904}}].

\bibitem{Evans:2016tlp}
T.~Evans, S.~Harnew, J.~Libby, S.~Malde, J.~Rademacker, and G.~Wilkinson, {\it {Improved determination of the $D \to K^-\pi^+\pi^+\pi^-$ coherence factor and associated hadronic parameters from a combination of $e^+e^-\to \psi(3770)\to c\bar{c}$ and $pp \to c \bar{c} X$ data}},  {\em Phys. Lett. B} {\bf 757} (2016) 520--527, [\href{http://arxiv.org/abs/1602.07430}{{\tt arXiv:1602.07430}}]. [Erratum: Phys.Lett.B 765, 402--403 (2017)].

\bibitem{BESIII:2021eud}
{\bf BESIII} Collaboration, M.~Ablikim et~al., {\it {Measurement of the $D \to K^-\pi^+\pi^+\pi^-$ and $D \to K^-\pi^+\pi^0$ coherence factors and average strong-phase differences in quantum-correlated ${D\bar{D}}$ decays}},  {\em JHEP} {\bf 05} (2021) 164, [\href{http://arxiv.org/abs/2103.05988}{{\tt arXiv:2103.05988}}].

\bibitem{BESIII:2022qkh}
{\bf BESIII} Collaboration, M.~Ablikim et~al., {\it {Improved measurement of the strong-phase difference $\delta _D^{K\pi }$ in quantum-correlated $D{\bar{D}}$ decays}},  {\em Eur. Phys. J. C} {\bf 82} (2022), no.~11 1009, [\href{http://arxiv.org/abs/2208.09402}{{\tt arXiv:2208.09402}}].

\bibitem{Pajero:2022vev}
T.~Pajero, {\it {Recent advances in charm mixing and CP violation at LHCb}},  {\em Mod. Phys. Lett. A} {\bf 37} (2022), no.~24 2230012, [\href{http://arxiv.org/abs/2208.05769}{{\tt arXiv:2208.05769}}].

\bibitem{Belle:2006ipk}
{\bf Belle} Collaboration, L.~M. Zhang et~al., {\it {Improved constraints on $D^0 - \overline{D^0}$ mixing in $D^0 \to K^+ \pi^-$ decays at BELLE}},  {\em Phys. Rev. Lett.} {\bf 96} (2006) 151801, [\href{http://arxiv.org/abs/hep-ex/0601029}{{\tt hep-ex/0601029}}].

\bibitem{BaBar:2007kib}
{\bf BaBar} Collaboration, B.~Aubert et~al., {\it {Evidence for $D^0 - \overline{D^0}$ Mixing}},  {\em Phys. Rev. Lett.} {\bf 98} (2007) 211802, [\href{http://arxiv.org/abs/hep-ex/0703020}{{\tt hep-ex/0703020}}].

\bibitem{LHCb:2025kch}
{\bf LHCb} Collaboration, R.~Aaij et~al., {\it {Search for charge-parity violation in semileptonically tagged $D^0 \to K^+ \pi^-$ decays}},  {\em JHEP} {\bf 03} (2025) 149, [\href{http://arxiv.org/abs/2501.11635}{{\tt arXiv:2501.11635}}].

\bibitem{LHCb:2024hyb}
{\bf LHCb} Collaboration, R.~Aaij et~al., {\it {Measurement of $D^0-\overline{D}^0$ mixing and search for $CP$ violation with $D^0\rightarrow K^+\pi^-$ decays}},  \href{http://arxiv.org/abs/2407.18001}{{\tt arXiv:2407.18001}}.

\bibitem{E791:1999bzz}
{\bf E791} Collaboration, E.~M. Aitala et~al., {\it {Measurements of lifetimes and a limit on the lifetime difference in the neutral D meson system}},  {\em Phys. Rev. Lett.} {\bf 83} (1999) 32--36, [\href{http://arxiv.org/abs/hep-ex/9903012}{{\tt hep-ex/9903012}}].

\bibitem{FOCUS:2000kxx}
{\bf FOCUS} Collaboration, J.~M. Link et~al., {\it {A Measurement of lifetime differences in the neutral D meson system}},  {\em Phys. Lett. B} {\bf 485} (2000) 62--70, [\href{http://arxiv.org/abs/hep-ex/0004034}{{\tt hep-ex/0004034}}].

\bibitem{CLEO:2001lgl}
{\bf CLEO} Collaboration, S.~E. Csorna et~al., {\it {Lifetime differences, direct CP violation and partial widths in D0 meson decays to K+ K- and pi+ pi-}},  {\em Phys. Rev. D} {\bf 65} (2002) 092001, [\href{http://arxiv.org/abs/hep-ex/0111024}{{\tt hep-ex/0111024}}].

\bibitem{Belle:2009xzl}
{\bf Belle} Collaboration, A.~Zupanc et~al., {\it {Measurement of y(CP) in D0 meson decays to the K0(S) K+ K- final state}},  {\em Phys. Rev. D} {\bf 80} (2009) 052006, [\href{http://arxiv.org/abs/0905.4185}{{\tt arXiv:0905.4185}}].

\bibitem{BaBar:2012bho}
{\bf BaBar} Collaboration, J.~P. Lees et~al., {\it {Measurement of $D^0-\bar{D}^0$ Mixing and CP Violation in Two-Body $D^0$ Decays}},  {\em Phys. Rev. D} {\bf 87} (2013), no.~1 012004, [\href{http://arxiv.org/abs/1209.3896}{{\tt arXiv:1209.3896}}].

\bibitem{BESIII:2015ado}
{\bf BESIII} Collaboration, M.~Ablikim et~al., {\it {Measurement of $y_{CP}$ in $D^0-\overline{D}^0$ oscillation using quantum correlations in $e^+e^-\to D^0\overline{D}^0$ at $\sqrt{s}$ = 3.773 GeV}},  {\em Phys. Lett. B} {\bf 744} (2015) 339--346, [\href{http://arxiv.org/abs/1501.01378}{{\tt arXiv:1501.01378}}].

\bibitem{Belle:2015etc}
{\bf Belle} Collaboration, M.~Stari\v{c} et~al., {\it {Measurement of $D^0 - \bar{D}^0$ mixing and search for CP violation in $D^0 \to K^+ K^-, \pi^+ \pi^-$ decays with the full Belle data set}},  {\em Phys. Lett. B} {\bf 753} (2016) 412--418, [\href{http://arxiv.org/abs/1509.08266}{{\tt arXiv:1509.08266}}].

\bibitem{LHCb:2018zpj}
{\bf LHCb} Collaboration, R.~Aaij et~al., {\it {Measurement of the Charm-Mixing Parameter $y_{CP}$}},  {\em Phys. Rev. Lett.} {\bf 122} (2019), no.~1 011802, [\href{http://arxiv.org/abs/1810.06874}{{\tt arXiv:1810.06874}}].

\bibitem{Belle:2019xha}
{\bf Belle} Collaboration, M.~Nayak et~al., {\it {Measurement of the charm-mixing parameter $y_{CP}$ in $D^{0}\to K^{0}_{S}\omega$ decays at Belle}},  {\em Phys. Rev. D} {\bf 102} (2020), no.~7 071102, [\href{http://arxiv.org/abs/1912.10912}{{\tt arXiv:1912.10912}}].

\bibitem{LHCb:2022gnc}
{\bf LHCb} Collaboration, R.~Aaij et~al., {\it {Measurement of the charm mixing parameter~$y_{CP} - y_{CP}^{K\pi}$ using two-body $D^0$ meson decays}},  {\em Phys. Rev. D} {\bf 105} (2022), no.~9 092013, [\href{http://arxiv.org/abs/2202.09106}{{\tt arXiv:2202.09106}}].

\bibitem{CDF:2014wyb}
{\bf CDF} Collaboration, T.~A. Aaltonen et~al., {\it {Measurement of indirect CP-violating asymmetries in $D^0\to K^+K^-$ and $D^0\to \pi^+\pi^-$ decays at CDF}},  {\em Phys. Rev. D} {\bf 90} (2014), no.~11 111103, [\href{http://arxiv.org/abs/1410.5435}{{\tt arXiv:1410.5435}}].

\bibitem{LHCb:2015xyd}
{\bf LHCb} Collaboration, R.~Aaij et~al., {\it {Measurement of indirect $CP$ asymmetries in $D^0\rightarrow K^-K^+$ and $D^0\rightarrow \pi^-\pi^+$ decays using semileptonic $B$ decays}},  {\em JHEP} {\bf 04} (2015) 043, [\href{http://arxiv.org/abs/1501.06777}{{\tt arXiv:1501.06777}}].

\bibitem{LHCb:2017ejh}
{\bf LHCb} Collaboration, R.~Aaij et~al., {\it {Measurement of the $C\!P$ violation parameter $A_\Gamma$ in $D^0 \to K^+K^-$ and $D^0 \to \pi^+\pi^-$ decays}},  {\em Phys. Rev. Lett.} {\bf 118} (2017), no.~26 261803, [\href{http://arxiv.org/abs/1702.06490}{{\tt arXiv:1702.06490}}].

\bibitem{LHCb:2019dom}
{\bf LHCb} Collaboration, R.~Aaij et~al., {\it {Updated measurement of decay-time-dependent CP asymmetries in $D^0 \to K^+K^-$ and $D^0 \to \pi^+\pi^-$ decays}},  {\em Phys. Rev. D} {\bf 101} (2020), no.~1 012005, [\href{http://arxiv.org/abs/1911.01114}{{\tt arXiv:1911.01114}}].

\bibitem{LHCb:2021vmn}
{\bf LHCb} Collaboration, R.~Aaij et~al., {\it {Search for time-dependent $CP$ violation in $D^0 \to K^+ K^-$ and $D^0 \to \pi^+ \pi^-$ decays}},  {\em Phys. Rev. D} {\bf 104} (2021), no.~7 072010, [\href{http://arxiv.org/abs/2105.09889}{{\tt arXiv:2105.09889}}].

\bibitem{LHCb:2016nxk}
{\bf LHCb} Collaboration, R.~Aaij et~al., {\it {Measurement of $CP$ asymmetry in $D^0\rightarrow K^-K^+$ decays}},  {\em Phys. Lett. B} {\bf 767} (2017) 177--187, [\href{http://arxiv.org/abs/1610.09476}{{\tt arXiv:1610.09476}}].

\bibitem{LHCb:2022lry}
{\bf LHCb} Collaboration, R.~Aaij et~al., {\it {Measurement of the Time-Integrated CP Asymmetry in $D^0 \to K^+K^-$ Decays}},  {\em Phys. Rev. Lett.} {\bf 131} (2023), no.~9 091802, [\href{http://arxiv.org/abs/2209.03179}{{\tt arXiv:2209.03179}}].

\bibitem{LHCb:2014kcb}
{\bf LHCb} Collaboration, R.~Aaij et~al., {\it {Measurement of $CP$ asymmetry in $D^0 \rightarrow K^- K^+$ and $D^0 \rightarrow \pi^- \pi^+$ decays}},  {\em JHEP} {\bf 07} (2014) 041, [\href{http://arxiv.org/abs/1405.2797}{{\tt arXiv:1405.2797}}].

\bibitem{LHCb:2016csn}
{\bf LHCb} Collaboration, R.~Aaij et~al., {\it {Measurement of the difference of time-integrated CP asymmetries in $D^0 \rightarrow K^{-} K^{+} $ and $D^0 \rightarrow \pi^{-} \pi^{+} $ decays}},  {\em Phys. Rev. Lett.} {\bf 116} (2016), no.~19 191601, [\href{http://arxiv.org/abs/1602.03160}{{\tt arXiv:1602.03160}}].

\bibitem{LHCb:2019hro}
{\bf LHCb} Collaboration, R.~Aaij et~al., {\it {Observation of CP Violation in Charm Decays}},  {\em Phys. Rev. Lett.} {\bf 122} (2019), no.~21 211803, [\href{http://arxiv.org/abs/1903.08726}{{\tt arXiv:1903.08726}}].

\bibitem{BaBar:2007tfw}
{\bf BaBar} Collaboration, B.~Aubert et~al., {\it {Search for CP violation in the decays $D^0 \to K^{-} K^{+}$ and $D^0 \to \pi^{-} \pi^{+}$}},  {\em Phys. Rev. Lett.} {\bf 100} (2008) 061803, [\href{http://arxiv.org/abs/0709.2715}{{\tt arXiv:0709.2715}}].

\bibitem{Belle:2008ddg}
{\bf Belle} Collaboration, M.~Staric et~al., {\it {Measurement of CP asymmetry in Cabibbo suppressed $D^0$ decays}},  {\em Phys. Lett. B} {\bf 670} (2008) 190--195, [\href{http://arxiv.org/abs/0807.0148}{{\tt arXiv:0807.0148}}].

\bibitem{DiCanto:2012ufu}
{\bf CDF} Collaboration, A.~Di~Canto, {\it {CP Violation in charm decays at CDF}},  {\em Nuovo Cim. C} {\bf 036} (2013), no.~01 26--28, [\href{http://arxiv.org/abs/1208.2517}{{\tt arXiv:1208.2517}}].

\bibitem{CDF:2011ejf}
{\bf CDF} Collaboration, T.~Aaltonen et~al., {\it {Measurement of CP--violating asymmetries in $D^0\to\pi^+\pi^-$ and $D^0\to K^+K^-$ decays at CDF}},  {\em Phys. Rev. D} {\bf 85} (2012) 012009, [\href{http://arxiv.org/abs/1111.5023}{{\tt arXiv:1111.5023}}].

\bibitem{UTfit:2012ich}
{\bf UTfit} Collaboration, A.~J. Bevan et~al., {\it {The UTfit Collaboration Average of D Meson Mixing Data: Spring 2012}},  {\em JHEP} {\bf 10} (2012) 068, [\href{http://arxiv.org/abs/1206.6245}{{\tt arXiv:1206.6245}}].

\bibitem{UTfit:2014hez}
{\bf UTfit} Collaboration, A.~J. Bevan et~al., {\it {The UTfit collaboration average of D meson mixing data: Winter 2014}},  {\em JHEP} {\bf 03} (2014) 123, [\href{http://arxiv.org/abs/1402.1664}{{\tt arXiv:1402.1664}}].

\bibitem{Kagan:2020vri}
A.~L. Kagan and L.~Silvestrini, {\it {Dispersive and absorptive $CP$ violation in $D^0- \overline{D^0}$ mixing}},  {\em Phys. Rev. D} {\bf 103} (2021), no.~5 053008, [\href{http://arxiv.org/abs/2001.07207}{{\tt arXiv:2001.07207}}].

\bibitem{Belle:2006cuz}
{\bf Belle} Collaboration, K.~Abe et~al., {\it {Study of $B^{\pm} \to D_{CP}K^{\pm}$ and $D^*_{CP}K^{*\pm}$ decays}},  {\em Phys. Rev. D} {\bf 73} (2006) 051106, [\href{http://arxiv.org/abs/hep-ex/0601032}{{\tt hep-ex/0601032}}].

\bibitem{Belle:2011ac}
{\bf Belle} Collaboration, Y.~Horii et~al., {\it {Evidence for the Suppressed Decay $B^- \to DK^-, D \to K^+\pi^-$}},  {\em Phys. Rev. Lett.} {\bf 106} (2011) 231803, [\href{http://arxiv.org/abs/1103.5951}{{\tt arXiv:1103.5951}}].

\bibitem{Belle:2013dtr}
{\bf Belle} Collaboration, M.~Nayak et~al., {\it {Evidence for the suppressed decay $B^- \rightarrow DK^-$, $D\rightarrow K^+\pi^- \pi^0$}},  {\em Phys. Rev. D} {\bf 88} (2013), no.~9 091104, [\href{http://arxiv.org/abs/1310.1741}{{\tt arXiv:1310.1741}}].

\bibitem{LHCb:2015dlc}
{\bf LHCb} Collaboration, R.~Aaij et~al., {\it {Study of $B^{-}\to DK^-\pi^+\pi^-$ and $B^-\to D\pi^-\pi^+\pi^-$ decays and determination of the CKM angle $\gamma$}},  {\em Phys. Rev. D} {\bf 92} (2015), no.~11 112005, [\href{http://arxiv.org/abs/1505.07044}{{\tt arXiv:1505.07044}}].

\bibitem{LHCb:2017egy}
{\bf LHCb} Collaboration, R.~Aaij et~al., {\it {Measurement of $CP$ observables in $B^{\pm} \rightarrow D K^{*\pm}$ decays using two- and four-body $D$ final states}},  {\em JHEP} {\bf 11} (2017) 156, [\href{http://arxiv.org/abs/1709.05855}{{\tt arXiv:1709.05855}}]. [Erratum: JHEP 05, 067 (2018)].

\bibitem{LHCb:2016bsl}
{\bf LHCb} Collaboration, R.~Aaij et~al., {\it {Constraints on the unitarity triangle angle $\gamma$ from Dalitz plot analysis of $B^0 \to D K^+ \pi^-$ decays}},  {\em Phys. Rev. D} {\bf 93} (2016), no.~11 112018, [\href{http://arxiv.org/abs/1602.03455}{{\tt arXiv:1602.03455}}]. [Erratum: Phys.Rev.D 94, 079902 (2016)].

\bibitem{LHCb:2020hdx}
{\bf LHCb} Collaboration, R.~Aaij et~al., {\it {Measurement of CP observables in $B^\pm \to D^{(*)} K^\pm$ and $B^\pm \to D^{(*)} \pi^\pm$ decays using two-body $D$ final states}},  {\em JHEP} {\bf 04} (2021) 081, [\href{http://arxiv.org/abs/2012.09903}{{\tt arXiv:2012.09903}}].

\bibitem{LHCb:2020vut}
{\bf LHCb} Collaboration, R.~Aaij et~al., {\it {Measurement of CP observables in B$^{\pm}$ \textrightarrow{} DK$^{\pm}$ and B$^{\pm}$ \textrightarrow{} D\ensuremath{\pi}$^{\pm}$ with D \textrightarrow{} $ {K}_{\mathrm{S}}^0{K}^{\pm }{\pi}^{\mp } $ decays}},  {\em JHEP} {\bf 06} (2020) 058, [\href{http://arxiv.org/abs/2002.08858}{{\tt arXiv:2002.08858}}].

\bibitem{LHCb:2021mmv}
{\bf LHCb} Collaboration, R.~Aaij et~al., {\it {Constraints on the CKM angle $\gamma$ from $B^\pm\rightarrow Dh^\pm$ decays using $D\rightarrow h^\pm h^{\prime\mp}\pi^0$ final states}},  {\em JHEP} {\bf 07} (2022) 099, [\href{http://arxiv.org/abs/2112.10617}{{\tt arXiv:2112.10617}}].

\bibitem{Belle:2023lha}
{\bf Belle, Belle-II} Collaboration, I.~Adachi et~al., {\it {Measurement of CP asymmetries and branching-fraction ratios for B$^{\pm}$\textrightarrow{} DK$^{\pm}$ and D\ensuremath{\pi}$^{\pm}$ with D \textrightarrow{}$ {K}_{\textrm{S}}^0 $K$^{\pm}$\ensuremath{\pi}$^{\mp}$ using Belle and Belle II data}},  {\em JHEP} {\bf 09} (2023) 146, [\href{http://arxiv.org/abs/2306.02940}{{\tt arXiv:2306.02940}}].

\bibitem{Belle:2023yoe}
{\bf Belle, Belle-II} Collaboration, I.~Adachi et~al., {\it {Measurement of branching-fraction ratios and CP asymmetries in B$^{\pm}$ \textrightarrow{} D$_{CP\pm}$K$^{\pm}$ decays at Belle and Belle II}},  {\em JHEP} {\bf 05} (2024) 212, [\href{http://arxiv.org/abs/2308.05048}{{\tt arXiv:2308.05048}}].

\bibitem{LHCb:2024oco}
{\bf LHCb} Collaboration, R.~Aaij et~al., {\it {Study of CP violation in B$^{0}$\textrightarrow{} DK$^{*}$(892)$^{0}$ decays with D \textrightarrow{} K\ensuremath{\pi}(\ensuremath{\pi}\ensuremath{\pi}), \ensuremath{\pi}\ensuremath{\pi}(\ensuremath{\pi}\ensuremath{\pi}), and KK final states}},  {\em JHEP} {\bf 05} (2024) 025, [\href{http://arxiv.org/abs/2401.17934}{{\tt arXiv:2401.17934}}].

\bibitem{LHCb:2022nng}
{\bf LHCb} Collaboration, R.~Aaij et~al., {\it {Measurement of the CKM angle $\gamma$ with $ B^\pm \to D[K^\mp \pi^\pm \pi^\pm \pi^\mp] h^\pm$ decays using a binned phase-space approach}},  {\em JHEP} {\bf 07} (2023) 138, [\href{http://arxiv.org/abs/2209.03692}{{\tt arXiv:2209.03692}}].

\bibitem{LHCb:2024ett}
{\bf LHCb} Collaboration, R.~Aaij et~al., {\it {Measurement of the CKM angle $\gamma$ in $B^{\pm} \to D K^*(892)^{\pm}$ decays}},  \href{http://arxiv.org/abs/2410.21115}{{\tt arXiv:2410.21115}}.

\bibitem{Belle:2010xyn}
{\bf Belle} Collaboration, A.~Poluektov et~al., {\it {Evidence for direct CP violation in the decay $B\to D^{(*)}K$, $D \to K_S \pi^+ \pi^-$ and measurement of the CKM phase $\phi_3$}},  {\em Phys. Rev. D} {\bf 81} (2010) 112002, [\href{http://arxiv.org/abs/1003.3360}{{\tt arXiv:1003.3360}}].

\bibitem{Belle:2019uav}
{\bf Belle} Collaboration, P.~K. Resmi et~al., {\it {First measurement of the CKM angle $\phi_3$ with $B^{\pm}\to D(K_{\rm S}^0\pi^+\pi^-\pi^0)K^{\pm}$ decays}},  {\em JHEP} {\bf 10} (2019) 178, [\href{http://arxiv.org/abs/1908.09499}{{\tt arXiv:1908.09499}}].

\bibitem{LHCb:2020yot}
{\bf LHCb} Collaboration, R.~Aaij et~al., {\it {Measurement of the CKM angle $\gamma$ in $B^\pm\to D K^\pm$ and $B^\pm \to D \pi^\pm$ decays with $D \to K_\mathrm S^0 h^+ h^-$}},  {\em JHEP} {\bf 02} (2021) 169, [\href{http://arxiv.org/abs/2010.08483}{{\tt arXiv:2010.08483}}].

\bibitem{Belle:2021efh}
{\bf Belle, Belle-II} Collaboration, F.~Abudin\'en et~al., {\it {Combined analysis of Belle and Belle II data to determine the CKM angle \ensuremath{\phi}$_{3}$ using B$^{+}$ \textrightarrow{} D($ {K}_S^0 $h$^{+}$h$^{-}$)h$^{+}$ decays}},  {\em JHEP} {\bf 02} (2022) 063, [\href{http://arxiv.org/abs/2110.12125}{{\tt arXiv:2110.12125}}]. [Erratum: JHEP 12, 034 (2022)].

\bibitem{LHCb:2023yjo}
{\bf LHCb} Collaboration, R.~Aaij et~al., {\it {A study of $C\!P$ violation in the decays $B^\pm\to[K^+K^-\pi^+\pi^-]_D h^{\pm}$ ($h = K, \pi$) and $B^\pm\to[\pi^+\pi^-\pi^+\pi^-]_D h^{\pm}$}},  {\em Eur. Phys. J. C} {\bf 83} (2023), no.~6 547, [\href{http://arxiv.org/abs/2301.10328}{{\tt arXiv:2301.10328}}].

\bibitem{LHCb:2023lib}
{\bf LHCb} Collaboration, R.~Aaij et~al., {\it {A model-independent measurement of the CKM angle \ensuremath{\gamma} in partially reconstructed B$^{\pm}$ \textrightarrow{} D$^{*}$h$^{\pm}$ decays with D \textrightarrow{} $ {K}_S^0 $h$^{+}$h$^{-}$ (h = \ensuremath{\pi}, K)}},  {\em JHEP} {\bf 02} (2024) 118, [\href{http://arxiv.org/abs/2311.10434}{{\tt arXiv:2311.10434}}].

\bibitem{LHCb:2023kpr}
{\bf LHCb} Collaboration, R.~Aaij et~al., {\it {Measurement of the CKM angle \ensuremath{\gamma} using the B$^{\pm}$\textrightarrow{} D$^{*}$h$^{\pm}$ channels}},  {\em JHEP} {\bf 12} (2023) 013, [\href{http://arxiv.org/abs/2310.04277}{{\tt arXiv:2310.04277}}].

\bibitem{LHCb:2023ayf}
{\bf LHCb} Collaboration, R.~Aaij et~al., {\it {Measurement of the CKM angle $\gamma $ in the ${{{B} ^0} \rightarrow {D} {{K} ^{*0}}}$ channel using self-conjugate ${D} \rightarrow {{K} ^0_{\textrm{S}}} h^+ h^-$ decays}},  {\em Eur. Phys. J. C} {\bf 84} (2024), no.~2 206, [\href{http://arxiv.org/abs/2309.05514}{{\tt arXiv:2309.05514}}].

\bibitem{LHCb:2018zap}
{\bf LHCb} Collaboration, R.~Aaij et~al., {\it {Measurement of $CP$ violation in $B^{0}\rightarrow D^{\mp}\pi^{\pm}$ decays}},  {\em JHEP} {\bf 06} (2018) 084, [\href{http://arxiv.org/abs/1805.03448}{{\tt arXiv:1805.03448}}].

\bibitem{LHCb:2020qag}
{\bf LHCb} Collaboration, R.~Aaij et~al., {\it {Measurement of the CKM angle $\gamma$ and $B^0_s$-$\bar{B}^0_s$ mixing frequency with $B^0_s \rightarrow D_s^\mp h^\pm \pi^\pm \pi^\mp$ decays}},  {\em JHEP} {\bf 03} (2021) 137, [\href{http://arxiv.org/abs/2011.12041}{{\tt arXiv:2011.12041}}].

\bibitem{LHCB-PAPER-2024-020}
{\bf LHCb} Collaboration, R.~Aaij et~al., {\it \textit{Measurement of CP asymmetry in $B^0_s \to D_s^{\mp}K^{\pm}$ decays}},  2024.
\newblock {LHCb-PAPER-2024-020}.

\bibitem{BaBar:2005jis}
{\bf BaBar} Collaboration, B.~Aubert et~al., {\it {Measurement of time-dependent CP-violating asymmetries and constraints on $\sin(2\beta+\gamma)$ with partial reconstruction of $B \to D^{*\mp} \pi^\pm$ decays}},  {\em Phys. Rev. D} {\bf 71} (2005) 112003, [\href{http://arxiv.org/abs/hep-ex/0504035}{{\tt hep-ex/0504035}}].

\bibitem{BaBar:2006slj}
{\bf BaBar} Collaboration, B.~Aubert et~al., {\it {Measurement of time-dependent CP asymmetries in $B^0 \to D^{(*)\pm} \pi^{\mp}$ and $B^0 \to D^{\pm} \rho^{\mp}$ decays}},  {\em Phys. Rev. D} {\bf 73} (2006) 111101, [\href{http://arxiv.org/abs/hep-ex/0602049}{{\tt hep-ex/0602049}}].

\bibitem{Belle:2006lts}
{\bf Belle} Collaboration, F.~J. Ronga et~al., {\it {Measurements of CP violation in $B^0 \rightarrow D^{*-} \pi^+$ and $B^0 \rightarrow D^{-} \pi^+$ decays}},  {\em Phys. Rev. D} {\bf 73} (2006) 092003, [\href{http://arxiv.org/abs/hep-ex/0604013}{{\tt hep-ex/0604013}}].

\bibitem{Belle:2011dhx}
{\bf Belle} Collaboration, S.~Bahinipati et~al., {\it {Measurements of time-dependent CP asymmetries in $B \to D^{*\mp} \pi^{\pm}$ decays using a partial reconstruction technique}},  {\em Phys. Rev. D} {\bf 84} (2011) 021101, [\href{http://arxiv.org/abs/1102.0888}{{\tt arXiv:1102.0888}}].

\bibitem{BaBar:2007dro}
{\bf BaBar} Collaboration, B.~Aubert et~al., {\it {Measurement of CP Violation Parameters with a Dalitz Plot Analysis of $B^\pm \to D(\pi^+ \pi^- \pi^0) K^\pm$}},  {\em Phys. Rev. Lett.} {\bf 99} (2007) 251801, [\href{http://arxiv.org/abs/hep-ex/0703037}{{\tt hep-ex/0703037}}].

\bibitem{BaBar:2008qcq}
{\bf BaBar} Collaboration, B.~Aubert et~al., {\it {Measurement of Ratios of Branching Fractions and CP-Violating Asymmetries of $B^\pm \to D^{*} K^\pm$ Decays}},  {\em Phys. Rev. D} {\bf 78} (2008) 092002, [\href{http://arxiv.org/abs/0807.2408}{{\tt arXiv:0807.2408}}].

\bibitem{BaBar:2009dzx}
{\bf BaBar} Collaboration, B.~Aubert et~al., {\it {Measurement of CP violation observables and parameters for the decays $B^{\pm} \to D K^{*\pm}$}},  {\em Phys. Rev. D} {\bf 80} (2009) 092001, [\href{http://arxiv.org/abs/0909.3981}{{\tt arXiv:0909.3981}}].

\bibitem{BaBar:2010hvw}
{\bf BaBar} Collaboration, P.~del Amo~Sanchez et~al., {\it {Measurement of CP observables in $B^{\pm} \to D_{CP} K^{\pm}$ decays and constraints on the CKM angle $\gamma$}},  {\em Phys. Rev. D} {\bf 82} (2010) 072004, [\href{http://arxiv.org/abs/1007.0504}{{\tt arXiv:1007.0504}}].

\bibitem{BaBar:2010otv}
{\bf BaBar} Collaboration, P.~del Amo~Sanchez et~al., {\it {Search for $b \to u$ transitions in $B^- \to D K^-$ and $D^* K^-$ decays}},  {\em Phys. Rev. D} {\bf 82} (2010) 072006, [\href{http://arxiv.org/abs/1006.4241}{{\tt arXiv:1006.4241}}].

\bibitem{BaBar:2011rud}
{\bf BaBar} Collaboration, J.~P. Lees et~al., {\it {Search for $b \to u$ Transitions in $B^\pm \to [K^\mp \pi^\pm \pi^0]_D K^\pm$ Decays}},  {\em Phys. Rev. D} {\bf 84} (2011) 012002, [\href{http://arxiv.org/abs/1104.4472}{{\tt arXiv:1104.4472}}].

\bibitem{CDF:2009wnr}
{\bf CDF} Collaboration, T.~Aaltonen et~al., {\it {Measurements of Branching Fraction Ratios and CP Asymmetries in $B^{\pm} \to D_{CP} K^{\pm}$ Decays in Hadron Collisions}},  {\em Phys. Rev. D} {\bf 81} (2010) 031105, [\href{http://arxiv.org/abs/0911.0425}{{\tt arXiv:0911.0425}}].

\bibitem{CDF:2011xrp}
{\bf CDF} Collaboration, T.~Aaltonen et~al., {\it {Measurements of branching fraction ratios and CP-asymmetries in suppressed $B^- \to D(\to K^+ \pi^-)K^-$ and $B^- \to D(\to K^+ \pi^-)\pi^-$ decays}},  {\em Phys. Rev. D} {\bf 84} (2011) 091504, [\href{http://arxiv.org/abs/1108.5765}{{\tt arXiv:1108.5765}}].

\bibitem{Belle:2006lys}
{\bf Belle} Collaboration, A.~Poluektov et~al., {\it {Measurement of $\phi_3$ with Dalitz plot analysis of $B^+ \to D^{(*)} K^{(*)+}$ decay}},  {\em Phys. Rev. D} {\bf 73} (2006) 112009, [\href{http://arxiv.org/abs/hep-ex/0604054}{{\tt hep-ex/0604054}}].

\bibitem{BaBar:2010uep}
{\bf BaBar} Collaboration, P.~del Amo~Sanchez et~al., {\it {Evidence for direct CP violation in the measurement of the Cabibbo-Kobayashi-Maskawa angle $\gamma$ with $B^{\pm} \to D^{(*)} K^{(*)\pm}$ decays}},  {\em Phys. Rev. Lett.} {\bf 105} (2010) 121801, [\href{http://arxiv.org/abs/1005.1096}{{\tt arXiv:1005.1096}}].

\bibitem{Belle:2015roy}
{\bf Belle} Collaboration, K.~Negishi et~al., {\it {First model-independent Dalitz analysis of $B^0 \to DK^{*0}$, $D\to K_S^0\pi^+\pi^-$ decay}},  {\em PTEP} {\bf 2016} (2016), no.~4 043C01, [\href{http://arxiv.org/abs/1509.01098}{{\tt arXiv:1509.01098}}].

\bibitem{LHCb:2021dcr}
{\bf LHCb} Collaboration, R.~Aaij et~al., {\it {Simultaneous determination of CKM angle $\gamma$ and charm mixing parameters}},  {\em JHEP} {\bf 12} (2021) 141, [\href{http://arxiv.org/abs/2110.02350}{{\tt arXiv:2110.02350}}].

\bibitem{LHCb-CONF-2024-004}
{\bf LHCb} Collaboration, {\it {Simultaneous determination of the CKM angle $\gamma$ and parameters related to mixing and $CP$ violation in the charm sector}},  tech. rep., CERN, Geneva, 2024.

\bibitem{Belle-II:2024eob}
{\bf Belle-II, Belle} Collaboration, I.~Adachi et~al., {\it {Determination of the CKM angle $\phi_{3}$ from a combination of Belle and Belle II results}},  \href{http://arxiv.org/abs/2404.12817}{{\tt arXiv:2404.12817}}.

\bibitem{HFLAV:2022pwe}
Y.~Amhis et~al., {\it {Averages of $b$-hadron, $c$-hadron, and $\tau$-lepton properties as of 2021}},  {\em Phys. Rev. D} {\bf 107} (2023) 052008, [\href{http://arxiv.org/abs/2206.07501}{{\tt arXiv:2206.07501}}].

\bibitem{Nir:1992uv}
Y.~Nir, {\it {CP violation}},  {\em Conf. Proc. C} {\bf 9207131} (1992) 81--136.

\bibitem{Branco:1999fs}
G.~C. Branco, L.~Lavoura, and J.~P. Silva, {\it {CP Violation}},  {\em Int.Ser.Monogr.Phys.} {\bf 103} (1999) 1--536.

\bibitem{Grossman:2009mn}
Y.~Grossman, Y.~Nir, and G.~Perez, {\it {Testing New Indirect CP Violation}},  {\em Phys.Rev.Lett.} {\bf 103} (2009) 071602, [\href{http://arxiv.org/abs/0904.0305}{{\tt arXiv:0904.0305}}].

\bibitem{Kagan:2009gb}
A.~L. Kagan and M.~D. Sokoloff, {\it {On Indirect CP Violation and Implications for D0 - anti-D0 and B(s) - anti-B(s) mixing}},  {\em Phys.Rev.} {\bf D80} (2009) 076008, [\href{http://arxiv.org/abs/0907.3917}{{\tt arXiv:0907.3917}}].

\bibitem{LHCb:2024jpt}
{\bf LHCb} Collaboration, R.~Aaij et~al., {\it {Search for time-dependent $CP$ violation in $D^0 \rightarrow \pi^+ \pi^- \pi^0$ decays}},  \href{http://arxiv.org/abs/2405.06556}{{\tt arXiv:2405.06556}}.

\bibitem{Rama:2013voa}
M.~Rama, {\it {Effect of $D-\bar{D}$ mixing in the extraction of gamma with $B^- \to D^0 K^-$ and $B^- \to D^0 \pi^-$ decays}},  {\em Phys. Rev. D} {\bf 89} (2014), no.~1 014021, [\href{http://arxiv.org/abs/1307.4384}{{\tt arXiv:1307.4384}}].

\bibitem{Gronau:1991dp}
M.~Gronau and D.~Wyler, {\it {On determining a weak phase from CP asymmetries in charged B decays}},  {\em Phys. Lett. B} {\bf 265} (1991) 172--176.

\bibitem{Gronau:1990ra}
M.~Gronau and D.~London, {\it {How to determine all the angles of the unitarity triangle from $B_d^0 \to D K_s$ and $B_s^0 \to D^0$}},  {\em Phys. Lett.} {\bf B253} (1991) 483--488.

\bibitem{Atwood:1996ci}
D.~Atwood, I.~Dunietz, and A.~Soni, {\it {Enhanced CP violation with $B \to K D^0 (\bar{D}^0)$ modes and extraction of the CKM angle gamma}},  {\em Phys. Rev. Lett.} {\bf 78} (1997) 3257--3260, [\href{http://arxiv.org/abs/hep-ph/9612433}{{\tt hep-ph/9612433}}].

\bibitem{Atwood:2000ck}
D.~Atwood, I.~Dunietz, and A.~Soni, {\it {Improved Methods for Observing CP Violation in $B^\pm\to KD$~and Measuring the CKM Phase $\gamma$}},  {\em Phys. Rev. D} {\bf 63} (2001) 036005, [\href{http://arxiv.org/abs/hep-ph/0008090}{{\tt hep-ph/0008090}}].

\bibitem{Giri:2003ty}
A.~Giri, Y.~Grossman, A.~Soffer, and J.~Zupan, {\it {Determining gamma using $B^{+-} \to DK^{+-}$ with multibody D decays}},  {\em Phys. Rev.} {\bf D68} (2003) 054018, [\href{http://arxiv.org/abs/hep-ph/0303187}{{\tt hep-ph/0303187}}].

\bibitem{Belle:2004bbr}
{\bf Belle} Collaboration, A.~Poluektov et~al., {\it {Measurement of $\phi_3$ with Dalitz plot analysis of $B^{\pm} \to D^{(*)} K^{\pm}$ decay}},  {\em Phys. Rev. D} {\bf 70} (2004) 072003, [\href{http://arxiv.org/abs/hep-ex/0406067}{{\tt hep-ex/0406067}}].

\bibitem{BPGGSZ_Belle}
A.~Bondar, ``{Proceedings of BINP special analysis meeting on Dalitz analysis}.'' 2002.

\bibitem{Dunietz:2000cr}
I.~Dunietz, R.~Fleischer, and U.~Nierste, {\it {In pursuit of new physics with $B_s$ decays}},  {\em Phys. Rev. D} {\bf 63} (2001) 114015, [\href{http://arxiv.org/abs/hep-ph/0012219}{{\tt hep-ph/0012219}}].

\bibitem{Artuso:2015swg}
M.~Artuso, G.~Borissov, and A.~Lenz, {\it {CP violation in the $B_s^0$ system}},  {\em Rev. Mod. Phys.} {\bf 88} (2016), no.~4 045002, [\href{http://arxiv.org/abs/1511.09466}{{\tt arXiv:1511.09466}}]. [Addendum: Rev.Mod.Phys. 91, 049901 (2019)].

\bibitem{ICHEP24-Bona}
{\bf UT\textit{fit}} Collaboration, M.~Bona, {\it Updates on unitarity triangle fits},  2024.

\bibitem{Ciuchini:2005mg}
M.~Ciuchini, M.~Pierini, and L.~Silvestrini, {\it {The Effect of penguins in the $B_d \to J / \Psi K^0$ CP asymmetry}},  {\em Phys. Rev. Lett.} {\bf 95} (2005) 221804, [\href{http://arxiv.org/abs/hep-ph/0507290}{{\tt hep-ph/0507290}}].

\bibitem{palma_silvestgammaddbar_2024}
R.~Di~Palma and L.~Silvestrini, {\it silvest/{GammaDDbar}: {Code} used for {Summer24} {DDbar} mixing and gamma global analysis},  Aug., 2024.

\bibitem{Caldwell:2008fw}
A.~Caldwell, D.~Kollar, and K.~Kroninger, {\it {BAT: The Bayesian Analysis Toolkit}},  {\em Comput.Phys.Commun.} {\bf 180} (2009) 2197--2209, [\href{http://arxiv.org/abs/0808.2552}{{\tt arXiv:0808.2552}}].

\bibitem{LHCb-CONF-2022-003}
{\bf LHCb} Collaboration, {\it {Simultaneous determination of the CKM angle $\gamma$ and parameters related to mixing and CP violation in the charm sector}},  tech. rep., CERN, Geneva, 2022.

\bibitem{DiPalma:2024bzk}
{\bf UTfit Collaboration} Collaboration, R.~Di~Palma, {\it {Simultaneous determination of the charm mixing and CP-violating parameters together with the CKM angle $\gamma$}},  {\em PoS} {\bf EPS-HEP2023} (2024) 344.

\end{thebibliography}\endgroup

\end{document}